\documentclass{jfm}
\usepackage{graphicx, graphics, subfigure,color}
\usepackage{epstopdf, epsfig}
\usepackage[colorlinks=true, citecolor=blue]{hyperref}
\usepackage{amsmath}

\begin{document}

\newcommand{\bl}[1]{\textcolor{red}{#1}}

\newtheorem{lemma}{Lemma}
\newtheorem{corollary}{Corollary}

\shorttitle{Stability of free viscous films} %for header on odd pages
\shortauthor{A. Choudhury et al.} %for header on even pages

%\title{Effect of variable surface viscosity on the stability of free viscous films}

\title{On the role of variable surface viscosity in free viscous films}

%\title{Stability analysis of free viscous film covered with surface active agents}

\author
 {
 Anjishnu Choudhury\aff{1},
  Venkatesh Kumar Paidi\aff{2} \corresp{Current address: Department of Chemical Engineering and Materials Science, University of Minnesota, Minneapolis, MN, USA},
  Sreeram K. Kalpathy\aff{2},
  \corresp{\email{sreeram@iitm.ac.in}}
  \and 
  Harish N. Dixit\aff{1}
  \corresp{\email{hdixit@iith.ac.in}}
  }

\affiliation
{
\aff{1}
Department of Mechanical and Aerospace Engineering, Indian Institute of Technology Hyderabad, Telangana - 502285, India.
\aff{2}
Department of Metallurgical and Materials Engineering, Indian Institute of Technology Madras, Chennai - 600036, India.
}

\maketitle

\begin{abstract}
The stability of a thin liquid film bounded by two free surfaces is examined in the presence of insoluble surface active agents. The surface active agents not only cause gradients in surface tension, but could also render surface viscosity to be significant, and  variable, as a function of their concentration. A set of three coupled nonlinear evolution equations are derived for the film height, concentration of surface active agents, and the horizontal liquid velocity which governs the dynamics of the free film. Surface tension varies linearly with surfactant concentration and two phenomenological models are used for variation of surface viscosity with concentration. In the first model suited for dilute concentration limit, surface viscosity varies linearly with surfactant concentration, while in the second model, surface viscosity varies nonlinearly with concentration and diverges at a critical concentration which is termed as the `jamming' limit. Linear stability analysis is carried out with both surface viscosity models and reveals the effect of various non-dimensional parameters, specifically the retarding nature of surface viscosity and Marangoni effects on film rupture. An analysis of the `jamming' limit of surfactant concentration reveals that $\Gamma_0^{(nl)}<3 \mathbb{D}/M$ is a sufficient criteria for instability of the system, where $\Gamma_0^{(nl)}$ is a normalized initial surface particle concentration, $\mathbb{D}$ is the disjoining pressure number and $\textit{M}$ is the Marangoni number. Nonlinear simulations suggest film profiles at rupture are also qualitatively different from those reported in earlier studies revealing that free films in the jamming limit are remarkably stable and their free surfaces behave like immobile interfaces consistent with experimental observations. It is shown that rupture times can be arbitrarily increased by tuning initial surfactant concentration on the interface offering a fluid dynamical route to stabilization of thin films. Furthermore, self-similar exponents extracted from our nonlinear simulations are used to explain topological differences between zero and constant surface viscosity in the vicinity of rupture.
\end{abstract}

\section{Introduction}
\label{introduction}
Surface active agents are often used to modify surface rheological properties of dispersed media such as liquid foams, emulsions, and soap films. The interactions of these particle-laden interfaces are of great interest in several industrial applications like foams and detergents, inks \citep{tad}, Pickering emulsions \citep{basav}, groundwater treatment \citep{ye}, and soil remediation \citep{mao}. Often surface active agents become important through their inadvertent presence, for instance, as contaminants in coating or printing processes. They typically alter the wetting characteristics and hydrodynamics of a system by giving rise to interfacial stresses and affecting the surface properties of the system. The relevant surface properties of interest are surface tension, Marangoni effects and concentration-dependent surface viscosity. A unified model that can characterize the interplay between these effects is necessary to truly predict the stability of such systems. As subject of the present work, we formulate and solve a mathematical model that addresses some of these interplays in a thin free liquid film, covered with insoluble surface-active agents at its free surface.

\begin{figure}
\centering
\label{fig1}\includegraphics[trim= 8mm 138mm 5mm 110mm ,clip, width=0.8\textwidth]{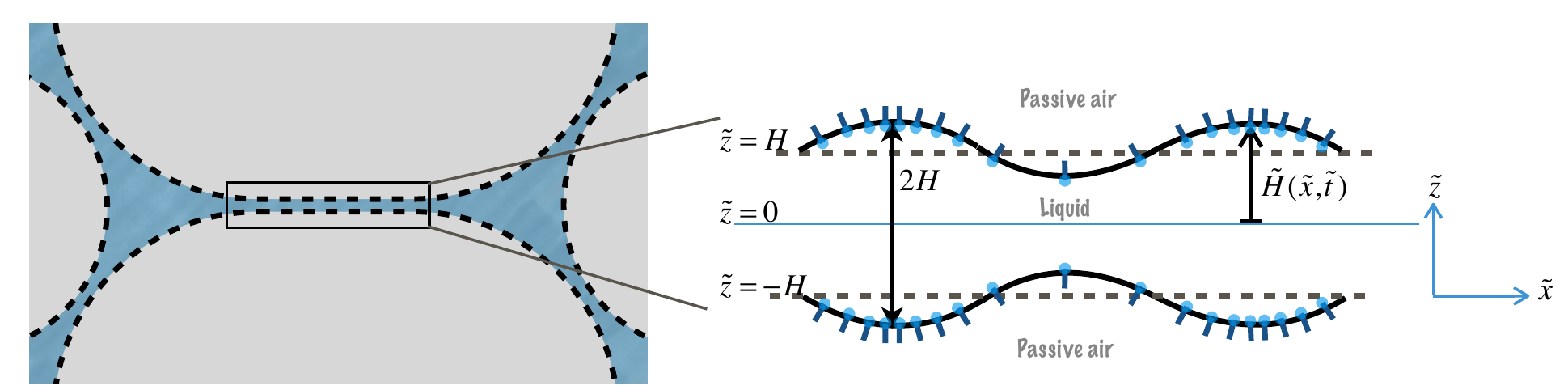}
\caption{Schematic of the problem geometry. A free-film either arises in a typical soap bubble in air or between two bubbles or droplets in an emulsion. Bubbles deform as they approach each other forming a flat free film between them.}
\end{figure}

The configuration of a free liquid film has direct practical relevance to soap bubbles and cosmetic foams. A soap bubble is a thin spherical shell of water with air at a slightly elevated pressure trapped inside. Drainage of the aqueous layer due to gravity would lead to thinning and van der Waals forces, eventually causes rupture. Similar dynamics are seen in Pickering emulsions \citep{ram, pick}. Though the process is conceptually simple, the physics involved in determining the precise bubble breakup time is complicated. The presence of surfactants create Marangoni stresses at the fluid-liquid interface which will delay the breakup time, \citep{gennes}. On the other hand, high pressure inside the bubble can promote rupture, often leading to catastrophic breakup. The idealised model studied here also has great relevance to bubbly suspensions and emulsions, though we do not make a distinction between the two in the current study. The model examined in the present work idealizes the space between two bubbles in an emulsion as a semi-infinite thin film of liquid in the longitudinal direction, with two free surfaces above and below it (See figure \ref{fig1}). The free surfaces are covered with a layer of surface active agents (hereafter simply referred to as `surfactants') that are insoluble in the bulk of the film. Though we use the word `surfactants' to describe any surface active agents, some of the findings of the study is expected to be relevant to interfacial flows with tiny colloidal particles on an interface.
We neglect gravitational effects in the current study owing to the very thin films ($O(100 nm)$) considered here. In such thin films, long range van der Waals forces are included, and for simplicity we ignore any stochastic effects such as Brownian motion. Our configuration differs from that of \citep{naire, naire2, naire3} who studied a vertical free film undergoing gravitational drainage, but owning to very small gravitational effects for such thin films, we believe our work may have relevance even to draining films.

%Parametric studies are reported by examining the dependence of film shape and rupture behaviour on the system parameters.
%%%%%%%%%%%%%%%%%%%%%%%%%%%%%%%%%%%%%%%%%%%%%%%%%%%%%%%%%%%%%%%%%%%
%%%%%%%%%%%%%%%%%%%%%%%%%%%%%%%%%%%%%%%%%%%%%%%%%%%%%%%%%%%%%%%%%%%

Before proceeding further, it is useful to highlight existing literature on modelling stability of thin free liquid films. Several prior investigations \citep{pre, hat, gal, davis, lister} have modelled the dynamics of thin free films in the framework of lubrication theory, accounting for effects of capillarity, evaporation, condensation, and bulk viscosity. Among these, \cite{hat} was the earliest to explore the role of surfactants, which illustrated the conflict between thermally- and surfactant-induced Marangoni flows. In their work, the dynamics of a free film were approximated by those of a wall film with half the thickness of a free film. A detailed presentation of the nonlinear evolution equations using long-wave theory for a clean free film was first given by \cite{davis}. The concept was later extended to include insoluble surfactants at the surface by \cite{dewit}, for which one arrives at a system of three coupled nonlinear evolution equations governing the film dynamics. Furthermore, this work was extended to soluble surfactants by \cite{cho}, but he ignored the role of surface viscosity. It was not until the work of \cite{oron} and later by \cite{naire, naire2} and \cite{naire3} that surface viscosity was included in these models. \cite{naire} considered the configuration of a vertically draining film while \cite{oron} studied a horizontal film without surfactants and Marangoni effects. A model incorporating Marangoni effects, surface viscosity, and bulk solubility of surfactants in horizontal free films was studied by \cite{matar}. However, in all these works, surface viscosity is always assumed to be constant throughout the surface and independent of surfactant concentration.

In the present work, we focus on horizontal non-draining films with surfactant-laden interfaces. We neglect the effect of solubility focussing primarily on surface rheology. In particular, we use a model framework that allows for a variation of surface viscosity as function of the surfactant concentration. Our work primarily stems from a need to describe systems where surfactants or colloidal particles exhibit large concentration variations on the interface. \cite{iva} show that at sufficiently high surfactant concentrations, the interface appears to form a rigid shell accompanied by an enhancement in surface viscosity. Rising spherical bubbles and sedimenting drops covered with surfactants are known to have a drag coefficient different from that of a clean interface. In the limit of Stokes flows, such systems were modelled using the `spherical cap' approximation \citep{leal} where a part of the drop surface is assumed to be rigid (immobile) while the remainder of the drop assumed to be devoid of surfactants \citep{sadh}. Diffusion of surfactants on the interface makes such an approximation unrealistic and one is expected to find a smooth variation in concentration continuously reducing from leeward to windward side of drop/bubble. Thus, in addition to Marangoni effects, we also introduce surface viscosity effects which depend on local concentration of surfactants as demonstrated in prior experiments by \cite{lopez}. This allows us to have a continuous variation in mobility on the interface consistent with diffusion effects of surfactants.  A nonlinear phenomenological model describing this dependence is incorporated in the governing equations to construct a robust framework. 

We adopt a scaling based on viscous forces balancing the inter-molecular van der Waals interactions. A lubrication approximation is taken in the equations owing to the small aspect ratio of the free films. It is found that, variable surface viscosity leads to new self-similar solutions for film evolution in the `dilute'-limit when surface active agents are sparsely distributed across the surface and the sensitivity of surface viscosity toward particle concentrations is weak. In addition, we also find that surfactants could potentially render a free film remarkably stable in the `jamming'-limit.

We first derive a set of governing equations by non-dimensionalizing the Navier-Stokes equations and interfacial boundary conditions in \S\ref{sec2}. We explore the linear stability of the system in \S\ref{sec3} and apply perturbation techniques to reveal an important criteria in \S\ref{sec:LSAjammed}. A brief discussion of numerical methods and the numerical solutions to the governing equations is provided in \S\ref{sec4}. This is followed by self similar solutions found in the `dilute' regime in \S\ref{sec:self}, and important findings in the `jamming' limit in \S\ref{sec:NLAjammed}. Finally, conclusions are made in \S\ref{sec5}, connecting our results to prior experiments and findings.

%We compare and contrast our results with those observed in earlier studies.
 %%%%%%%%%%%%%%%%%%%%%%%%%%%%%%%%%%%%%%%%%%%%%%%%%%%%%%%%%%%%%%%%%%%
%%%%%%%%%%%%%%%%%%%%%%%%%%%%%%%%%%%%%%%%%%%%%%%%%%%%%%%%%%%%%%%%%%%
%%%%%%%%%%%%%%%%%%%%%%%%%%%%%%%%%%%%%%%%%%%%%%%%%%%%%%%%%%%%%%%%%%%
%%%%%%%%%%%%%%%%%%%%%%%%%%%%%%%%%%%%%%%%%%%%%%%%%%%%%%%%%%%%%%%%%%%
%%%%%%%%%%%%%%%%%%%%%%%%%%%%%%%%%%%%%%%%%%%%%%%%%%%%%%%%%%%%%%%%%%%
%%%%%%%%%%%%%%%%%%%%%%%%%%%%%%%%%%%%%%%%%%%%%%%%%%%%%%%%%%%%%%%%%%%
\section{Problem formulation}\label{sec2}
\subsection{Problem geometry and governing equations}
Figure 1 shows the idealized two-dimensional problem setup. A thin film of incompressible
liquid, extending infinitely in the lateral direction, is bounded by a passive gas phase above
and below it. The free surfaces on both sides are covered with surfactants which
are assumed to be insoluble in the bulk of the film. These agents could represent either surfactants in the classical sense, significantly affecting the liquid surface tension, or serve as colloidal particulates which alter the surface rheology. The film has mean thickness $2H$ (dimensional), and the liquid has a bulk viscosity and density represented by $\mu$ and $\rho$ (both dimensional) respectively. The free surface is at a height $\tilde{h}(\tilde{x}, \tilde{t})$. Further, we let $\tilde{\sigma}_{0}$ denote
the dimensional mean surface tension of the liquid film when the surface concentration of surfactants $\tilde{\Gamma}$ is maintained at at a fixed reference concentration $\tilde{\Gamma}_{0}$. Table \ref{table:parameterset} gives some realistic dimensional values for the various physical parameters used in our study. 
\begin{table}
\begin{tabular}{p{3cm} p{6cm} c r}
\textit{Parameter} & \textit{Definition} & \textit{Estimate} & \textit{units} \\ [0.5ex] 
$H$ & Mean film thickness & $10^{-9}-10^{-7}$&$m$ \\ [1ex]
%\hline
$\mu$ & Liquid viscosity & $10^{-3}$&$kg\hspace{1mm}m^{-1}\hspace{0.5mm}s^{-1}$  \\[1ex]
%\hline
$\rho$ & Liquid density &  $10^{3} $&$kg\hspace{1mm}m^{-3}$ \\[1ex]
%\hline
$A$ & Hamaker constant & $10^{-21}-10^{-19}$&$J$  \\[1ex]
%\hline
$\tilde{\sigma_0}$ & Surface tension of clean film & $10^{-3}-10^{-2}$&$N\hspace{1mm}m^{-1}$ \\ [1ex]
%\hline
$\tilde{\Gamma_0}$ & Mean surfactant concentration & $10^{-6}$&$mol\hspace{1mm}m^{-2}$ \\ [1ex]
%\hline
$D_s$ & Surface diffusion coefficient & $10^{-12}-10^{-8} $&$m^2\hspace{1mm}s^{-1}$ \\ [1ex]
%\hline
$\tilde{\eta_0}$ & Surface viscosity & $10^{-12}-10^{-6}$&$kg\hspace{1mm}s^{-1}$ \\ [1ex] 

\end{tabular}
\caption{Estimates of relevant physical parameters}
\label{table:parameterset}
\end{table}

Following prior work \citep{dewit, davis, matar}, we will consider only the squeezing (or varicose) mode which is symmetric about the horizontal ($\tilde{x}$) axis, and is thus considered the most unstable mode.
The dimensional momentum balance in the $\tilde{x}$- and $\tilde{z}$ directions, and the continuity equation may be written as:
\begin{equation}
\label{nsx}
\rho (\tilde{u}_t + \tilde{u}\tilde{u}_x + \tilde{v}\tilde{u}_z) = -(\tilde{P}_{\tilde{x}} + \tilde{\Phi}_{\tilde{x}}) + \mu(\tilde{u}_{\tilde{x}\tilde{x}} + \tilde{u}_{\tilde{z}\tilde{z}}),
\end{equation}
\begin{equation}
\label{nsy}
\rho (\tilde{u}_t + \tilde{u}\tilde{v}_x + \tilde{v}\tilde{v}_z) = -(\tilde{P}_{\tilde{z}} + \tilde{\Phi}_{\tilde{z}}) + \mu(\tilde{v}_{\tilde{x}\tilde{x}} + \tilde{v}_{\tilde{z}\tilde{z}}),
\end{equation}
\begin{equation}
\label{con}
\tilde{u}_{\tilde{x}} + \tilde{v}_{\tilde{z}} = 0.
\end{equation}
Here, $\tilde{u}$ and $\tilde{v}$ denote the velocity components in $\tilde{x}$ and $\tilde{z}$ directions, while $\tilde{P}$ is the fluid pressure. We have included a disjoining pressure term $\tilde{\Phi}$ in our model, considering films to be thinner than $\sim 100 nm$, where van der Waals forces are prevalent \citep{oron, berg}. In the present work, we set $\tilde{\Phi}=A/6\pi(2\tilde{h})^3$ where $A$ is the dimensional Hamaker constant representing van der Waals attraction between the two free surfaces separated by the liquid. The attractive van der Waals forces will be the prime destabilizing factor for the film. The dynamics of the surface active species at the free surface neglecting their solubility in the bulk, is governed by a convection-diffusion equation \citep{matar}:
\begin{equation}\label{surf}
\frac{\partial\tilde{\Gamma}}{\partial\tilde{t}}+(\nabla_{s}\cdot\textbf{n})\tilde{\Gamma}(\textbf{n}\cdot\textbf{u}) + \nabla_s\cdot(u_s\tilde{\Gamma}) = D_s\nabla^2\tilde{\Gamma}.
\end{equation}
Here $D_s$ is surface diffusivity, treated as a constant, $\nabla_s$ is the surface gradient operator,
$u_s$ is the surface velocity field, and $\textbf{n}$ is the unit vector normal to the free surface. The boundary conditions at the free surface ($ \tilde{z} $= $\tilde{h}$($\tilde{x}$, $\tilde{t}$)) are the normal stress continuity
and the shear stress balance. The vector form of these equations in the presence of variable surface tension and surface viscosity effects are taken from the work by \cite{naire}, written as:
\begin{equation}
-\textbf{n}\cdot||\mathsfbi{T}||\cdot\textbf{n} = 2\kappa\sigma + 2\kappa\left(k^s + \mu^s\right)\nabla_s\cdot\textbf{u},
\end{equation}
\begin{equation}
-\textbf{t}\cdot||\mathsfbi{T}||\cdot\textbf{n} = \textbf{t}\cdot\nabla_s\sigma + \left(k^s + \mu^s\right)\textbf{t}\cdot\nabla_s\nabla_s\cdot\textbf{u} + \textbf{t}\cdot\nabla_s\left(k^s + \mu^s\right)\nabla_s\cdot\textbf{u},
\end{equation}
where, $\textbf{t}$ is the unit vector tangent to the free surface, $\mathsfbi{T}$ is the stress tensor and the dilatational $(k^s)$ and shear $(\mu^s)$ components of surface viscosities will hereafter be written in terms of a single parameter, $\tilde\eta=(k^s  + \mu^s)$ in the rest of the paper. A simplified version of the above equations in scalar form have been stated in Appendix A. The other conditions that complete the system are the kinematic condition at the free surface, and the symmetry
condition at the center line ($\tilde{z} = 0$). These are written as:
\begin{equation}\label{kin}
\tilde{h}_{\tilde{t}} + \tilde{u}\tilde{h}_{\tilde{x}} = \tilde{v} ~~~~~~~~~~~~~~~~\mathrm{(kinematic),}
\end{equation}
\begin{equation}
\label{sym}
\tilde{u}_{\tilde{z}} = 0, ~~~ and ~~~ \tilde{v} = 0 ~~~~~\mathrm{(squeezing~mode~symmetry).}
\end{equation}
%%%%%%%%%%%SCALINGS AND NON DIMENSIONALIZATION%%%%%%%%%%%%%%%%%%%%%%%%%%%%%%%%%%%%%%%%%%%%%%%%%%%%%%%%%%%%%%%%%%%%%%%%%%%%%%%%%%%%%%%%%%%%%%%%%%%%%%%%%%%%%%%%%%%%%SCALINGS AND NON DIMENSIONALIZATION%%%%%%%%%%%%%%%%%%%%%%%%%%%%%%%
\subsection{Scalings and non-dimensionalization}\label{sec:scalings}
The scalings for the different variables are explained next, where symbols with the tilde ($\sim$)
decoration denote dimensional versions. The spatial coordinates $\tilde{z}$ and $\tilde{x}$ are scaled as:
\begin{equation}
\label{scale}
z=\frac{\tilde{z}}{H}, ~~~~~~~~~ x=\frac{\tilde{x}}{L},
\end{equation}
where $L$ is the characteristic length along $\tilde{x}$ (for e.g., the wavelength of a typical interfacial perturbation). We further assume that $\epsilon=H/L\ll1$, i.e. a small aspect ratio film, which would permit us to employ the lubrication approximation. The velocity components and the fluid pressure are scaled as follows:
\begin{equation}
\label{velocityscalings}
u=\frac{\tilde{u}}{A/6\pi\mu\epsilon L^2}, ~~~~~~ v=\frac{\tilde{v}}{A/6\pi\mu\epsilon^{2} L^2}, ~~~~~~ P=\frac{\tilde{P}}{A/6\pi\epsilon L^3}.
\end{equation}
The scalings in (\ref{velocityscalings}) reflect a balance between viscous stresses and the disjoining pressure gradient caused due to van der Waals forces. We note that our choice of scalings is different from the ones used by \cite{matar} (extensional viscous stresses $\sim$ Marangoni stresses), and would be more apt for ultrathin films ($\le 100~nm$) \citep{oron, ting, berg}. The natural choice for the characteristic time scale is the ratio of the characteristic scales of velocity and length. Thus, the dimensionless time variable is expressed as:
\begin{equation}
t=\frac{\tilde{t}}{6\pi\mu\epsilon^3 L^3/A}.
\end{equation}
Since surface properties depend on surfactant concentration, suitable scalings are needed for the concentration, surface tension, and surface viscosity. The surfactant concentration is scaled in terms of the reference concentration $\tilde{\Gamma}_0$:
\begin{equation}
\Gamma(x,t)=\frac{\tilde{\Gamma}(\tilde{x},\tilde{t})}{\tilde{\Gamma}_0}.
\end{equation}
This reference concentration could, for instance, be an initial equilibrium surface concentration, or a critical concentration of colloidal surface active agents in a jammed-state interface ($\tilde{\Gamma}_{max}$). The actual choice of $\tilde{\Gamma}_0$ will have implications on the model used for variable surface viscosity effects which will be discussed later. The surface tension $\tilde{\sigma}$ is scaled with respect to its clean interface value (i.e. in the limit $\tilde{\Gamma}_0(x, t)\to0)$ whereas the surface viscosity, $\tilde\eta$, is scaled with $\tilde{\eta_0}$, the choice of which depends on the type of variation used for $\tilde{\eta}$ (see Appendix B),
\begin{equation}\label{ndm}
\sigma(x,t)=\frac{\tilde{\sigma}(\tilde{x},\tilde{t})}{\tilde{\sigma_0}}; ~~~~~~ \eta(x,t)=\frac{\tilde{\eta}(\tilde{x},\tilde{t})}{\tilde{\eta}_0}.
\end{equation}
\\%%%%%%%%%%%%%%%%%%%%%%%%%%%%%%%%%%%%%%%%%%%%%%%%%%%%%%%
Several dimensionless characteristic numbers appear upon rewriting \ref{nsx}-\ref{sym} using the scalings mentioned above. These are summarized in table \ref{table:dimnumbers} along with their typical order of magnitude estimates based on physical values given in table \ref{table:parameterset}. Among these, the parameters representing dominant effects are $\textit{Bo}$ (non-dimensional surface viscosity) and $\mathbb{D}$ (non-dimensional surface tension). Further, the Marangoni number $\textit{M}$ and the surface viscosity gradient parameter, $\beta$, represent the variation of surface tension and surface viscosity with surfactant concentration. A linear variation with concentration is assumed for surface tension:
\begin{equation}
\sigma=1-\textit{M}\Gamma(x,t).
\label{sigma}
\end{equation}
For the surface viscosity, a linear as well as a nonlinear model is used:
\begin{eqnarray}
\label{eta1}
\eta=1+\beta\left(\Gamma(x,t)-1\right),\\
\label{eta2}
\eta=\frac{1}{{[1-\Gamma(x,t)]}^\alpha}.
\end{eqnarray}
We use the linear model (\ref{eta1}) for an interface with dilute surfactants, thus the reference concentration, $\tilde{\Gamma}_0$ has a value in the dilute limit. For the nonlinear model (\ref{eta2}), the jammed state concentration value, $\tilde{\Gamma}_{max}$ of the system would intuitively be the suitable non-dimensionalizing factor for $\tilde{\Gamma}$ as the surface viscosity is expected to diverge in the jammed state, analogous to bulk viscosity \citep{maki, krig, que}. Recent studies in polymer blends also reveal such a dependence of surface viscosity on nano-particles concentration \citep{verm}. Note that the nonlinear phenomenological model (NPM) simplifies to a linear model in the limit of dilute surfactant concentration. Accordingly, the exponent $\alpha$ is related to the surface viscosity gradient parameter $\beta$ by the relation,
\begin{equation}
%\frac{\alpha}{\Gamma_{0\text{(nonlinear)}}}=\frac{\beta}{\Gamma_{0\text{(linear)}}},
%\tilde{\eta}^{(L)}_0\big[1+\beta\left(\Gamma^{(L)}_0 - 1\right)\big] = \tilde{\eta}^{(NL)}_0\big[1+\alpha\Gamma^{(NL)}_0\big]
\frac{\tilde{\Gamma}_0}{\tilde{\Gamma}_{max}}=\frac{\beta}{\alpha\left(1-\beta\right)}=\frac{\Gamma^{(nl)}_0}{\Gamma^{(l)}_0}
\label{etanonlinear}
\end{equation}
where $\Gamma^{(l)}_0$ and $\Gamma^{(nl)}_0$ denote the respective non-dimensional base state values in the linear and nonlinear models. Further details on the validity and derivation of the surface viscosity models are given in Appendix B. We also note that the NPM for surface viscosity reduces to a constant surface viscosity when $\Gamma \to 0$, in accordance with earlier studies on surface viscosity by \cite{scriv}.
%%%%%%%%%%%%%%%%%%%%%%%%%%%%%%%%%%%%%%%%%%%%%%%%%%%%%%%
\begin{table}
	\begin{tabular}{p{3cm}  p{5.5cm}   r}
\textit{Parameter} &\textit{Definition} & \textit{Order of magnitude estimates} \\ [0.5ex] 
 $\epsilon=\frac{\tilde{h}_0}{L}$ & Aspect ratio & $10^{-3} - 10^{-2}$ \\ [2ex] 
 $\textit{Bo} = \frac{\tilde{\eta_0}}{\mu h_0}$ & Boussinesq number & $10^{-1} - 10^1$  \\[2ex]
 $\mathbb{D}=\frac{A}{6\pi\tilde{\sigma}_0\epsilon L^2}$ & Disjoining pressure number & $10^{-3} - 10^{-2}$  \\[2ex]
 $\textit{M}=\frac{\Gamma_0}{\tilde{\sigma}_0}\left(\frac{\partial\tilde{\sigma}}{\partial\tilde{\Gamma}}\right)$ & Marangoni number & $0 - 5\times10^{-3}$  \\[2ex]
 $\beta=\frac{\Gamma_0}{\tilde{\eta}_0}\left(\frac{\partial\tilde{\eta}}{\partial\tilde{\Gamma}}\right)$ & Surface viscosity gradient number & $0 - 1$\\ [2ex]
 $\Pen=\frac{A}{6\pi\mu h_0 D_s}$ & Peclet number & $10^{-1}-10^2$ \\ [2ex]
 $\Rey=\frac{\rho A}{6\pi\mu^2\epsilon\tilde{h}_0}$ & Reynolds number & $10^{-2} - 1$ \\ [2ex]
\end{tabular}
	\caption{Estimates of non-dimensional system parameters evaluated using characteristic velocity scale, and dimensional estimates from Table \ref{table:parameterset}.}
\label{table:dimnumbers}
\end{table}
%%%%%%%%%%%%%%%%%%%%%%%%%%%%%%%%%%%%%%%%%%%%%%%%%%%%%%%

The disjoining pressure number, $\mathbb{D}$ in Table \ref{table:dimnumbers}, defined as $\mathbb{D}=A/6\pi\sigma_0\epsilon L^2$ has been deliberately scaled to contain $\epsilon$, so that capillary effects are retained in the normal stress boundary condition, as also described in Appendix \ref{nsapp}. It represents the ratio of intermolecular van der Waals interactions to surface tension forces. A disjoining pressure interpretation of the capillary effects is apt as it represents the overall wetting characteristics of the thin-film system, as has been done in an earlier work \citep{alle}. With respect to Marangoni and surface viscosity effects, we restrict ourselves to the default regime examined in prior works \citep{dewit, matar} in which both Marangoni effects and surface viscosity only appear in the first order correction of the tangential stress balance condition. Accordingly, if $\textit{M}$ is an $O(1)$ quantity, a weak Marangoni limit is defined by rescaling it as $\textit{M}=\epsilon^2\hat{\textit{M}}$, where the new O(1) parameter is $\hat{\textit{M}}$. Then, Marangoni effects would drop out in the lubrication limit when $O(\epsilon^2)$ terms are ignored. Following the definition of $\sigma$ in (\ref{sigma}), the weak Marangoni limit would imply, at leading order,
\begin{equation}\label{si}
\sigma=1,~~~~and~~~~ \sigma_x=0.
\end{equation}
For the surface viscosity, $\eta$ is retained as an O(1) quantity, but it would still appear only in the first (and higher) order corrections to the stress balance conditions as shown by \cite{matar}. To explore the strong surface viscosity regime, $\eta$ may be rescaled such that it is retained at leading order, and the boundary conditions are self-consistent. However, as $\epsilon$ (aspect ratio) can be made arbitrarily small, presenting the surface viscosity terms at $O(\epsilon)$ stress balance conditions (as in the present formulation) could still be deemed acceptable despite the diverging nature of NPM.
%%%%%%%%%%%%%%%%%%%%%%%%%%%%%%%%%%%%%%%%%%%%%%%%%%%%%%%%%%%%%%%
%%%%%%%%%%%%%%%%%%%%%%%%%%%%%%%%%%%%%%%%%%%%%%%%%%%%%%%%%%%%%%%

\subsection{Evolution equations}
With the scalings presented in \S\ref{sec:scalings}, the governing equations (\ref{nsx}-\ref{con}) reduce to the following when only the zeroth (leading) order terms are considered, denoted with the superscript $(0)$
\begin{eqnarray}
\label{nsx0}
u^{(0)}_{zz}=0,\\
\label{p0}
P^{(0)}_{z}=v^{(0)}_{zz},\\
u^{(0)}_{x}+v^{(0)}_{z}=0.
\end{eqnarray}
Using the symmetry conditions (\ref{sym}) at $z=0$ gives:
\begin{eqnarray}\label{vel}
u^{(0)}=c(x,t),\\
\label{vel2}
v^{(0)}=-c_{x}z
\end{eqnarray}
where $c(x,t)$ is still an unknown function independent of $z$. The leading order normal stress balance (see Appendix \ref{nsapp}) is written as:
\begin {equation}\label{nsbc}
-P^{(0)} = \mathbb{D}^{-1}h_{xx} + 2h_{x}u_{z} - 2v_z.
\end{equation}
Integrating (\ref{p0}) and comparing it with (\ref{nsbc}), and substituting (\ref{vel}-\ref{vel2}), we obtain:
\begin{equation}
P^{(0)} = -2c_x - \mathbb{D}^{-1}h^{(0)}_{xx}.
\end{equation}
The leading order tangential stress balance (see Appendix \ref{tsapp}), on the other hand, simplifies to: (using \ref{si})
\begin{equation}\label{tsbc}
u^{(0)}_z =\mathbb{D}^{-1}\sigma_{x}=-\epsilon^2\mathbb{D}^{-1}\textit{M}\Gamma^{(0)}_{x}=0. 
\end{equation}
This is also consistent with (\ref{vel}). 
The kinematic condition (\ref{kin}) and the surfactant transport equation (\ref{surf}) provide two nonlinear evolution equations for $h^{(0)}$ and $\Gamma^{(0)}$, with $c(x, t)$ still undetermined. As discussed by \citep{dewit} and \citep{matar}, the system can be completed only through a third equation obtained by considering the first order corrections of (\ref{nsx0}) and (\ref{tsbc}) (refer Appendix \ref{tsapp}, \ref{xmapp}). To these, we substitute (\ref{eta1}) or (\ref{eta2}) to account for variable viscosity. The final system of three nonlinear evolution equations read:
\begin{eqnarray}
\label{nlea}
 h_t + {(ch)}_x = 0,\\
\label{nleb}
 \Gamma_t + {(c\Gamma)}_x = \frac{\Gamma_{xx}}{\Pen},
\end{eqnarray}
\begin{eqnarray}
\label{nlec}
 \Rey(c_t + cc_x) - \mathbb{D}^{-1}h_{xxx} - \frac{3}{8h^4}h_x - 4c_{xx} = 4\frac{h_x c_x}{h} + \textit{Bo}(1+\beta(\Gamma-1))\frac{c_{xx}}{h} \nonumber\\
 +\textit{Bo}\beta\frac{c_x\Gamma_x}{h}-\mathbb{D}^{-1}\textit{M}\frac{\Gamma_x}{h}.\\
\mathrm{(for~linear~model~using~(\ref{eta1}))}\nonumber 
\end{eqnarray}
or,
\begin{eqnarray}
\label{nled}
\Rey(c_t + cc_x) - \mathbb{D}^{-1}h_{xxx} - \frac{3}{8h^4}h_x - 4c_{xx} = 4\frac{h_x c_x}{h} + \textit{Bo}\frac{c_{xx}}{(1-\Gamma)^\alpha h} \nonumber\\
+\alpha\textit{Bo}\frac{c_x\Gamma_x}{(1-\Gamma)^{\alpha+1}h}-\mathbb{D}^{-1}\textit{M}\frac{\Gamma_x}{h}. \\
\mathrm{(for~NPM~using~(\ref{eta2}))}\nonumber
\end{eqnarray}
Here, $\Rey~\&~\Pen$ are as defined in Table \ref{table:dimnumbers}. The superscript (0) has been omitted in (\ref{nlea} - \ref{nled}) and the remainder of the paper for easy readability. Equations (\ref{nlea} - \ref{nled}) reduce to those derived by \cite{dewit} in the limit $\textit{Bo} = 0$, and to those of \cite{matar} when we set $\beta = 0$, in the absence of surfactant solubility. 

%%%%%%%%%%%%%%%%%%%%%%%%%%%%%%%%%%%%%%%%%%%%%%%%%%%%%%%%%%%%%%%%%%%%%%%%%%%%%%%%%%%%%%%%%%%%%%%%%%%%%%%%%%%%%%%%%%%%%%%%%%%%%%%%%%%%%%%%%%%%%%%%%%%%
%%%%%%%%%%%%%%%%%%%%%%%%%%%%LINEAR STABILITY ANALYSIS%%%%%%%%%%%%%%%%%%%%%%%%%%%%%%%%%%%%%%%%%%%%%%%%%%%%%%%%%%%%%%%%%%%%%%%%%%%%%%%%%%%%%%%%%%%%%%%%%%%%%%%%%%%%%%%%%%%%%LINEAR STABILITY ANALYSIS%%%%%%%%%%%%%%%%%%%%%LINEAR STABILITY ANALYSIS%%%%%%%%%%%%%%%%%%%%%%%%%%%%%%%%%%%%%%%%%LINEAR STABILITY ANALYSIS%%%%%%%%%%%%%%%%%%%%%%%%%%%%%%%%%LINEAR STABILITY ANALYSIS%%%%%%%%%%%%%%%%%%%%%%%%%%%%%%%%%%%%%%%%%%%%%%%%%%
\section{Linear Stability Analysis}\label{sec3}
We linearize the system of three nonlinear equations (\ref{nlea}), (\ref{nleb}), and (\ref{nlec}) or (\ref{nled}) as the case be, by perturbing the dependent variables about a uniform base state as follows:
\begin{eqnarray}
\label{pert}
h(x,t)=\frac{1}{2} + \hat{h}e^{(ikx+st)},\\
\Gamma(x,t)=\Gamma_0 + \hat{\Gamma}e^{(ikx+st)},\\
c(x,t)=\hat{c}e^{(ikx+st)}.
\end{eqnarray}
Here $s$ is the growth rate and $k$ is the wavenumber of the perturbation. Using standard linear stability analysis, the following dispersion relations are obtained for the linear and nonlinear surface viscosity models:
\begin{eqnarray}
\label{disp}
s^3 + s^2k^2\left(\frac{1}{\Pen}+\frac{4}{\Rey}+\frac{2\textit{Bo}(1+\beta)(\Gamma_0^{(l)} - 1)}{\Rey}\right) \nonumber\\
+ sk^2\left(\frac{4k^2}{\Pen\Rey}+\frac{\mathbb{D}^{-1}k^2}{2\Rey}-\frac{3}{\Rey}+\frac{\mathbb{D}^{-1}\textit{M}\Gamma_0^{(l)} }{\Rey}+\frac{2\textit{Bo}(1+\beta)(\Gamma_0^{(l)} - 1)k^2}{\Pen\Rey}\right)\nonumber\\
+\frac{k^4}{\Pen\Rey}\left(\frac{\mathbb{D}^{-1}k^2}{2}-3\right)=0\\ 
\nonumber\\
\label{disp2}
s^3 + s^2k^2\left(\frac{1}{\Pen}+\frac{4}{\Rey}+\frac{2\textit{Bo}}{\Rey(1-\Gamma_0^{(nl)})^\alpha}\right) \nonumber\\
+ sk^2\left(\frac{4k^2}{\Pen\Rey}+\frac{\mathbb{D}^{-1}k^2}{2\Rey}-\frac{3}{\Rey}+\frac{\mathbb{D}^{-1}\textit{M}\Gamma_0^{(nl)}}{\Rey}+\frac{2\textit{Bo}k^2}{\Pen\Rey(1-\Gamma_0^{(nl)})^\alpha}\right)\nonumber\\
+\frac{k^4}{\Pen\Rey}\left(\frac{\mathbb{D}^{-1}k^2}{2}-3\right)=0
\end{eqnarray}
where, $\Gamma_0^{(l)}$ and $\Gamma_0^{(nl)}$ are the non-dimensional base states for surfactant concentration in the linear model and NPM respectively, as has already been discussed in deriving (\ref{etanonlinear}). The above dispersion relations are in agreement with earlier studies \citep{davis, matar, dewit} when constant surface viscosity is considered. The main novelty of the present work is a careful study of variable surface viscosity and these enter through the variables $\beta$ and $\alpha$ in \eqref{disp} and \eqref{disp2} respectively.

%%%%%%%%%%%%%%%%%%%%%%%%%%%%%%%%%%%%%%%%%%%%%%%%%%%%%%%%%%%%%%%%%%%%%%%%%%%%%%%%%%%%%%%%%%%%%%%%%%%%%%%%%%%%%%%%%%%%%%%%%%%%%%%%%%%%%%%%%%%%%%%%%
\subsection{Parametric studies}
\begin{figure}
\centering
%\subfigure[]{\label{fig:a}\includegraphics[trim= 10mm 3mm 23mm 11mm ,clip, width=0.45\textwidth]{varyingA.eps}}
\subfigure[]{\label{figure1a}\includegraphics[trim= 12mm 75mm 20mm 82mm ,clip, width=0.45\textwidth]{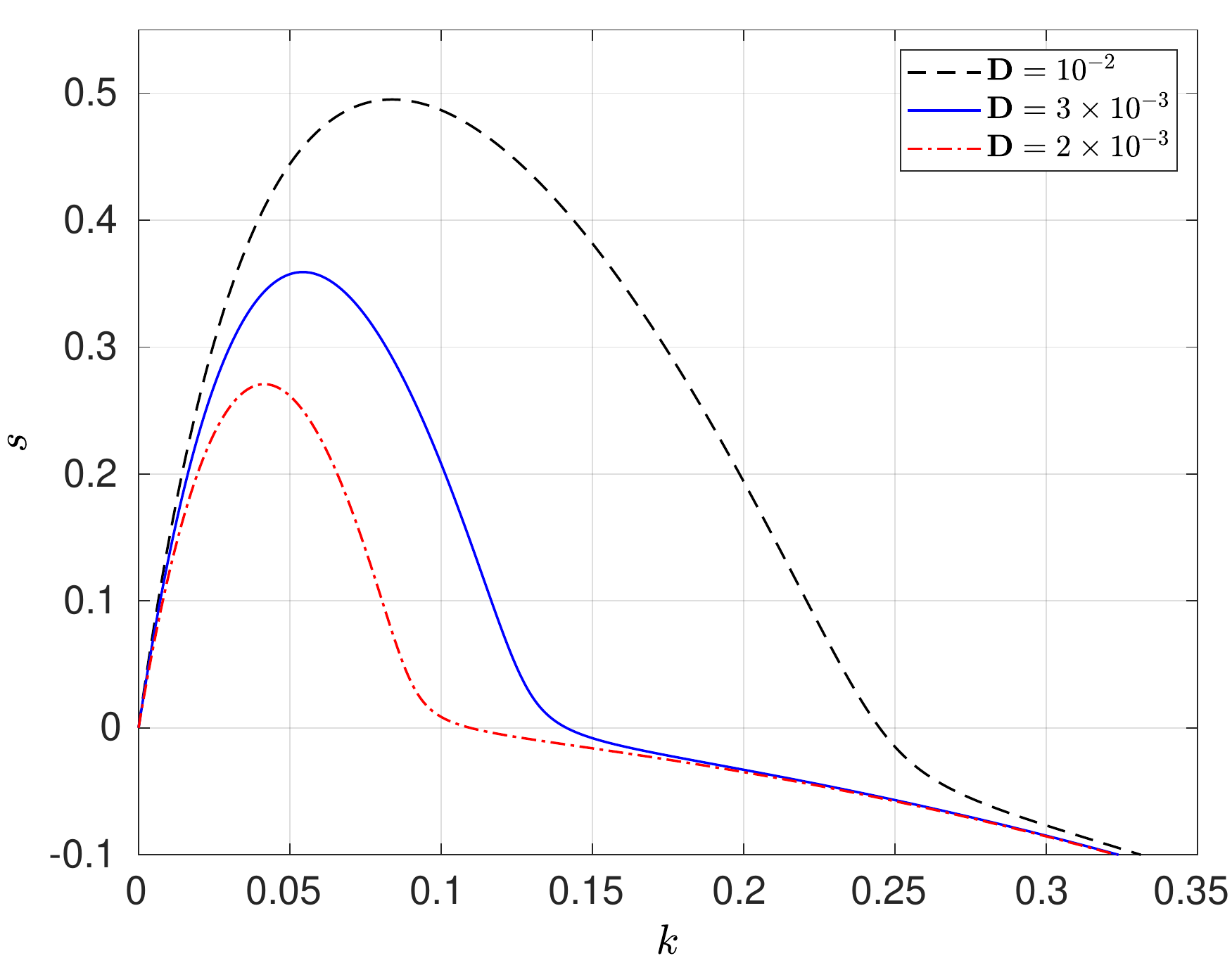}}
\subfigure[]{\label{figure1b}\includegraphics[trim= 8mm 75mm 20mm 82mm ,clip, width=0.45\textwidth]{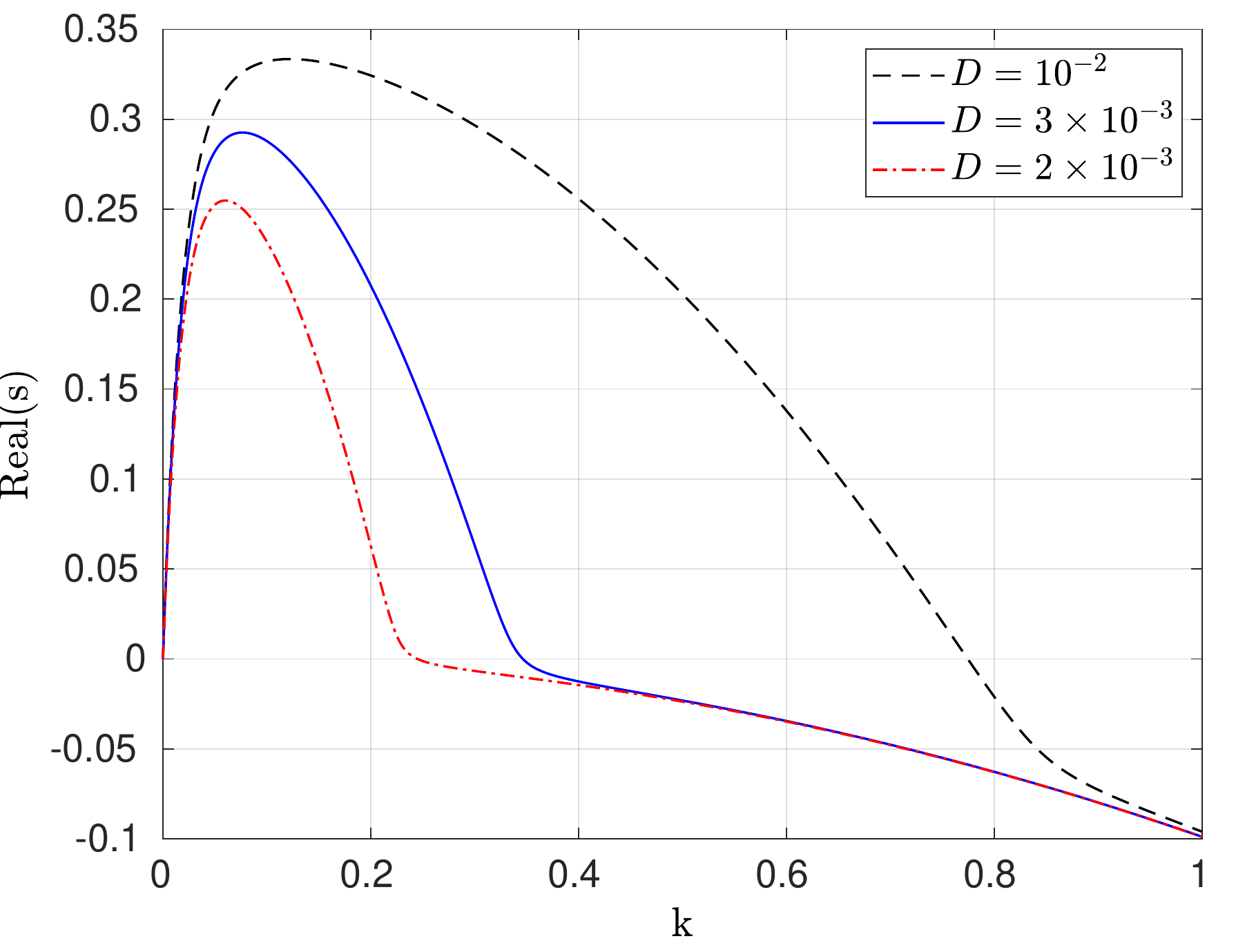}}
\subfigure[]{\label{fig1:c}\includegraphics[trim= 12mm 75mm 20mm 86mm ,clip, width=0.45\textwidth]{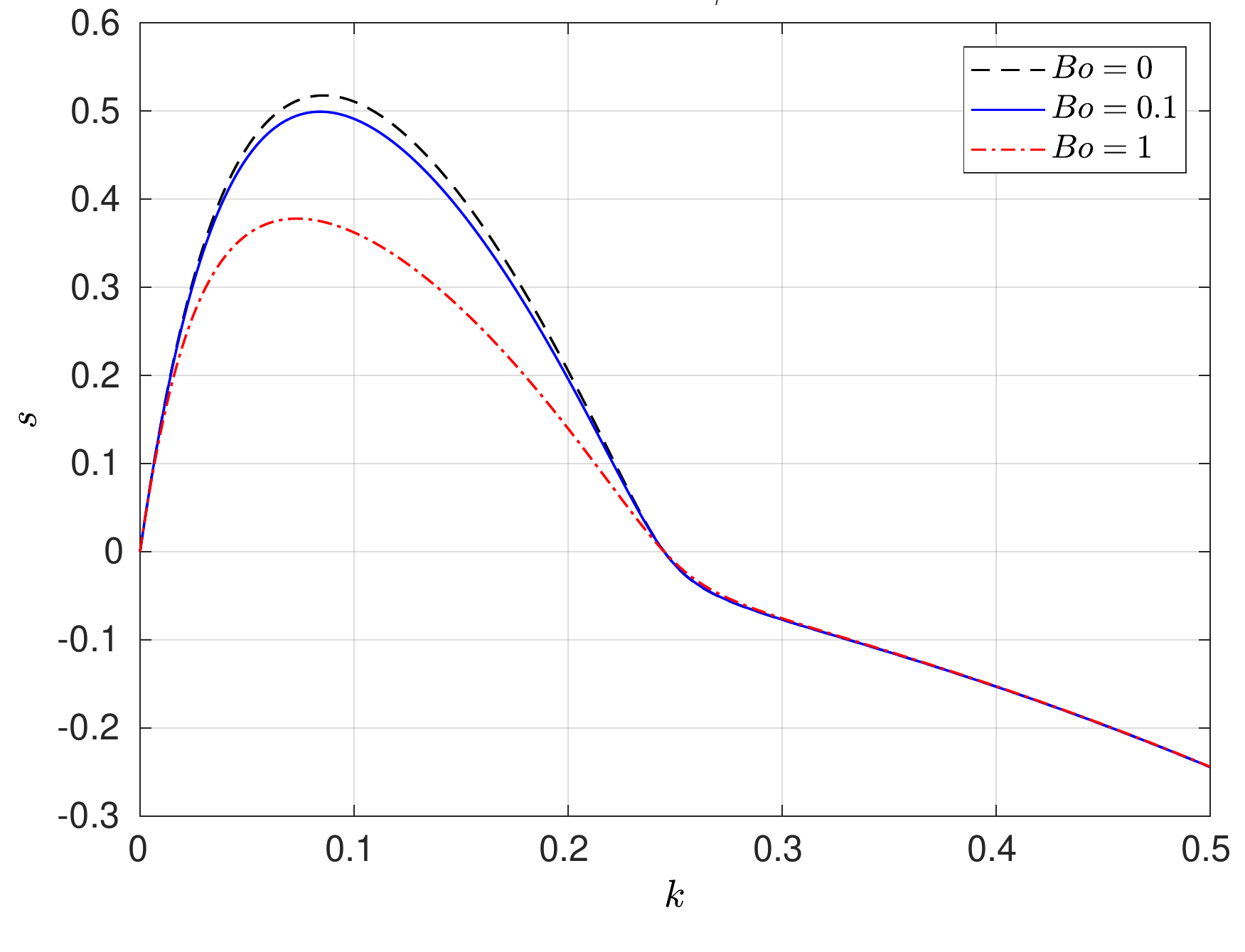}}
\subfigure[]{\label{fig1:d}\includegraphics[trim= 12mm 75mm 20mm 82mm ,clip, width=0.45\textwidth]{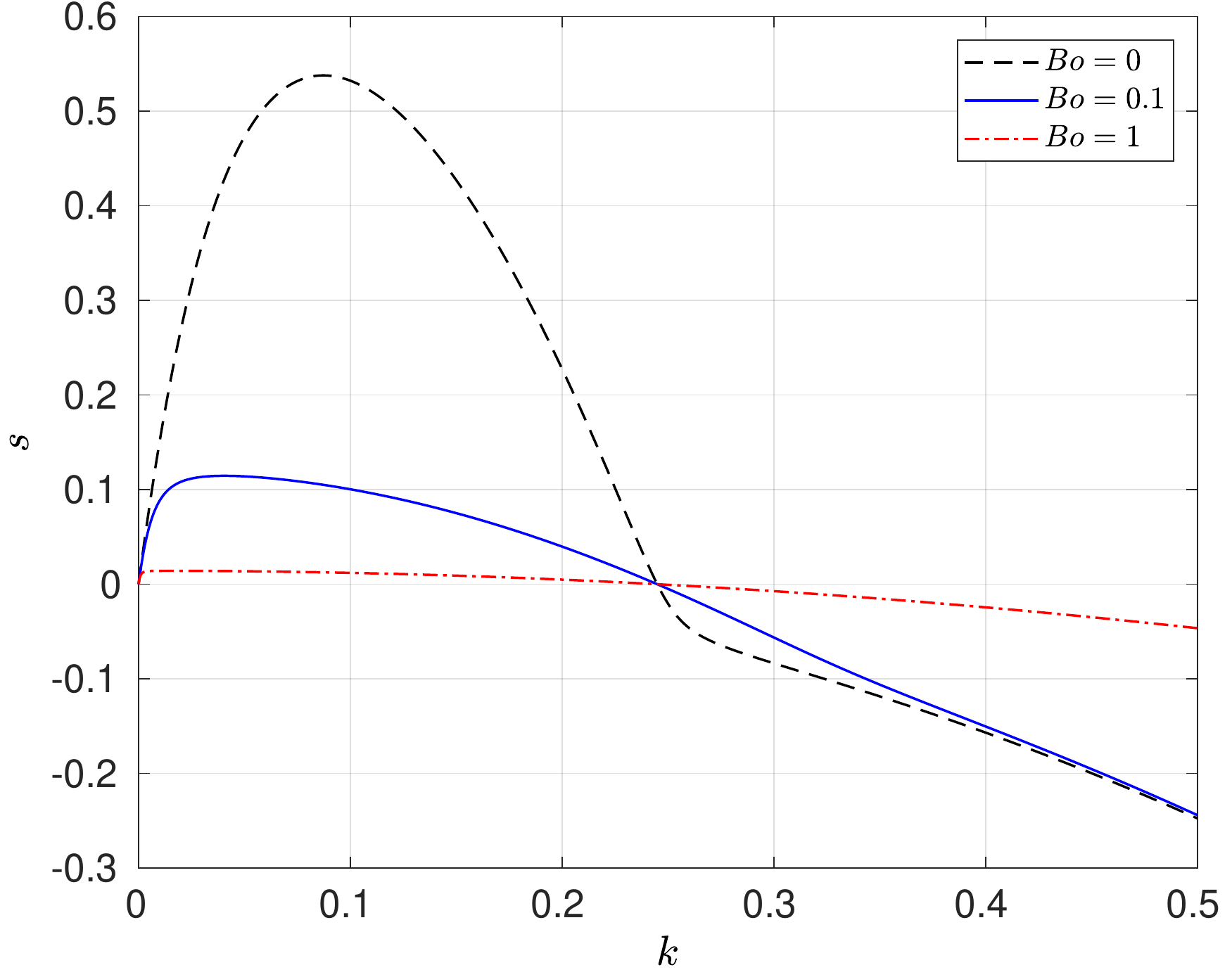}}
\subfigure[]{\label{fig1:e}\includegraphics[trim= 12mm 75mm 20mm 82mm ,clip, width=0.45\textwidth]{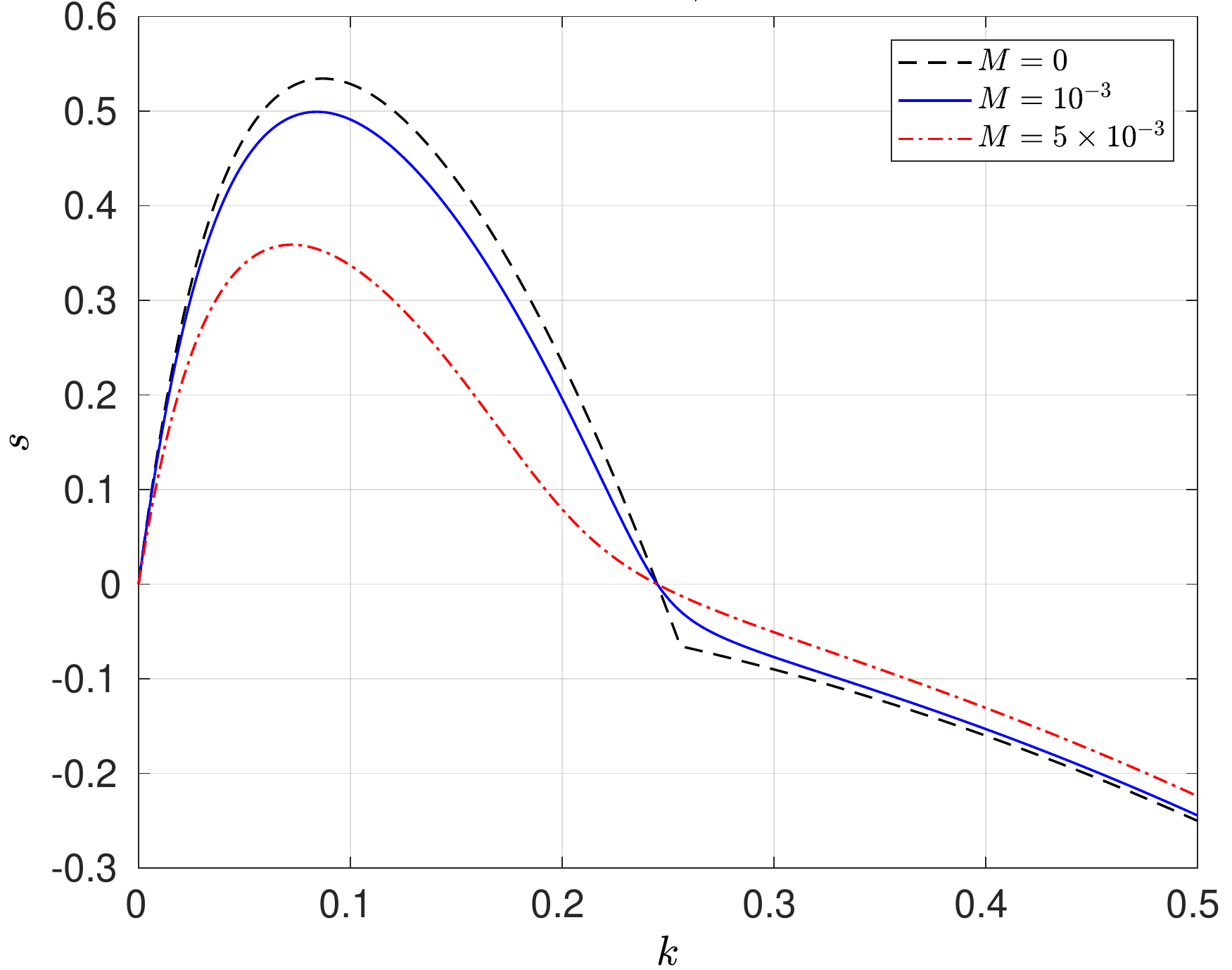}}
\subfigure[]{\label{fig1:f}\includegraphics[trim= 10mm 75mm 20mm 82mm ,clip, width=0.45\textwidth]{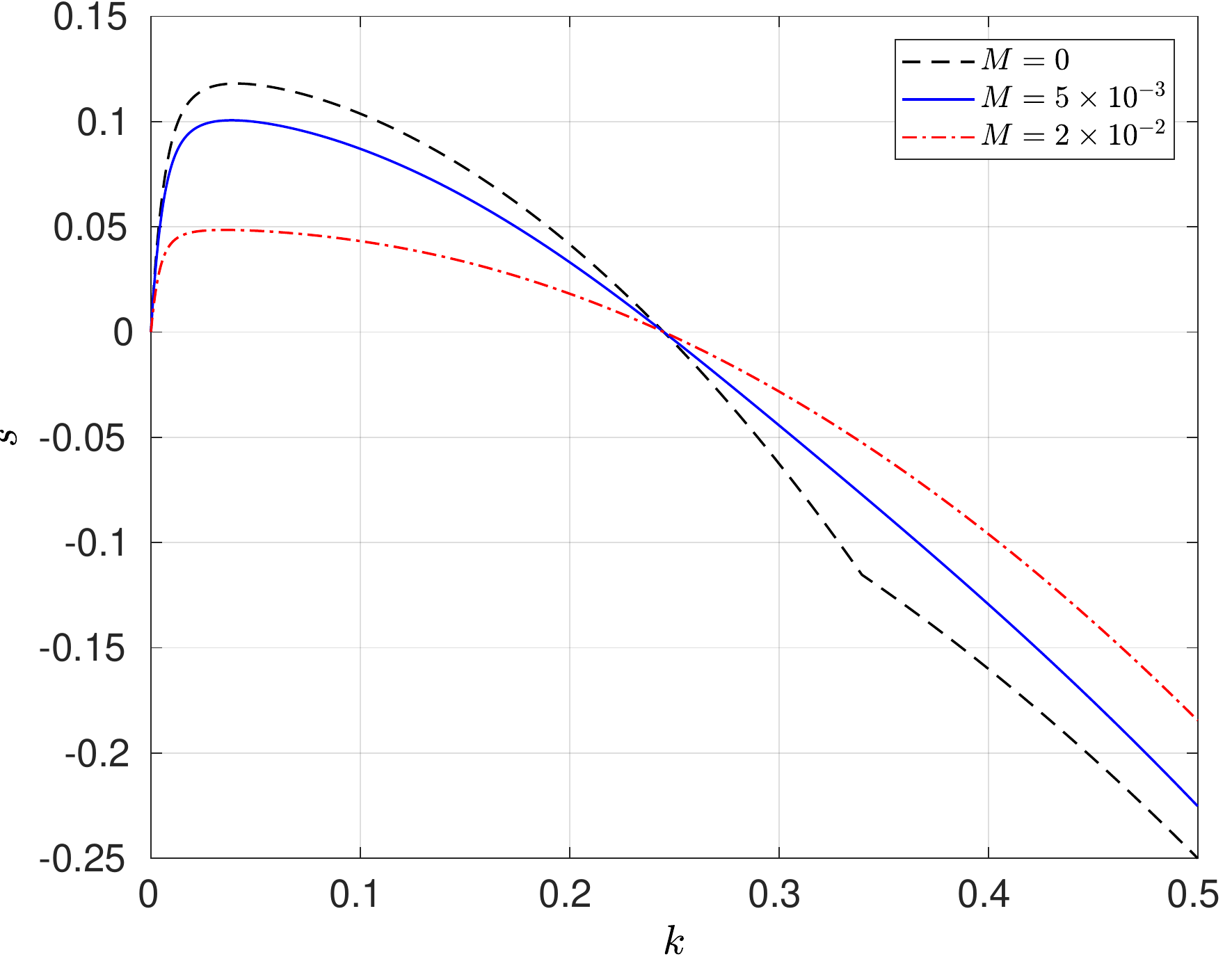}}
\caption{Dispersion curves for linear and nonlinear models. Left panels are for the linear model (equation \ref{disp}) with $\Gamma^{(l)}_0=2$ and the right panels are for NPM (equation \ref{disp2}) with $\Gamma^{(nl)}_0=0.9$ for varying $\mathbb{D}$, $Bo$ and $M$. (a,b) Varying $\mathbb{D}$ with $\textit{Bo}=10^{-1}, ~\textit{M}=10^{-3}$ fixed; (c,d) varying $\textit{Bo}$ with $\mathbb{D}^{-1}=100, ~\textit{M}=10^{-3}$ fixed; (e,f) varying $\textit{M}$ with $\textit{Bo}=10^{-1},~\mathbb{D}^{-1}=100$ fixed. Other parameters are fixed at $\Pen=1,~\Rey=10^{-2}, ~\beta=10^{-3},~\alpha=2$. As is clearly evident, cut-off wavenumber, $k_c$, is insensitive to changes in $Bo$ and $M$ and depends solely on $\mathbb{D}$. Figure \ref{fig1:d} shows that growth rate is highly sensitive to surface viscosity effects in NPM and has been analysed in detail in section \S\ref{sec:LSAjammed}.}
\end{figure}
\begin{figure}
\centering
%\subfigure[]{\label{fig2:a}\includegraphics[trim= 11mm 75mm 22mm 86mm ,clip, width=0.45\textwidth]{varyingBeta.pdf}}
\subfigure[]{\label{fig2:a}\includegraphics[trim= 12mm 75mm 20mm 82mm ,clip, width=0.45\textwidth]{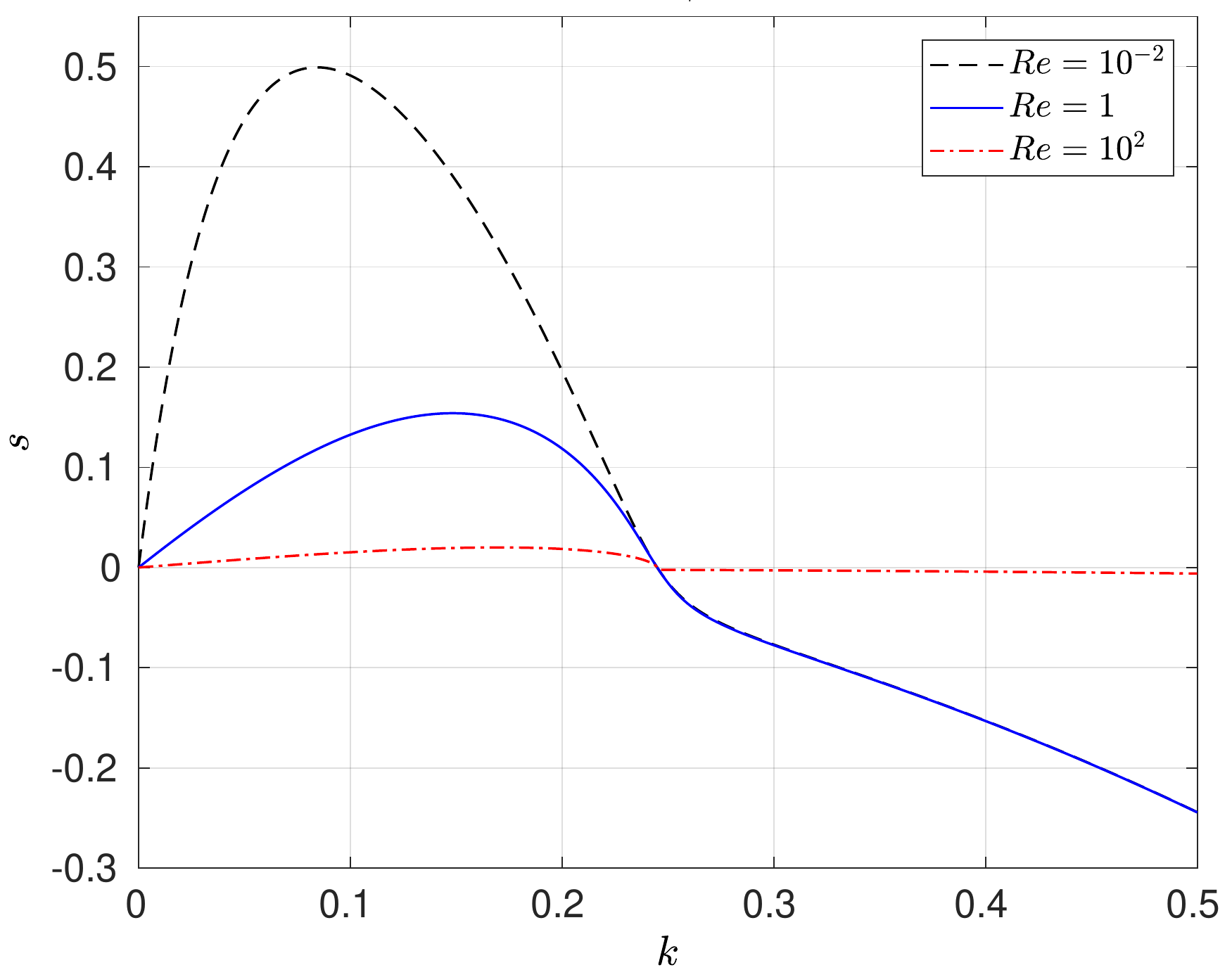}}
\subfigure[]{\label{fig2:b}\includegraphics[trim= 8mm 75mm 20mm 82mm ,clip, width=0.45\textwidth]{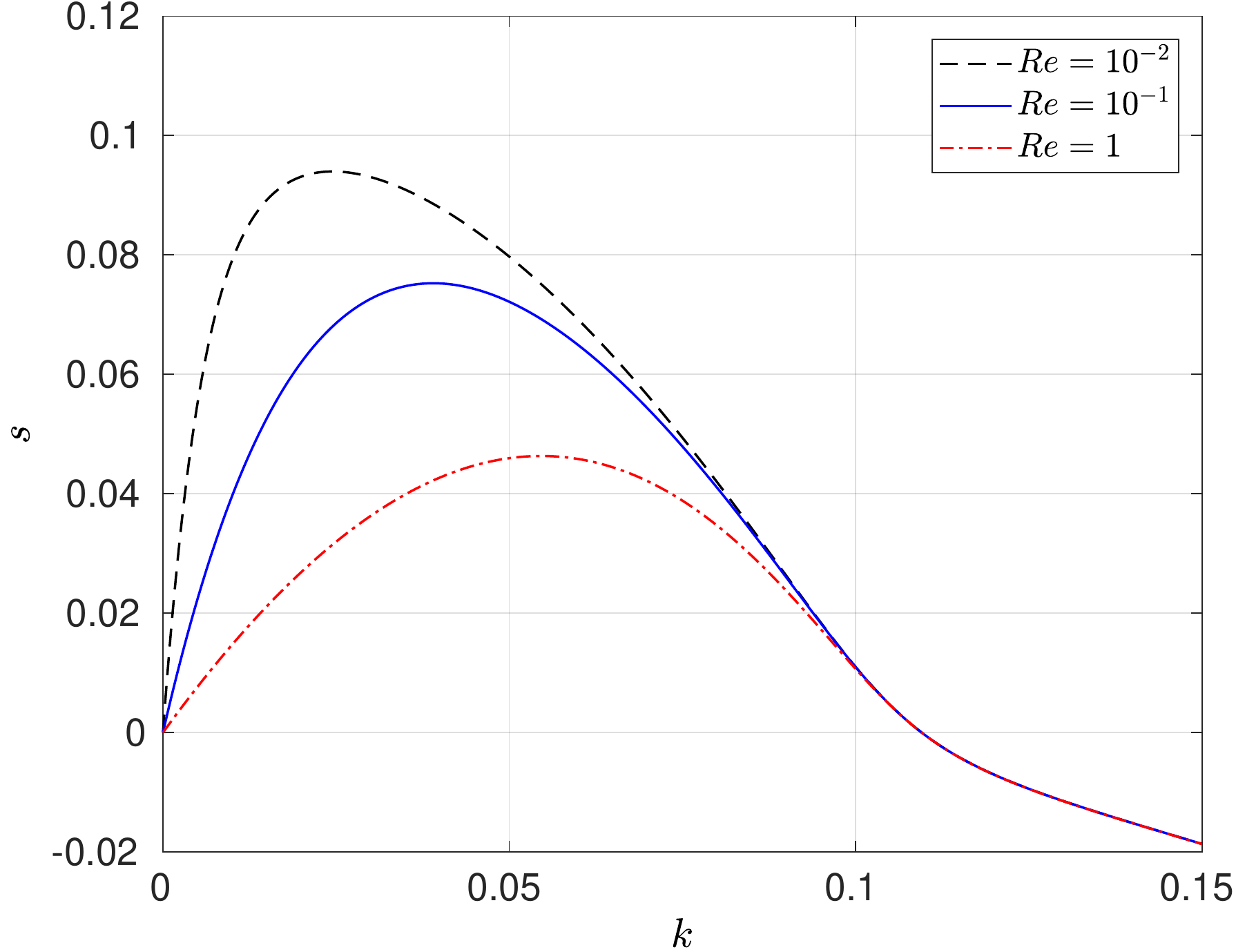}}
\subfigure[]{\label{fig2:c}\includegraphics[trim= 12mm 75mm 20mm 82mm ,clip, width=0.45\textwidth]{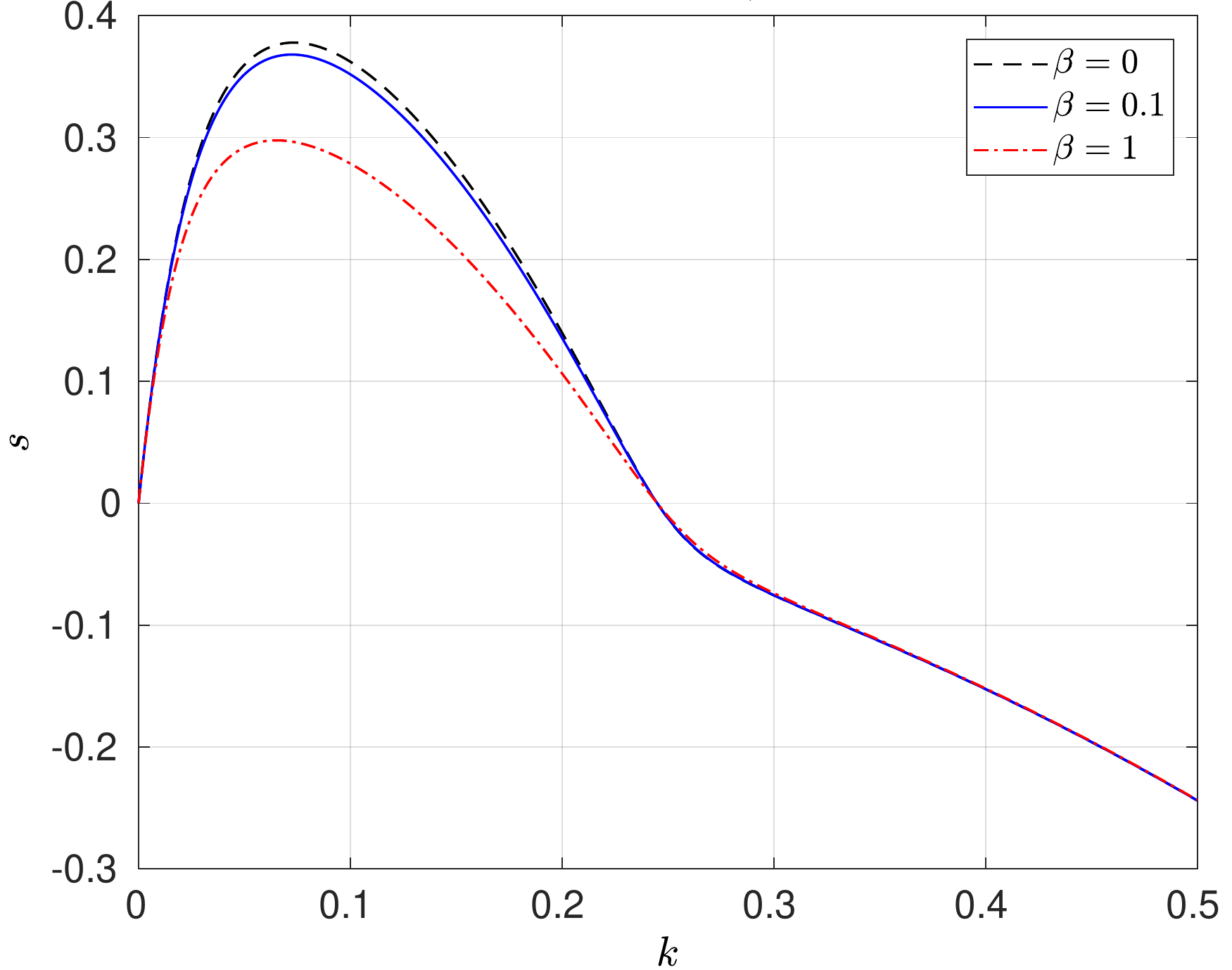}}
\subfigure[]{\label{fig2:d}\includegraphics[trim= 8mm 75mm 20mm 82mm ,clip, width=0.45\textwidth]{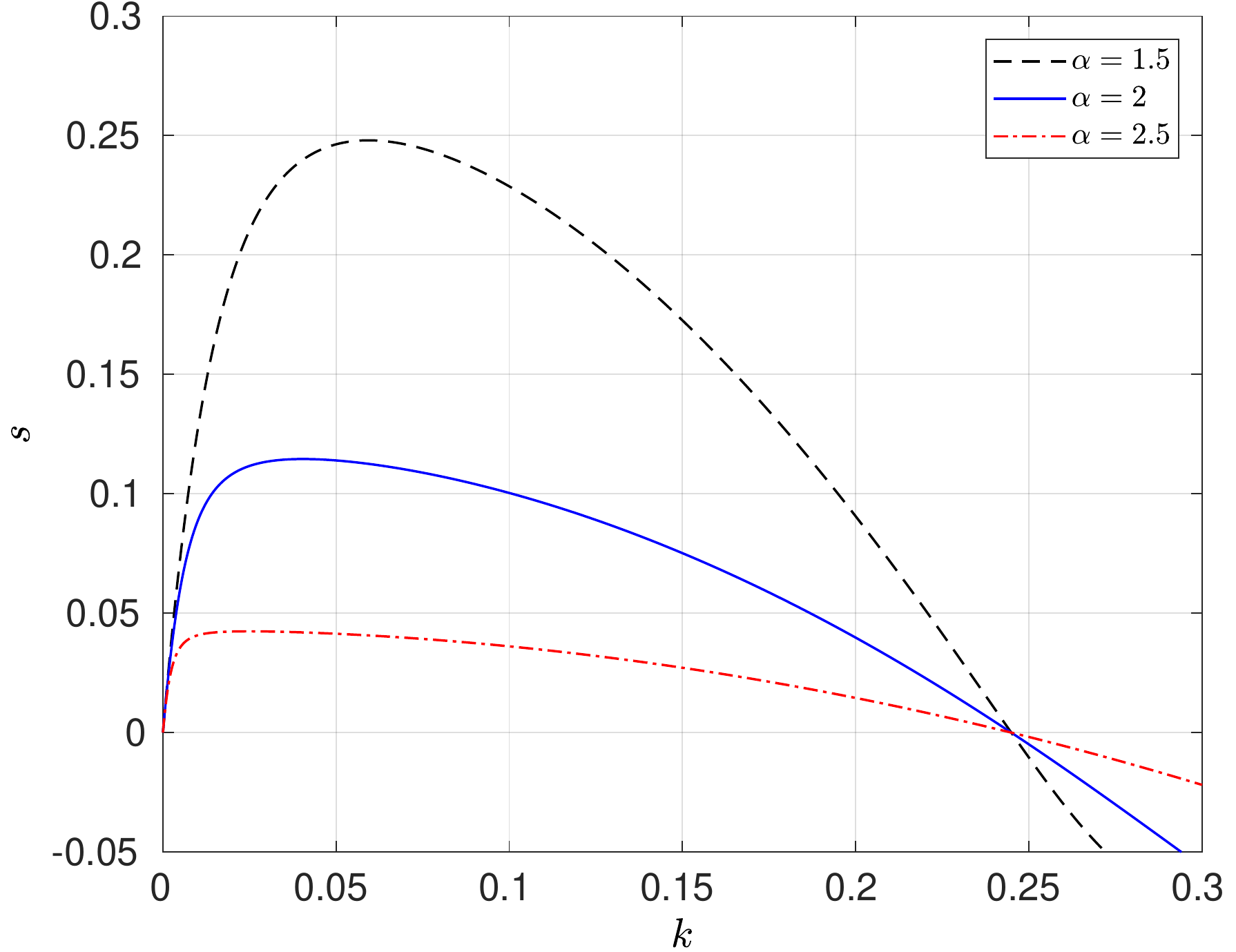}}
\caption{Dispersion curves for linear and nonlinear models. Left panels are for the linear model (equation \ref{disp}) with $\Gamma^{(l)}_0=2$ and the right panels are for NPM (equation \ref{disp2}) with $\Gamma^{(nl)}_0=0.9$. (a,b) Varying $\Rey$ with $\Pen=1,~\beta=10^{-3}$ fixed; (c,d) varying $\beta$ and $\alpha$ in the linear and NPM respectively with $\Pen=1,~\Rey=10^{-2}$ fixed. Other parameters are fixed at $\mathbb{D}=10^{-2}, ~\textit{Bo}=10^{-1},~\textit{M}=10^{-3}$.}
\end{figure}
The dispersion curves for the linear and NPM models are plotted in Fig. 2  and 3 for a range of parameter values chosen from table \ref{table:dimnumbers}. All the curves exhibit a fastest growing mode, $k_{max}$ and a cut-off wavenumber, $k_c$,
\begin{eqnarray}
\label{kc}
k_c=\sqrt{6\mathbb{D}},
\end{eqnarray}
above which the film is stable. Fig. \ref{figure1a} and \ref{figure1b} show the linear stability characteristics with variation in the parameter $\mathbb{D}$ for the two models. We set $\Gamma_0^{(nl)} = 0.9$ for the dispersion curves of NPM. Intuitively, a decrease in $\mathbb{D}$ results in greater opposition from surface tension towards film breakup, and a smaller driving force from intermolecular van der Waals attractions. Accordingly, it is seen that the maximum growth rates reduces with decreasing $\mathbb{D}$.
Figures \ref{fig1:c} - \ref{fig1:f} are the linear stability curves for different values of $\textit{Bo}$ and $\textit{M}$ for both the models. Both surface viscosity and the solutal Marangoni effect act as film stabilizers, as inferred from the growth rate plots. Note that, varying $\textit{Bo}$ has drastic effects in case of NPM (fig. \ref{fig1:d}). 

Figures \ref{fig2:a}-\ref{fig2:d} are plots of the dispersion relation for the other system parameters: $\Rey$ and $\Pen$ for both the models. The nature of the dispersion curves plotted for these parameters are also intuitive. Physically, it may be understood that a greater value of $\Rey$ implies more inertial forces relative to viscous forces. Though both the parameters act as retarding effects towards growth of instability, the contribution of the former is more significant, leading to smaller growth rates at higher $\Rey$. As regards to $\Pen$, a higher surface diffusivity of the surfactants (lower values of $\Pen$) facilitates higher gradients of surfactant-concentrations to destabilize the film evolution which can be clearly observed in fig. \ref{fig2:c}. The linear stability curves for NPM, governed by (\ref{disp2}) may seem qualitatively similar to those of linear model, but has a distinctive boundary layer structure (for $k \to O\left(1-\Gamma\right)^{\alpha}$) which is very different from that of the linear model. This is discussed in \S\ref{sec:LSAjammed} in detail. Both the models follow (\ref{kc}). Equations (\ref{disp}) and (\ref{disp2}) become identical when the parameters are chosen to satisfy (\ref{etanonlinear}) in the limit of $\tilde{\Gamma}\ll\tilde{\Gamma}_{max}$.

%%%%%%%%%%%%%%%%%%%%%%%%%%%%%%%%%%%%%%%%%%%%%%%%%%%%%%%%%%%%%%%%%%%%%%%%%
%%%%%%%%%%%%%%%%%%%%%%%%%%%%%%%%%%%%%%%%%%%%%%%%%%%%%%%%%%%%%%%%%%%%%%%%%
%%%%%%%%%%%%%%%%%%%%%%%%%%%%%%%%%%%%%%%%%%%%%%%%%%%%%%%%%%%%%%%%%%%%%%%%%
%%%%%%%%%%%%%%%%%%%%%%%%%%%%%%%%%%%%%%%%%%%%%%%%%%%%%%%%%%%%%%%%%%%%%%%%%
%%%%%%%%%%%%%%%%%%%%%%%%%%%%%%%%%%%%%%%%%%%%%%%%%%%%%%%%%%%%%%%%%%%%%%%%%
%%%%%%%%%%%%%%%%%%%%%%%%%%%%%%%%%%%%%%%%%%%%%%%%%%%%%%%%%%%%%%%%%%%%%%%%%
%%%%%%%%%%%%%%%%%%%%%%%%%%%%%%%%%%%%%%%%%%%%%%%%%%%%%%%%%%%%%%%%%%%%%%%%%
%%%%%%%%%%%%%%%%%%%%%%%%%%%%%%%%%%%%%%%%%%%%%%%%%%%%%%%%%%%%%%%%%%%%%%%%%
%%%%%%%%%%%%%%%%%%%%%%%%%%%%%%%%%%%%%%%%%%%%%%%%%%%%%%%%%%%%%%%%%%%%%%%%%
%%%%%%%%%%%%%%%%%%%%%%%%%%%%%%%%%%%%%%%%%%%%%%%%%%%%%%%%%%%%%%%%%%%%%%%%%
%%%%%%%%%%%%%%%%%%%%%%%%%%%%%%%%%%%%%%%%%%%%%%%%%%%%%%%%%%%%%%%%%%%%%%%%%
%%%%%%%%%%%%%%%%%%%%%%%%%%%%%%%%%%%%%%%%%%%%%%%%%%%%%%%%%%%%%%%%%%%%%%%%%
%%%%%%%%%%%%%%%%%%%%%%%%%%%%%%%%%%%%%%%%%%%%%%%%%%%%%%%%%%%%%%%%%%%%%%%%%
\subsection{`Jammed limit' analysis }\label{sec:LSAjammed}
In this section, we show that it is possible to obtain a simple perturbation solution to the dispersion relation \eqref{disp2} in the limit of large surfactant concentrations, i.e. for $\Gamma_0^{(nl)} \rightarrow 1$. 
In this limit, the surfactant concentration approaches the jamming limit and mobility of surfactant molecules reduces to zero. It will be shown later (see \S\ref{sec:NLAjammed}) that the tangential velocity indeed reduces to zero as $\Gamma_0^{(nl)} \rightarrow 1$. This limit of \eqref{disp2} is distinctly different from that of the linear model in \eqref{disp} which is valid in the limit of small $\Gamma_0^{(nl)}$. It is shown below that the nature of the cubic equation changes in the jamming limit which is not possible using the linear model. To proceed with the perturbation analysis, it is convenient to define a small parameter, $\delta = \left(1-\Gamma_0^{(nl)}\right)^{\alpha}$. As $\Gamma_0^{(nl)} \rightarrow 1$, $\delta \rightarrow 0$. Equation \eqref{disp2} can then be rewritten in terms of $\delta$:
\begin{eqnarray}
\label{pert1}
 \delta s^3 + s^2\left[\delta a_1(k) + a_2(k)\right] + s\left[\delta a_3(k) + a_4(k)\right]+ \delta a_5(k)=0,
\end{eqnarray}
where the coefficients $a_1$ to $a_5$ are functions of $k$ and can be obtained by comparing (\ref{pert1}) with (\ref{disp2}):
\begin{eqnarray}
\label{pertcoeff}
&&a_1(k) = \left( \frac{1}{\Pen} + \frac{4}{\Rey}\right)k^2,\\
&&a_2(k) = \frac{2Bo}{\Rey}k^2,\\
&&a_3(k)= \left( \frac{\mathbb{D}^{-1}\textit{M}\Gamma_0^{(nl)}}{\Rey}   - \frac{3}{\Rey} + \frac{4}{\Pen\Rey}k^2 + \frac{\mathbb{D}^{-1}}{2Re}k^2\right)k^2,\\
&&a_4(k)= \frac{2Bo}{\Pen\Rey}k^4 ,\\
&&a_5(k) = \left(\frac{\mathbb{D}^{-1}k^2}{2} - 3 \right)\frac{k^4}{\Pen\Rey}.
\end{eqnarray}

\begin{figure}
\centering
\subfigure[]{\label{pert:a}\includegraphics[trim= 9mm 75mm 20mm 82mm ,clip, width=0.45\textwidth]{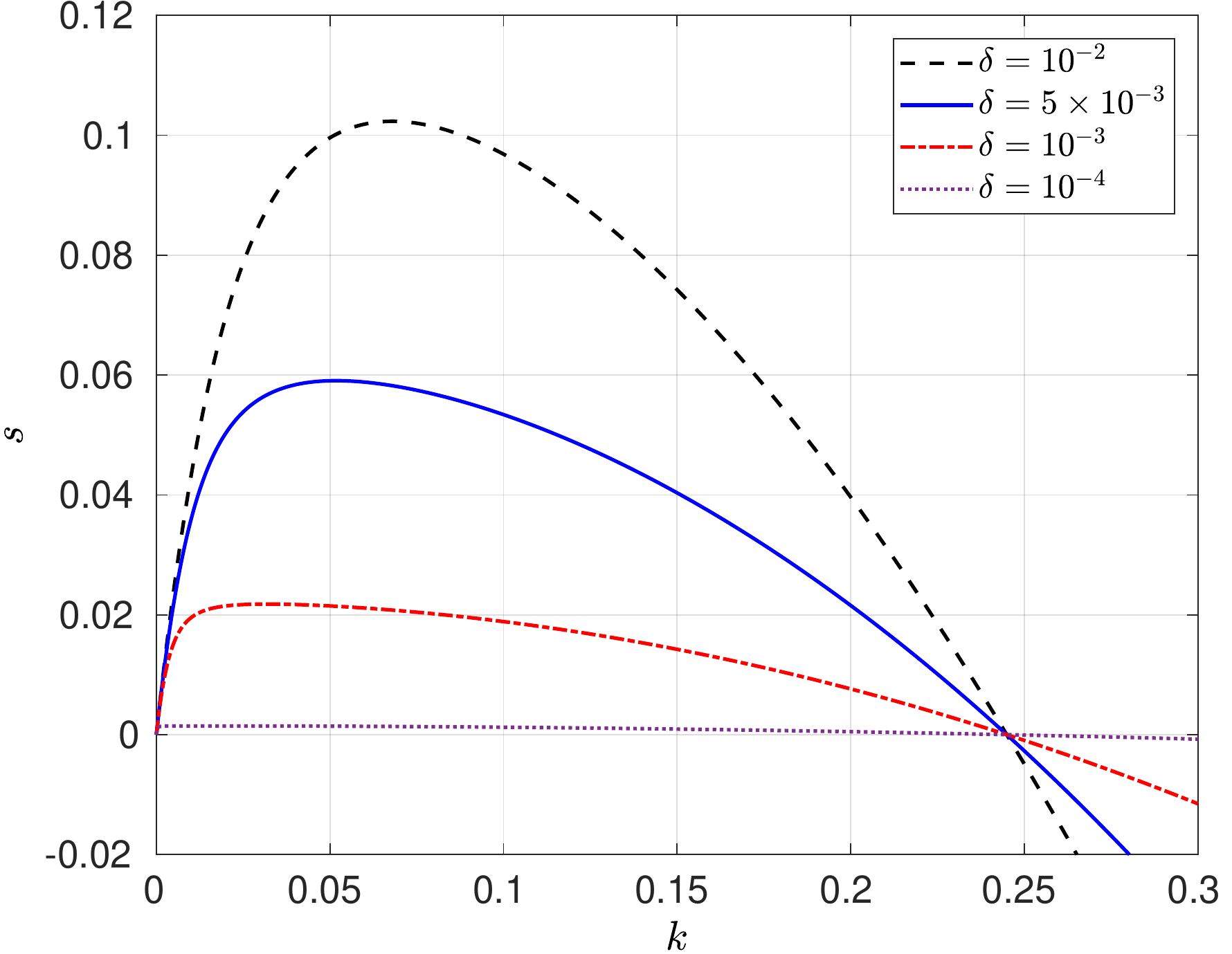}}
\subfigure[]{\label{pert:b}\includegraphics[trim= 12mm 75mm 20mm 82mm ,clip, width=0.45\textwidth]{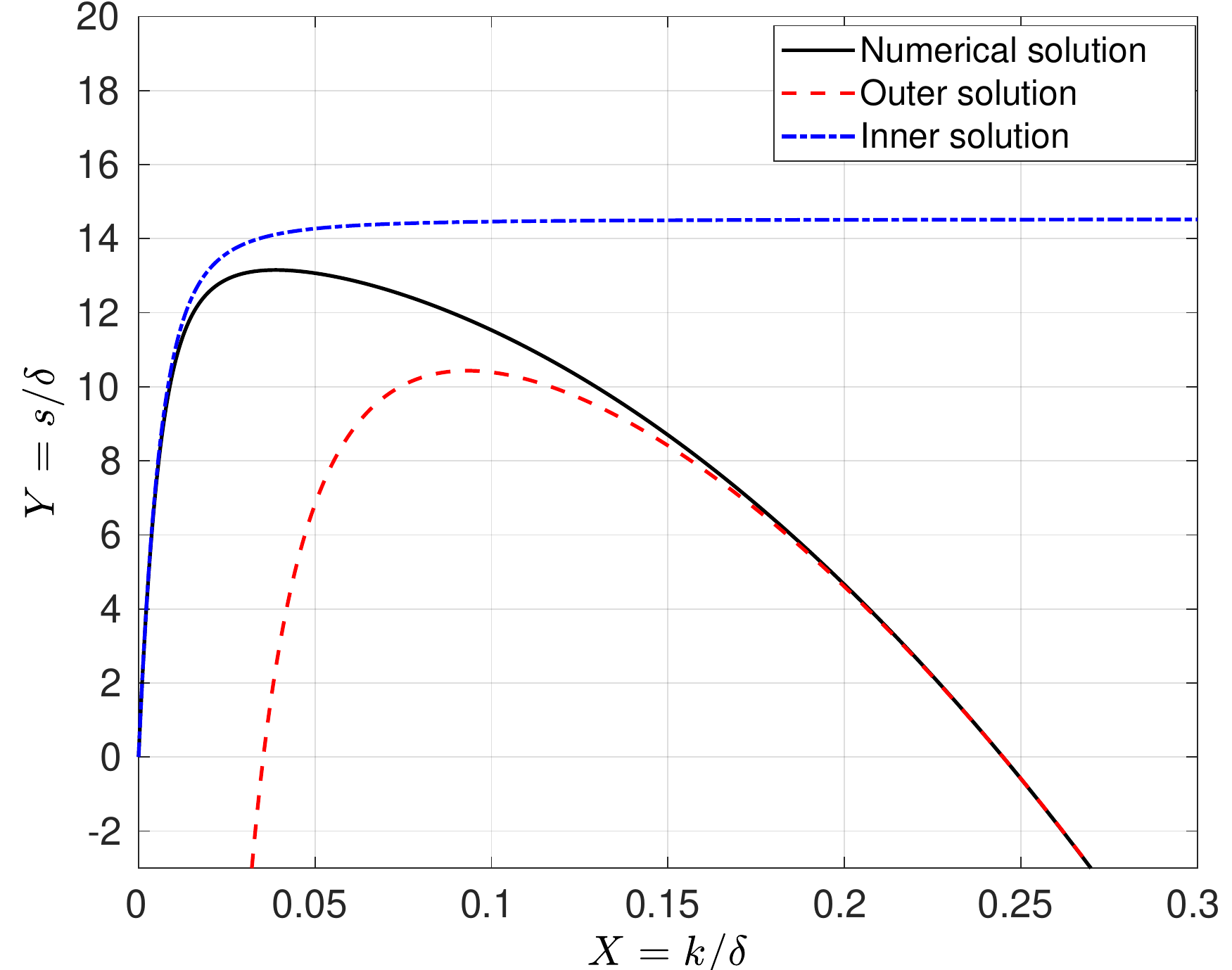}}
\caption{ (a) Unstable root for dispersion relation \eqref{pert1} with varying $\delta=(1-\Gamma_0^{(nl)})^{\alpha}$. Growth rate scales with $\delta$ and $s(k)$ curve exhibits a sharp gradient near $k\rightarrow 0$. (b) Comparison of perturbation solutions showing inner,  $Y^{(i)}$, (dot-dashed) and outer,  $Y^{(o)}$, (dashed) lines along with full dispersion relation from equation \eqref{pert2} compared with numerical roots (solid). Other parameters are fixed at $\mathbb{D}^{-1}=100, \textit{M}=10^{-3},\Pen=1,\Rey=10^{-2}, \beta=10^{-3},\alpha=2$}
\end{figure}

In the limit $\delta \rightarrow 0$, Equation (\ref{pert1}) is a standard singular perturbation problem for growth rate $s$. Fig. \ref{pert:a} shows plots of the sole positive root of (\ref{pert1}) for different values of $\delta$. It is apparent that the growth rate $s$ scales with $\delta$ and singular root appears in the region $k \le \delta$. We therefore define a rescaled growth parameter $Y = s/\delta$, so that equation (\ref{pert1}) now becomes
\begin{eqnarray}
\label{pert2}
 \delta^3 Y^3 + Y^2\left[\delta^2 a_1(k) + \delta a_2(k)\right] + Y\left[\delta a_3(k) + a_4(k)\right]+ a_5(k)=0,
\end{eqnarray}
A standard expansion in $\delta$ in the form $Y^{(o)} = Y_0^{(o)} + \delta Y_1^{(o)} + ... $ results in two regular roots in the outer region, i.e. $k \gg \delta$. Focussing on the root which is unstable, we obtain
\begin{eqnarray}
 && Y_0^{(o)} = -\frac{a_5}{a_4},\\
 && Y_1^{(o)} = -\frac{a_2 a_5^2}{a_4^3} + \frac{a_3 a_5}{a_4^2}.
\end{eqnarray}
To obtain the singular root of the inner region $0\le k\le\delta$ which has a boundary layer-like structure, we define a rescaled variable $X = k/\delta$, where $X \sim \it{O}(1)$. The solution to the inner region in terms of this rescaled variable at leading order is:
\begin{equation}
\label{isol}
  Y_0^{(i)} = -\frac{b_2}{2} + \frac{1}{2} \sqrt{b_2^2 - 4 b_3}
\end{equation}
where $\displaystyle b_2=\frac{2Bo}{\Rey}X^2;~~b_3= \left(\frac{\mathbb{D}^{-1} \textit{M}\Gamma_0}{\Rey} -\frac{3}{\Rey}\right)X^2$. Figure \ref{pert:b} shows a comparison of inner solution, $Y_0^{(i)}$ (dot-dashed line), outer solution, $Y_0^{(o)}$ (dashed line) and the numerical root, $Y$ (solid line) obtained from (\ref{pert2}). Examining (\ref{isol}), it can be shown that the necessary condition for stable growth rates ($Y_0^{(i)}<0$) is $b_3 > 0$. Hence we arrive at a necessary condition for obtaining stable modes in the space $0 \le k \le \delta$:
\begin{equation}
\label{isol2}
  \Gamma_0^{(nl)} > \frac{3}{\mathbb{D}^{-1}\textit{M}}
\end{equation}
We can therefore define a critical concentration, $\Gamma_c = 3/\mathbb{D}^{-1}\textit{M}$. Any value of $\Gamma_0^{(nl)}>\Gamma_c$ should result in inherently stable films. Also note that since $\Gamma_0^{(nl)}$ is, by definition, bounded by: $0<\Gamma_0^{(nl)}<1$, we can hypothesize that no amount of surfactants can stabilize the system if the product of the system parameters, $\mathbb{D}^{-1}\textit{M}<3$. Examining the system parameters in real physical systems, we find that $\mathbb{D}^{-1}\textit{M}\sim \it{O}\left(10^{-1}\right)$ in real systems suggesting that though foams can be stable for long durations, indefinite stability is not guaranteed. This is consistent with the observation in \cite{kloek} who studied bubble dissolution with complex interfacial and bulk rheologies.
%%%%%%%%%%%%%%%%
%%%%%%%%%%%%%%%%%%%%%%%%%%%%%%%%
It is easy to derive a simple relationship between the amplitude of tangential velocity at the interface, $\hat{c}$ and height perturbation, $\hat{h}$ using the kinematic condition \eqref{nlea}:
\begin{equation}
 \label{eq:c_vs_s}
 \hat{c} = \frac{is}{k h_0}\hat{h}
\end{equation}
The above relationship suggests that the tangential velocity at the interface, $c(x,t)$, has a phase-difference with the interface height, $h(x,t)$ since $\hat{c}$ is a complex number. Since interface height is symmetric about the rupture location, tangential velocity is anti-symmetric. This result is unsurprising since as the interface height reduces towards rupture, fluid is driven away from the rupture location. An important outcome of the above analysis is that growth rate scales with the small parameter $\delta$ in the jamming limit. This dependence can be written in the simple form
\begin{equation}
 s = \left(1-\Gamma_0^{(nl)}\right)^{\alpha} F(\Rey, \Pen, \mathbb{D}, M, Bo),
\end{equation}
where we have used the definition of $\delta$ and functional dependence on other parameters, $F$, is not reproduced for simplicity. Substituting $s$ into \eqref{eq:c_vs_s}, we get
\begin{equation}
 \label{eq:c_vs_Gamma}
 \hat{c} = \frac{iF\left(1-\Gamma_0^{(nl)}\right)^{\alpha}}{k h_0}\hat{h}
\end{equation}
As $\Gamma_0^{(nl)} \rightarrow 1$, it is clear that $c(x,t) \rightarrow 0$, thus rendering the interface immobile in the jammed limit. Both the anti-symmetric nature of tangential velocity and approach to immobile limit are confirmed through full numerical solutions of equations \eqref{nlea}, \eqref{nleb} and \eqref{nled} in section \S\ref{sec4}.

%%%%%%%%%%%%%%%%%%%%%%%%%%%%%%%%%%%%%%%%%%%%%%%%%%%%%%%%%%%%%%%%%%%%%%%%%%%%%%%%%%%%%%%%%%%%%%%%%%%%%%%%%%%%%%%%%%%%%%%%%%%%%%%%%%%%%%%%%%%%%%%%%%%%
%%%%%%%%%%%%%%%%%%%%%%%%%%%%NONLINEAR DYNAMICS%%%%%%%%%%%%%%%%%%%%%%%%%%%%%%%%%%%%%%%%%%%%%%%%%%%%%%%%%%%%%%%%%%%%%%%%%%%%%%%%%%%%%%%%%%%%%%%%%%%%%%%%%%%%%%%%%%%%%NONLINEAR DYNAMICS%%%%%%%%%%%%%%%%%%%%%NONLINEAR DYNAMICS%%%%%%%%%%%%%%%%%%%%%%%%%%%%%%%%%%%%%%%%%NONLINEAR DYNAMICS%%%%%%%%%%%%%%%%%%%%%%%%%%%%%%%%%NONLINEAR DYNAMICS%%%%%%%%%%%%%%%%%%%%%%%%%%%%%%%%%%%%%%%%%%%%%%%%%%
\section{Nonlinear evolution}\label{sec4}
\subsection{Numerical methods}
The nonlinear evolution and dynamics of the free-film was studied by numerically solving the set of partial differential equations \eqref{nlea} - \eqref{nled} for a range of parameter values. A Fourier pseudo spectral (FPS) method is used to solve these PDEs, since it is most suited to handle periodic boundary conditions. Discretization of the spatial derivatives in the Fourier domain reduces the set of PDEs to ordinary differential equations \citep{fps}. Time marching is done using an implicit Adams-Moulton method (trapezoidal rule) using MATLAB function \textit{fsolve} to solve the linearized equations in fourier space. An adaptive time stepping, $\Delta t \propto \kappa^{-1}$(curvature) is also implemented to capture the dynamics at later times, when the evolution proceeds much faster. The initial conditions used in our simulations are as follows: 
\begin{eqnarray}
\label{ic}
h(x,0)= \frac{1}{2} + 0.01\cos\left(\frac{2\pi x}{\lambda}\right),\\
\Gamma(x,0)= \Gamma_0 + \Delta\hat{\Gamma}\cos\left(\frac{2\pi x}{\lambda}\right),\\
c(x,0)=0,
\end{eqnarray}
where $\lambda=2\pi/k_{max}$ is the fastest growing wavelength, determined from linear stability analysis. The perturbation amplitude for surfactant concentration, $\Delta\hat{\Gamma}$ is set using:
\begin{equation}
\Delta\hat{\Gamma}=\frac{s_{max}\Gamma_0\Delta\hat{h}}{h_0\left(s_{max}+k^2/\Pen\right)},
\end{equation} 
which is obtained from linear stability analysis. Here, we use $\Gamma_0^{(l)} = 2 $ and $\Gamma_0^{(nl)} \in [0,0.95]$ for the linear and nonlinear models respectively. Further, we set $h_0 = 0.5,~\Delta\hat{h}=0.01$, with $s_{max}$ being the maximum value of the growth rate predicted by the linear theory, at $k=k_{max}$. The time steps used in the initial stages are of the order $\Delta t \sim O(10^{-1})$ and reduces to $\Delta t \sim O(10^{-3})$ at the latter stages of our numerical simulations. The spatial domain is discretised into 256 nodal points ($\Delta x = 1/256$) in all the simulations, ensuring grid-independence of the results.  While exploring self-similarity, the grid size and time step values are further refined. We set $\Delta x = 1/4096$ and use adaptive time-stepping with lower end values of the order $\Delta t \sim O(10^{-5})$. We assume the film to have ruptured when the minimum height of the film, $h_{min} \le 10^{-4}$.

%%%%%%%%%%%%%%%%%%%%%%%%%%%%%%%%%%%%%%%%%%%%%%%%%%%%%%%%%%%%
%%%%%%%%%%%%%%%%%%%%%%%%%%%%%%%%%%%%%%%%%%%%%%%%%%%%%%%%%%%%
%%%%%%%%%%%%%%%%%%%%%%%%%%%%%%%%%%%%%%%%%%%%%%%%%%%%%%%%%%%%
%%%%%%%%%%%%%%%%%%%%%%%%%%%%%%%%%%%%%%%%%%%%%%%%%%%%%%%%%%%%
%%%%%%%%%%%%%%%%%%%%%%%%%%%%%%%%%%%%%%%%%%%%%%%%%%%%%%%%%%%%
%%%%%%%%%%%%%%%%%%%%%%%%%%%%%%%%%%%%%%%%%%%%%%%%%%%%%%%%%%%%
%%%%%%%%%%%%%%%%%%%%%%%%%%%%%%%%%%%%%%%%%%%%%%%%%%%%%%%%%%%%
%%%%%%%%%%%%%%%%%%%%%%%%%%%%%%%%%%%%%%%%%%%%%%%%%%%%%%%%%%%%
%%%%%%%%%%%%%%%%%%%%%%%%%%%%%%%%%%%%%%%%%%%%%%%%%%%%%%%%%%%%
%%%%%%%%%%%%%%%%%%%%%%%%%%%%%%%%%%%%%%%%%%%%%%%%%%%%%%%%%%%%
\subsection{`Dilute limit' analysis}
\subsubsection{Evolution and parametric studies}
\begin{figure}
\centering
\subfigure[]{\label{fig3:a}\includegraphics[trim= 12mm 82mm 24mm 86mm ,clip, width=0.45\textwidth]{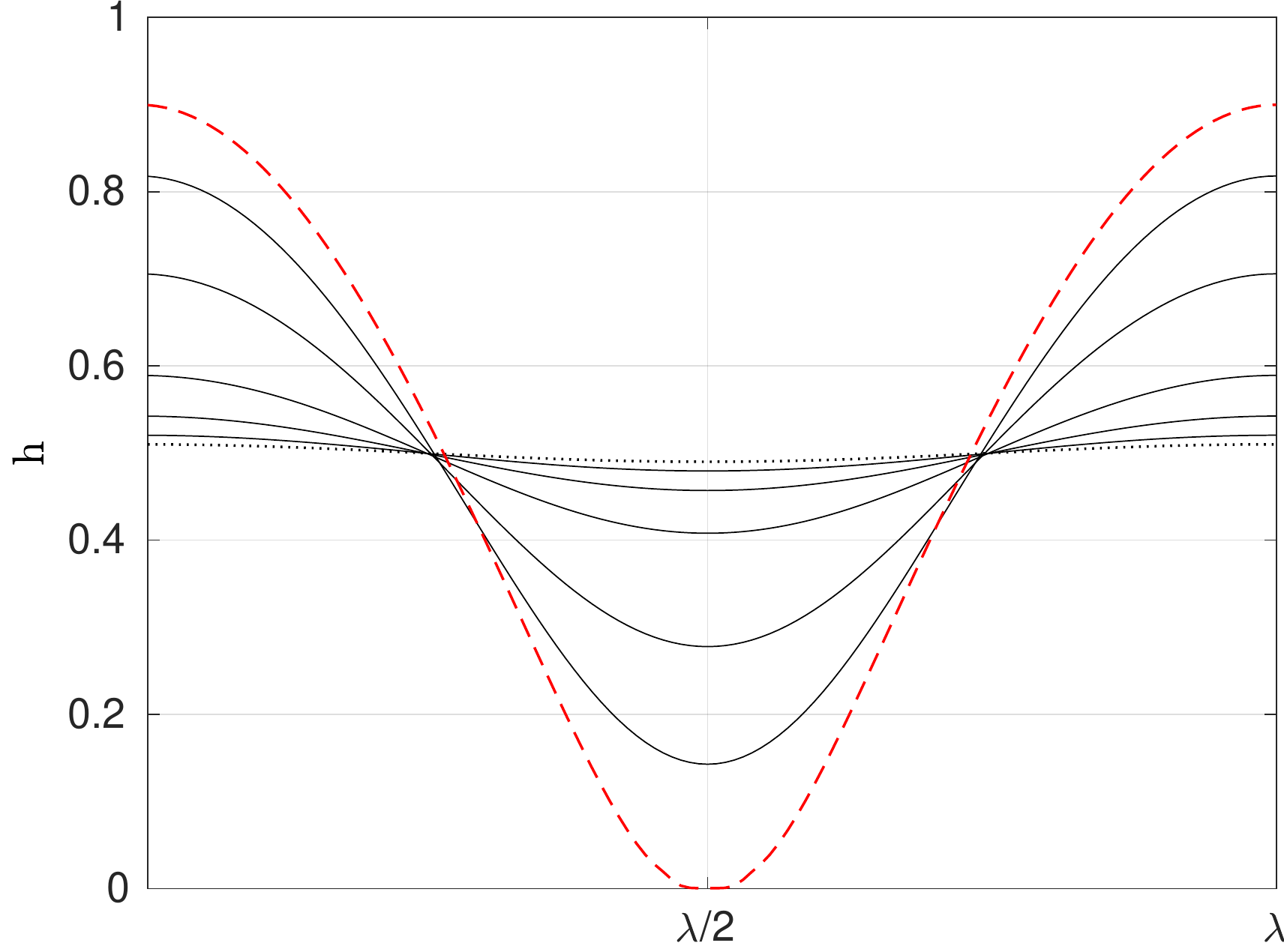}}
\subfigure[]{\label{fig3:b}\includegraphics[trim= 12mm 84mm 24mm 84mm ,clip, width=0.45\textwidth]{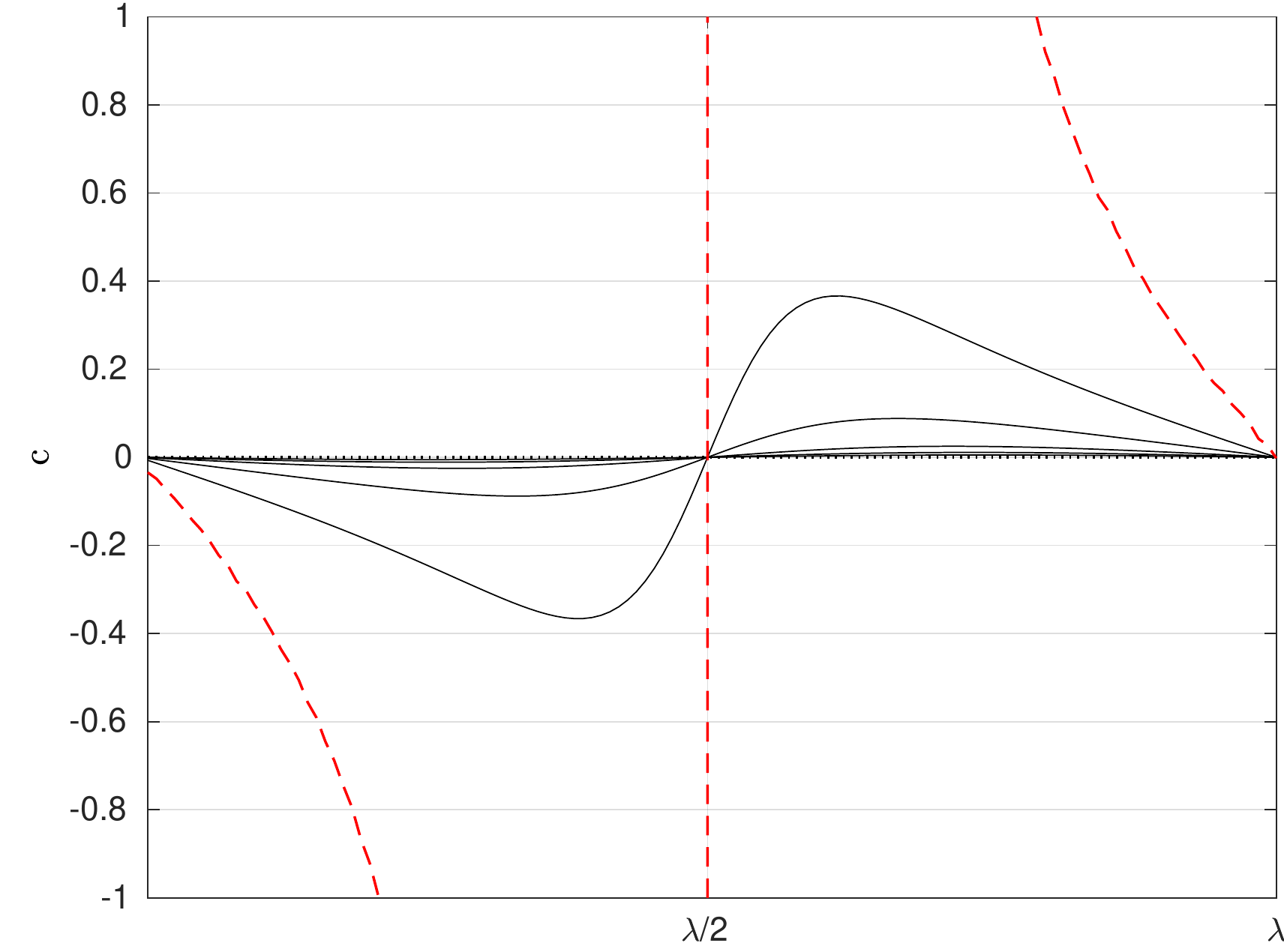}}
\subfigure[]{\label{fig3:c}\includegraphics[trim= 12mm 80mm 24mm 84mm ,clip, width=0.45\textwidth]{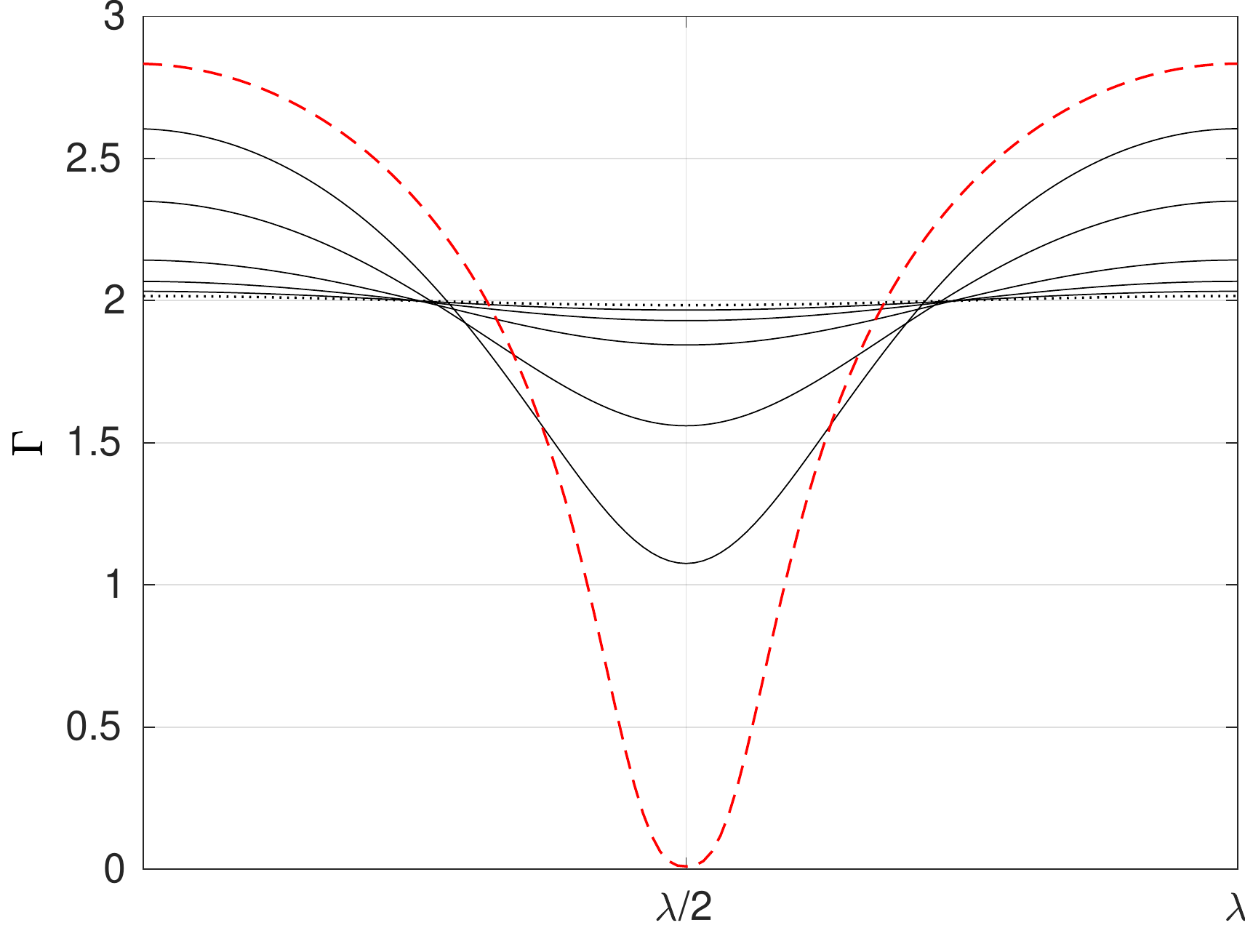}}
\subfigure[]{\label{fig3:d}\includegraphics[trim= 12mm 79mm 24mm 84mm ,clip, width=0.45\textwidth]{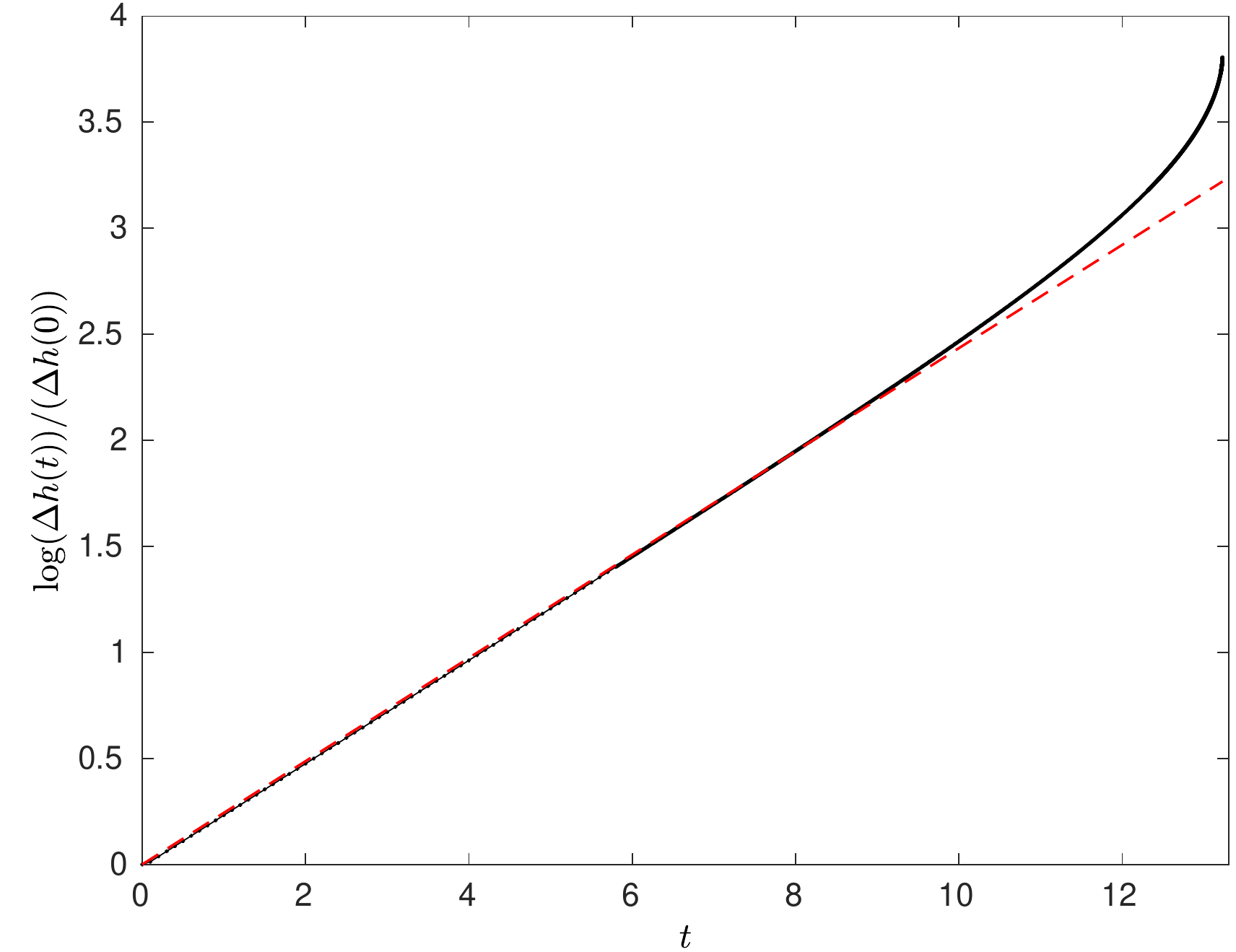}}
\caption{Spatiotemporal evolution of interface height (a), tangential velocity (b) and surfactant concentration (c) for the linear model obtained by solving equations \eqref{nlea}-\eqref{nlec} at various times $t=0,3,6,9,12,13,13.22$. Initial condition at $t=0$ is shown with a dotted curve and rupture profile shown with dashed curve occurs at $t=13.22$. The other parameters used are $\mathbb{D}^{-1}=50,\textit{M}=10^{-3}= \beta,\textit{Bo}=1,\Pen=20,\Rey=2$. (d) Comparison of numerically obtained growth rate and linear theory predictions for the same set of parameters. Here $\Delta h(t)$ is the difference between the maximum and minimum values of $h(x,t)$ at any time $t$ (see \cite{atul2019}).}
\label{fig:nonlinear_spatiotemporal}
\end{figure}
The nonlinear dynamics using the linear model (\ref{eta1}) for surface viscosity in a film with dilute concentration of surfactants are discussed first. We remind that this is a special limit of NPM as discussed in \S\ref{sec:scalings}. Figure 5 shows spatiotemporal evolution profiles for (a) film height $h$, (b) concentration $\Gamma$, (c) velocity at the interface, $c$ for a typical case: $\mathbb{D}=0.02,~\textit{M}=10^{-3}, ~\beta=10^{-3},~\textit{Bo}=1,~\Pen=20,~\Rey=2$. The profiles have been plotted on spatial coordinates over a domain length $\lambda$. Given that the boundary conditions are periodic, successive crests/troughs in the profiles are separated by $\lambda$, consistent with expectations of the spinodal dewetting mechanism. The numerically obtained growth rate is also validated with the predictions of linear theory, as is shown in fig. \ref{fig3:d}. At short times, the linear theory predictions are in agreement with the nonlinear solutions, but deviate at later times closer to rupture, when the nonlinear terms become significant. Parametric studies on film evolution in the `dilute' regime reveal that qualitative features of evolution follow the trends observed in earlier studies \citep{matar,davis}. The rapid formation of sharp ruptures in the profiles of $h$ and $\Gamma$ indicate the possibility of self-similar solutions close to rupture.
\begin{figure}
\subfigure[]{\label{fig6:a}\includegraphics[trim= 15mm 76mm 18mm 84mm ,clip, width=0.49\textwidth]{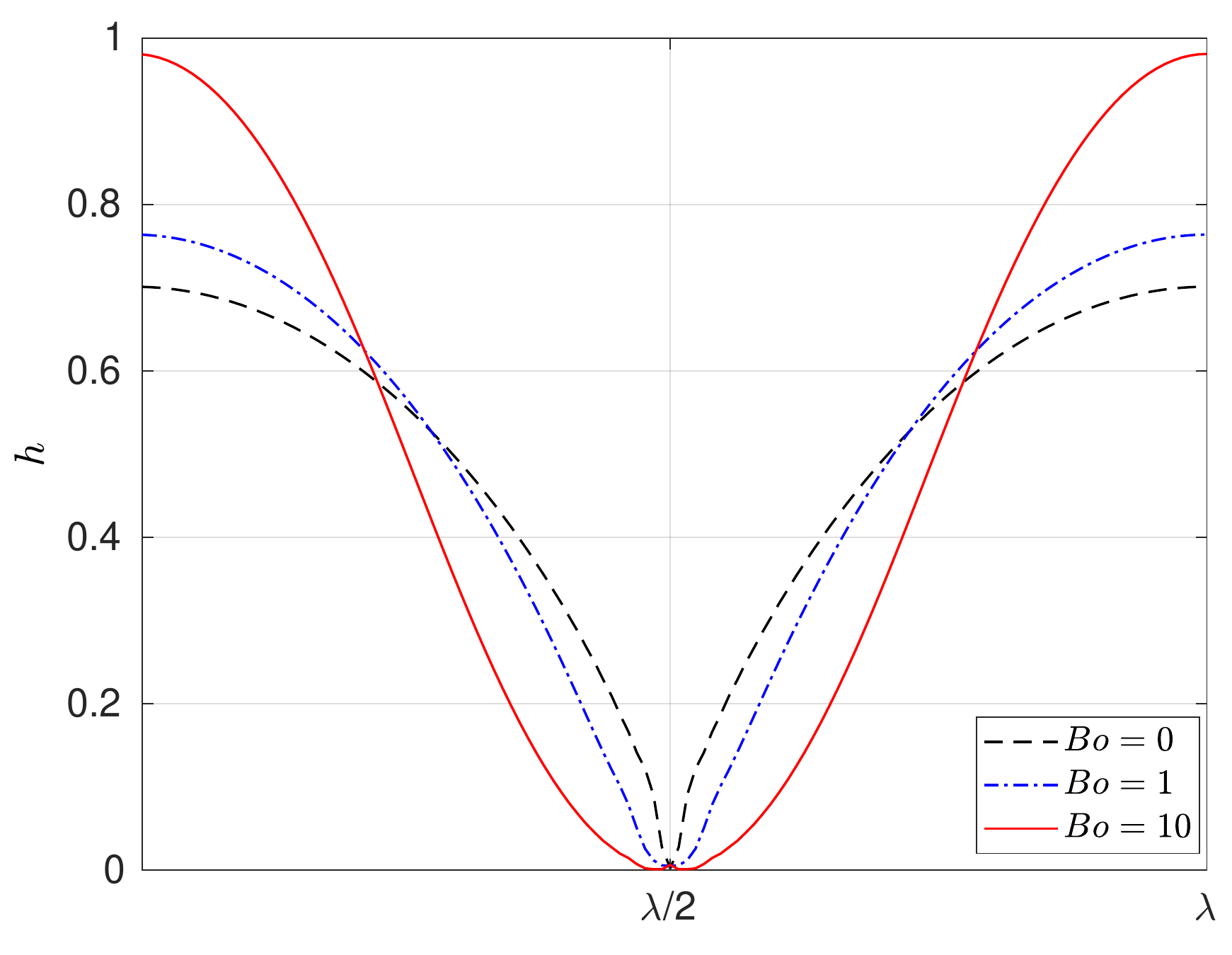}}
\subfigure[]{\label{fig6:b}\includegraphics[trim= 12mm 76mm 22mm 84mm ,clip, width=0.49\textwidth ]{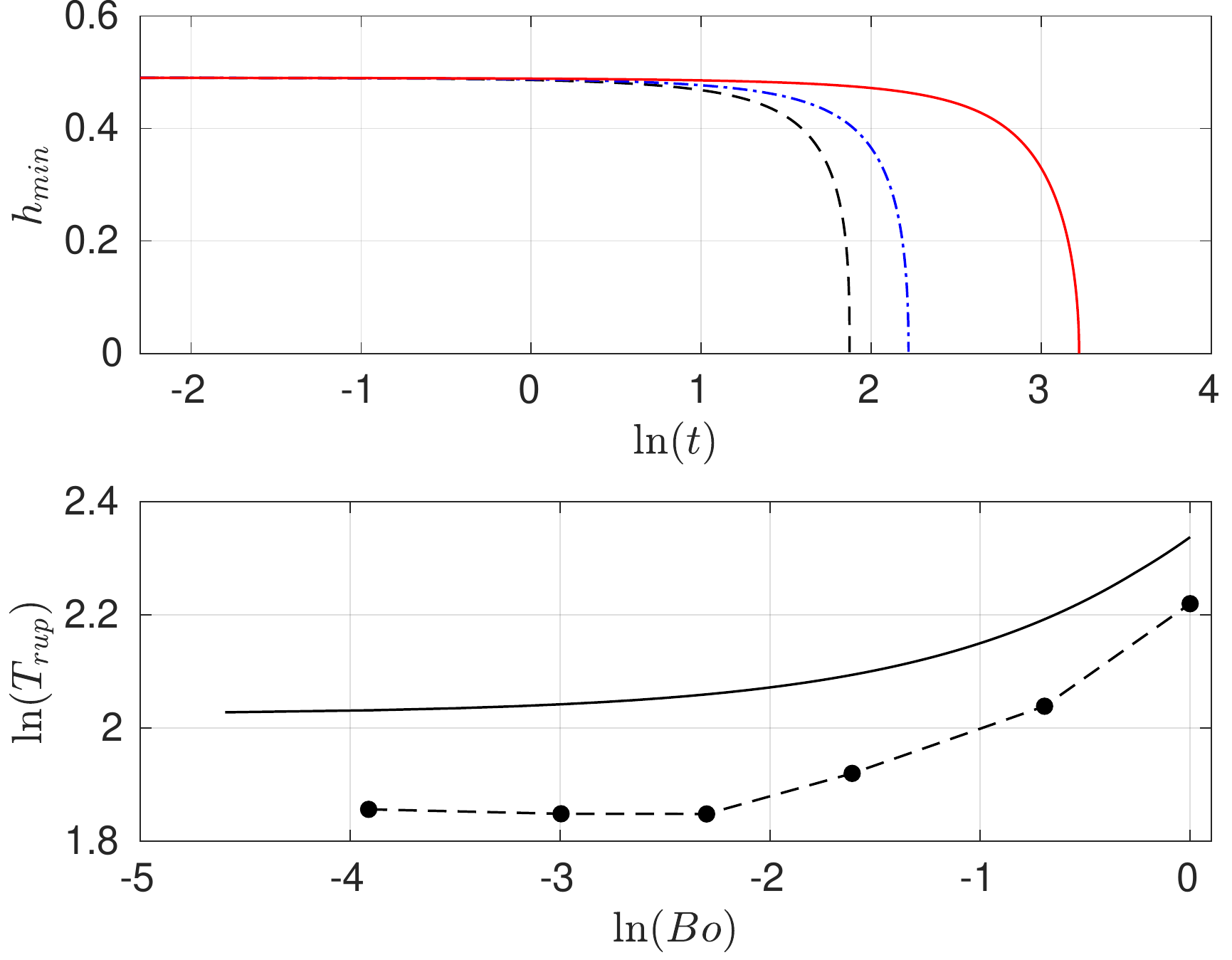}}
\subfigure[]{\label{fig6:c}\includegraphics[trim= 12mm 76mm 20mm 84mm ,clip, width=0.49\textwidth]{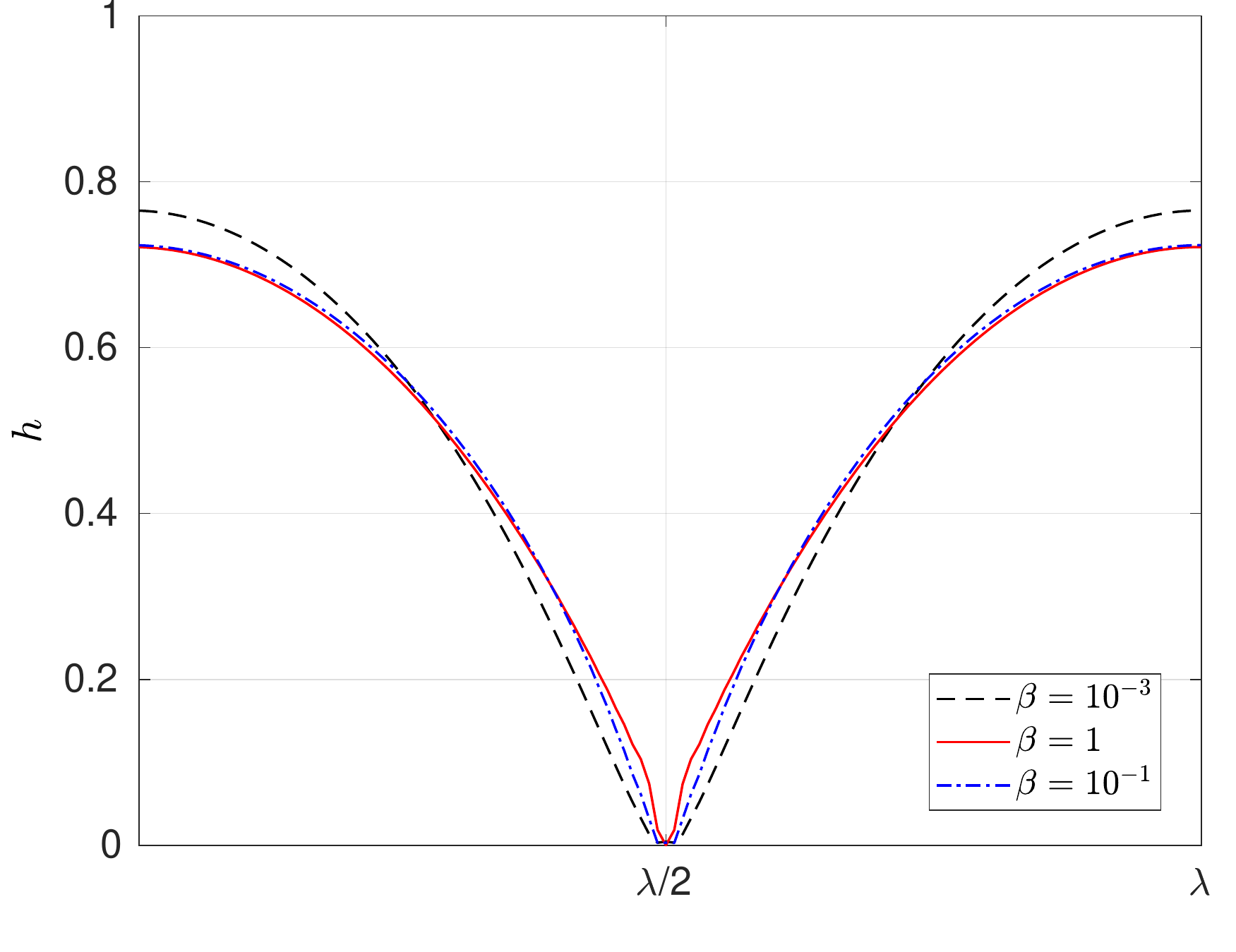}}
\subfigure[]{\label{fig6:d}\includegraphics[trim= 12mm 76mm 24mm 84mm ,clip, width=0.49\textwidth]{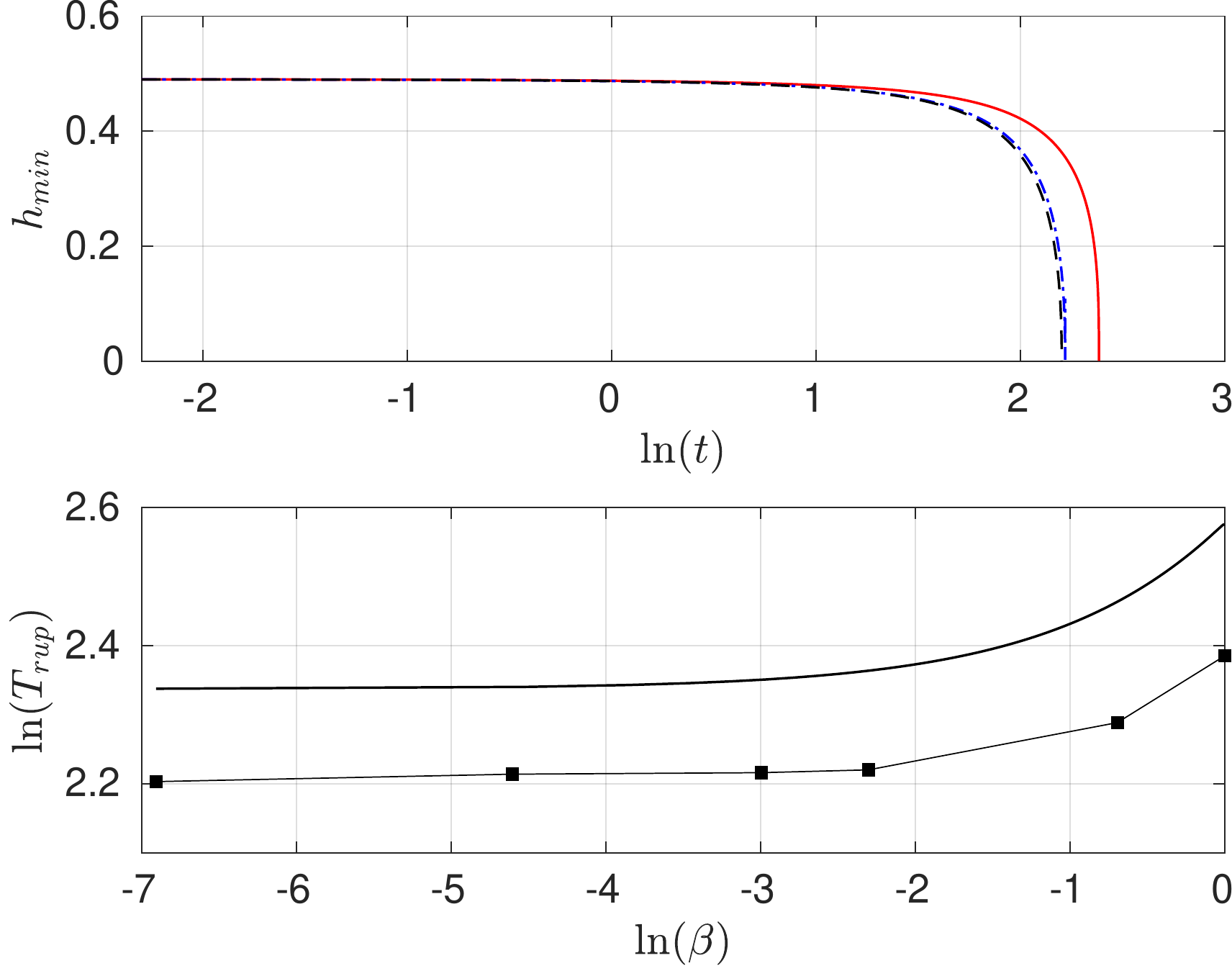}}
\caption{Nonlinear results from the linear model obtained by solving \eqref{nlea}-\eqref{nlec}. (a) Effect of varying $\textit{Bo}$ on height profiles at the time of rupture with $\beta=10^{-3}$ fixed. $Bo=0$ corresponds to the case of zero surface viscosity and exhibits a cusp-like solution. Increasing $Bo$ causes the interface to flatten near rupture. (b) Evolution of minimum film thickness ($h_{min}$) with time and comparison of rupture time predictions from linear stability analysis (solid) and nonlinear simulations (circles) for varying $\textit{Bo}$ as used in (a). (c) Effect of varying $\beta$ on height profiles at the time of rupture with $\textit{Bo}=1$. As $\beta$ increases, the film profile near rupture approaches a cusp-like solution. (d) Evolution of minimum film thickness ($h_{min}$) with time and comparison of rupture time predictions from linear stability analysis (solid) and nonlinear simulations (squares) for varying $\beta$. Other parameter values are fixed at $\mathbb{D}=10^{-2},~\Pen=1,~\Rey=10^{-2},~\textit{M}=10^{-3}$.}
\end{figure}

Figure 6 shows the effect of surface viscosity ($\textit{Bo}$) and its linear concentration dependence (represented by $\beta$) on the interface profile and film evolution kinetics. For smaller values of $\textit{Bo}$, the surface viscosity contribution is lesser, and a sharp, pointed rupture is observed \citep{davis,dewit}. Increasing $\textit{Bo}$ flattens the cusps formed in the vicinity of rupture. Surface viscosity may be interpreted as a diffusivity of surface velocity or momentum, hence it neutralizes velocity gradients at the surface. Therefore, there is greater advection of the liquid onto either side near the rupture location resulting in a flatter rupture profile. On the other hand, increasing the parameter $\beta$ produces a more cusp-like rupture, as shown in figure \ref{fig6:c}. This is understandable since surface viscosity from \eqref{eta1} at rupture location reduces to simply becoming $\eta|_{x=x_r}\approx 1-\beta$ since $\Gamma \ll 1$ near rupture. Therefore increasing $\beta$ reduces the local value of surface viscosity near rupture location. It has to be noted that the maximum value of $\beta$ cannot exceed unity as a result of positivity of $\eta$. The upper limit of $\beta=1$ causes the non-dimensional surface viscosity to become identical to the non-dimensional surface concentration, i.e. $\eta|_{\beta=1} = \Gamma$. Since $\Gamma \rightarrow 0$ at the rupture location, $\eta$ too vanishes at the rupture location. This causes the film profile to bear resemblance to that of a zero surface viscosity case as is clearly evident with the cusp-like profile at $\beta=1$ in figure \ref{fig6:c}. This profile matches very well with the case of zero surface viscosity, $\textit{Bo}=0$ as shown in figure \ref{fig6:a}. This is further discussed in \S\ref{sec:self}. The rupture times appear to be of the same order of magnitude for a wide range of values of $\beta$ as seen in Fig. \ref{fig6:d}. 

%%%%%%%%%%%%%%%%%%%%%%%%%%%%%%%%%%%%%%%%%%%%%%%%%%%%%%%%%%%%%%%%%%%%%%%%%
%%%%%%%%%%%%%%%%%%%%%%%%%%%%%%%%%%%%%%%%%%%%%%%%%%%%%%%%%%%%%%%%%%%%%%%%%
%%%%%%%%%%%%%%%%%%%%%%%%%%%%%%%%%%%%%%%%%%%%%%%%%%%%%%%%%%%%%%%%%%%%%%%%%
%%%%%%%%%%%%%%%%%%%%%%%%%%%%%%%%%%%%%%%%%%%%%%%%%%%%%%%%%%%%%%%%%%%%%%%%%
%%%%%%%%%%%%%%%%%%%%%%%%%%%%%%%%%%%%%%%%%%%%%%%%%%%%%%%%%%%%%%%%%%%%%%%%%
%%%%%%%%%%%%%%%%%%%%%%%%%%%%%%%%%%%%%%%%%%%%%%%%%%%%%%%%%%%%%%%%%%%%%%%%%
%%%%%%%%%%%%%%%%%%%%%%%%%%%%%%%%%%%%%%%%%%%%%%%%%%%%%%%%%%%%%%%%%%%%%%%%%
%%%%%%%%%%%%%%%%%%%%%%%%%%%%%%%%%%%%%%%%%%%%%%%%%%%%%%%%%%%%%%%%%%%%%%%%%
%%%%%%%%%%%%%%%%%%%%%%%%%%%%%%%%%%%%%%%%%%%%%%%%%%%%%%%%%%%%%%%%%%%%%%%%%
%%%%%%%%%%%%%%%%%%%%%%%%%%%%%%%%%%%%%%%%%%%%%%%%%%%%%%%%%%%%%%%%%%%%%%%%%
%%%%%%%%%%%%%%%%%%%%%%%%%%%%%%%%%%%%%%%%%%%%%%%%%%%%%%%%%%%%%%%%%%%%%%%%%
%%%%%%%%%%%%%%%%%%%%%%%%%%%%%%%%%%%%%%%%%%%%%%%%%%%%%%%%%%%%%%%%%%%%%%%%%
\subsubsection{Self-similar solutions}\label{sec:self}

%%%%%%%%%%%%%%
\begin{figure}
\centering
\subfigure[]{\label{selfplot2}\includegraphics[trim= 12mm 80mm 20mm 96mm ,clip, width=0.45\textwidth]{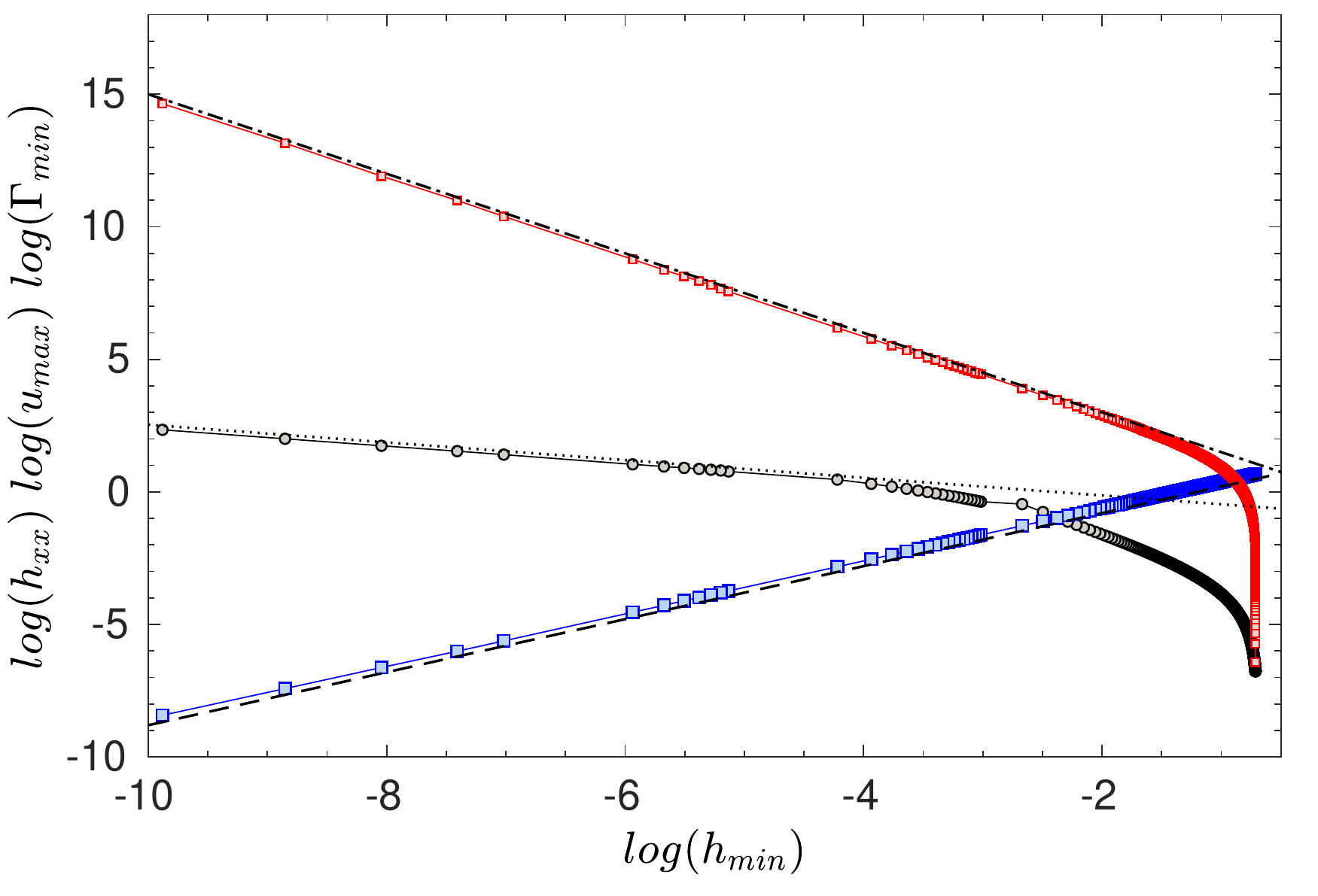}}
\subfigure[]{\label{hvk2}\includegraphics[trim= 15mm 80mm 20mm 92mm ,clip, width=0.45\textwidth]{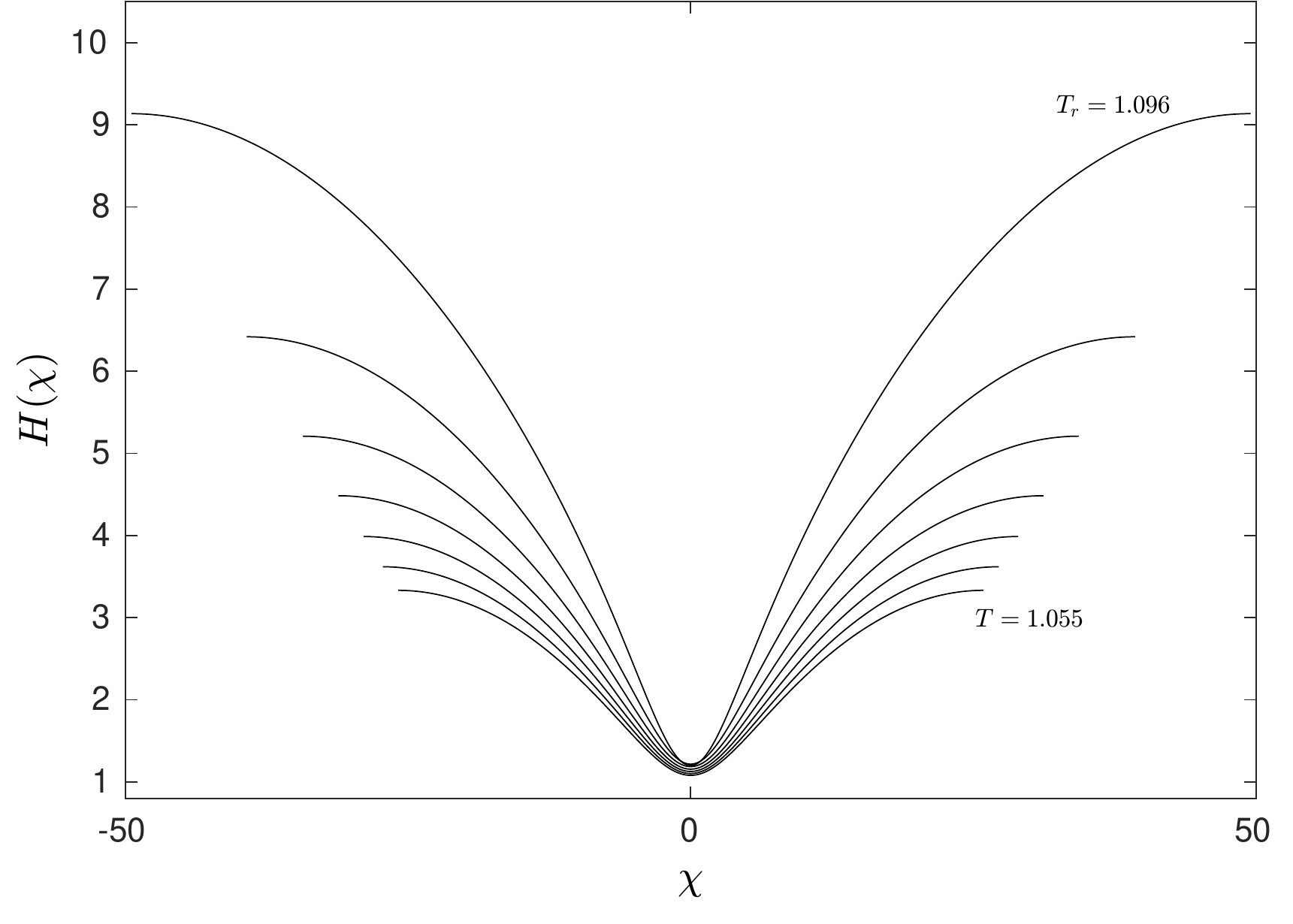}}
\subfigure[]{\label{fig:blu}\includegraphics[trim= 12mm 80mm 20mm 90mm ,clip, width=0.45\textwidth]{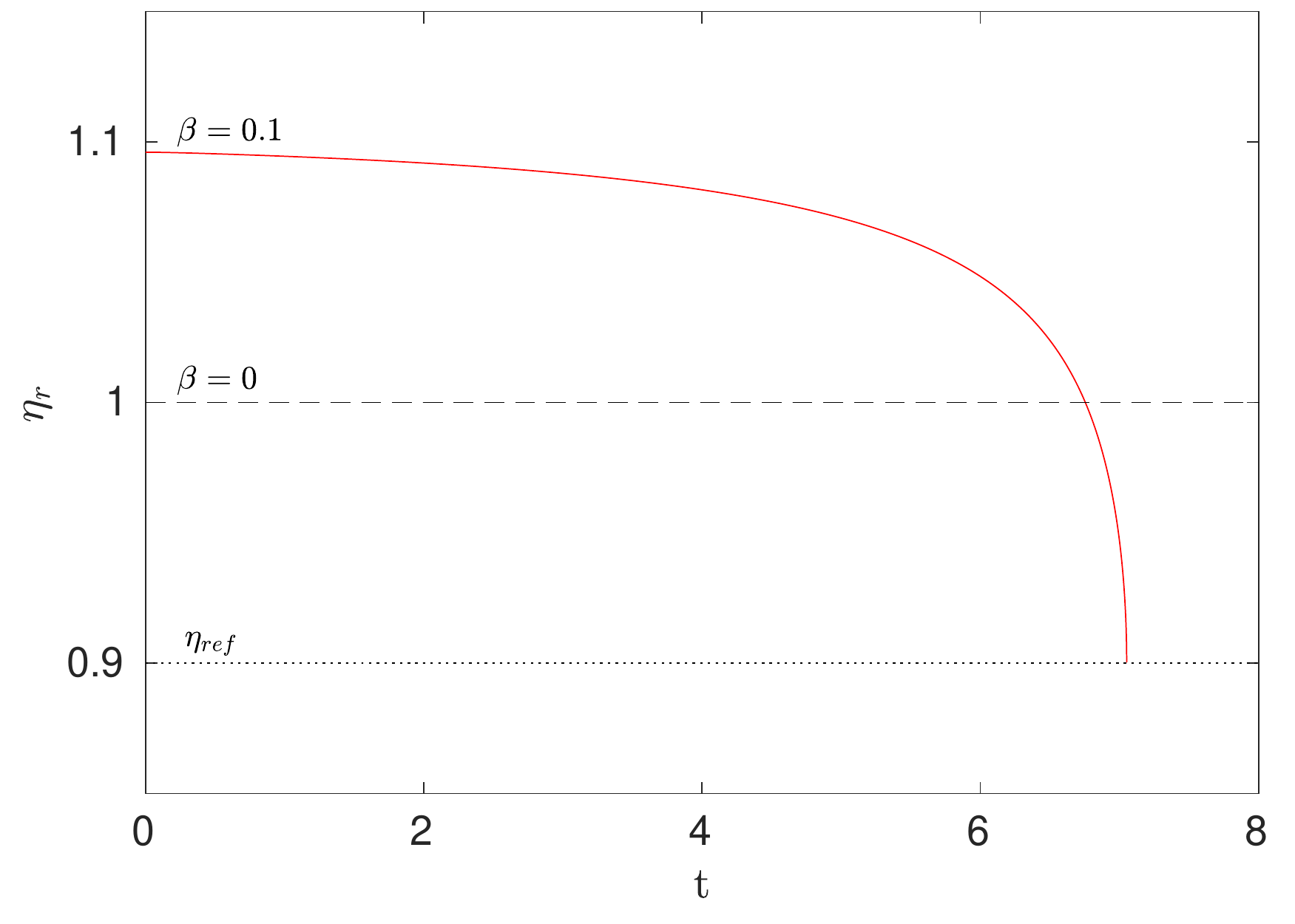}}
\subfigure[]{\label{hvk1}\includegraphics[trim= 15mm 80mm 20mm 90mm ,clip, width=0.45\textwidth]{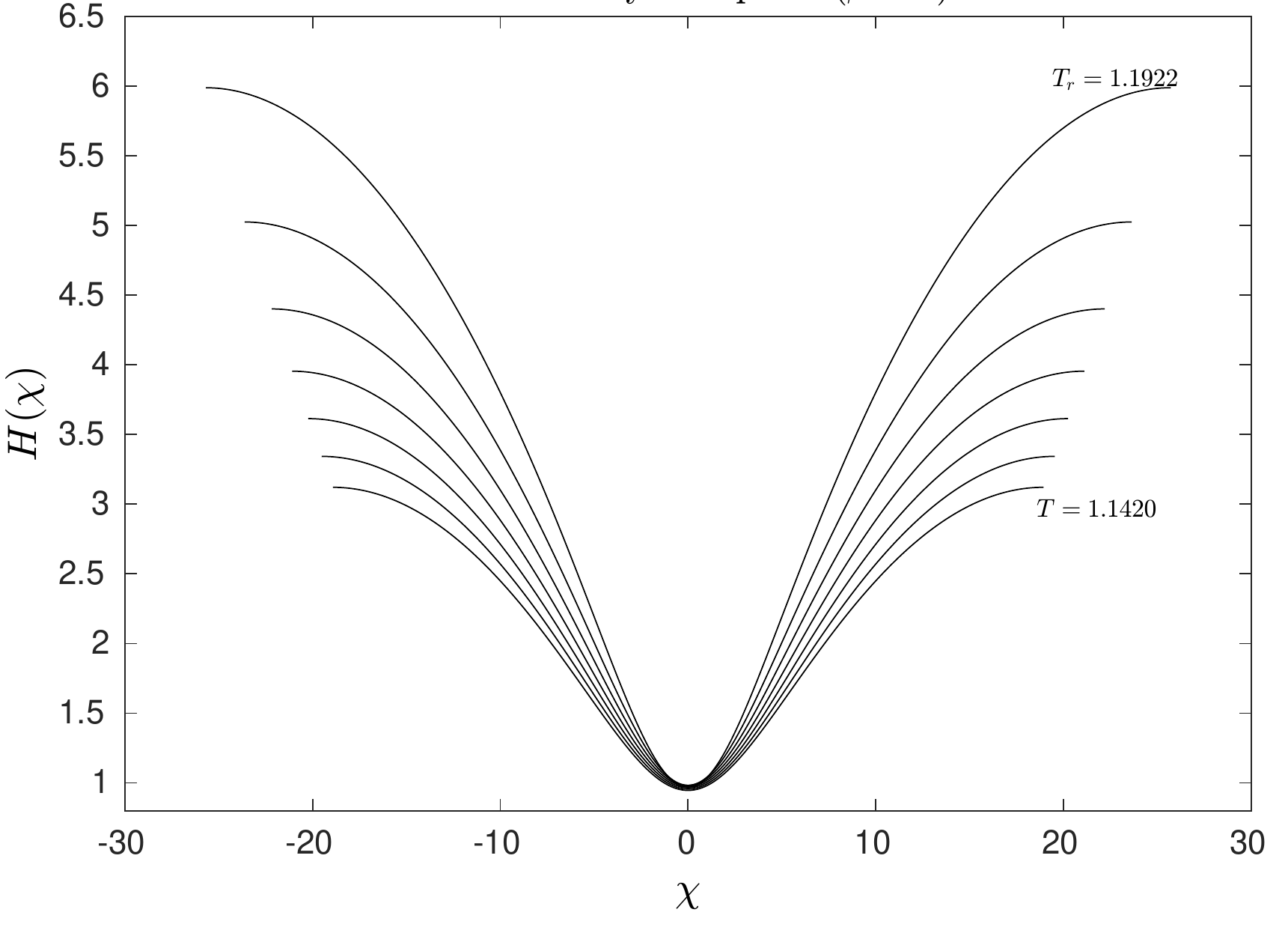}}
\caption{Results from self-similar analysis. (a) Plot of $(h^{min}_{xx})$ (circles), $(c_{max})$ (diamonds), and $(\Gamma_{min})$ (squares) vs $(h_{min})$ plotted in the log-log scale for $\beta = 0.5$ reveals power-law dependence. Relevant scaling exponents are extracted from the slopes of these profiles (see text for details). In the limit of small $h_{min}$, the slopes assume the values of -1.5 (dot-dashed line), -0.3 (dotted line) and 1 (dashed line). (b) Self-similar film profiles at the rupture location close to the break-up time $T_r = 1.096$ for $\beta=0.5$. (c) Variation of surface viscosity with time at rupture location for linear model with $\beta=0.1$ (solid) and constant surface viscosity (dashed), $\beta=0$. The value of $\eta_{ref}$ is also shown in the figure and it denotes the value of reference surface viscosity on a clean interface. At rupture location, surfactant concentration goes to zero causing $\eta$ to approach $\eta_{ref}$. Note that surface viscosity at rupture location for $\beta=0.1$ is lower than that for a case with constant surface viscosity $\beta=0$. (d) Self-similar film profiles at the rupture location close to the break-up time $T_r = 1.192$ for the case of constant surface viscosity, $\beta=0$. Interface profiles in (b) exhibit a higher value of curvature than in (d) owing to lower surface viscosity as shown in (c). Other parameters are $\Pen=1,\textit{M}=10^{-1},\Rey=10^{-2},\mathbb{D}=10^{-3},\textit{Bo}=1$.}
\label{fig:self-similarity}
\end{figure}
%%%%%%%%%%%%%%%%%%%%%%%%
Next, we investigate the presence of self-similar behaviour in free film profiles in the vicinity of rupture location. Equations \eqref{nlea}-\eqref{nlec} do not lead to perfect self-similar solutions, but contain a few terms with weak time dependence. We assume solutions for the unknowns $h(x,t)$, $\Gamma(x,t)$ and $c(x,t)$ in the form:
%%%
\begin{eqnarray}
\label{self}
&& h(x,t)=\tau^i H(\chi), \qquad \qquad c(x,t)=\tau^k U(\chi), \nonumber \\
&& \Gamma(x,t)=\tau^l G(\chi), \qquad \qquad \chi(x,t)=(x - x_r)/\tau^j,
\end{eqnarray}
%%%
where $\chi$ is the similarity variable and $\tau=t_r - t$, $x_r$ and $t_r$ being the spatial location and time of rupture, respectively. Substituting (\ref{self}) into (\ref{nlea})-(\ref{nlec}) and subsequent simplification yields:
\begin{eqnarray}
\label{self2}
&&H_\chi j\chi - iH + \mathbf{\tau}^{1+k-j}(UH)_\chi = 0 , \\ 
\label{self3}
&& -lG + jG_\chi \chi + \tau^{1+k-j}(UG)_\chi = \tau^{1-2j}\frac{G_{\chi\chi}}{\Pen} ,\\ 
\label{self4}
&& \Rey\left(\tau^{2j-1}(j\chi U_\chi) + \tau^{i+j+k}UU_\chi\right) - \mathbb{D}^{-1}\tau^{2i-k-j}H_{\chi\chi\chi} - \tau^{-2i-k+j}\frac{3H_\chi}{8H^4} - 4 \tau^i U_{\chi\chi} \nonumber \\ 
&& = \frac{4 \tau^i H_\chi U_\chi}{H} + \tau^{2i+l}\textit{Bo}\beta\left(\frac{U_\chi G_\chi}{H} + \frac{GU_{\chi\chi}}{H}\right) + \textit{Bo}(1-\beta)\frac{U_{\chi\chi}}{H} - \tau^{j+l-k}\mathbb{D}^{-1}\textit{M}\frac{G_\chi}{H}.
\end{eqnarray}

To obtain scalings followed by $h$, $\Gamma$ and $c$, we follow the approach used by \cite{matar} and \cite{lister}. These scalings allow us to extract the exponents $i, j, k, l$ without knowledge of rupture time \emph{a priori}. The natural logarithm of the curvature $(h^{min}_{xx})$, velocity $(c_{max})$ and surfactant concentration on the surface $(\Gamma_{min})$ at $x_r$ are plotted against the natural logarithm of film thickness $(h_{min})$ at $x_r$. Three algebraic equations in terms of $i,j,k,l$ can be generated by comparing the slopes of $(h^{min}_{xx})$, $(c_{max})$ and $(\Gamma_{min})$, shown in figure \ref{selfplot2}, with relations obtained from equation \eqref{self}. A number of choices are possible for the fourth equation. For self-similarity, we require the exponent of $\tau$ in all the terms of equations \eqref{self2}-\eqref{self4} to vanish. An algebraic equation can be written by setting the powers of $\tau$ to zero in the above equations and each such equation can be solved along with the three algebraic equations obtained from numerical simulation. This leads to multiple solutions for $i,j,k,l$. Most of the solutions can be discarded on physical grounds. For example, since $h(x,t),\Gamma(x,t) \rightarrow 0$ during rupture as evident in figure \ref{fig:nonlinear_spatiotemporal}, we require $i>0$ and $l > 0$. Similarly, tangential velocity, $c(x,t)$ rapidly increases on either side of rupture location since fluid is rapidly driven away from the rupture location as interface height reduces to zero. This requires $k<0$ in \eqref{self}. We can also argue on similar grounds that $j>0$ in the definition of similarity variable $\chi$. Of the many solutions that are possible for $i,j,k,l$, a large number of solutions are ruled out by the above constraints on these variables. It has to be noted that the above procedure does not eliminate $\tau$ dependence in all the terms since the equations do not admit perfect self-similarity as was noted in earlier works. The terms which retain $\tau$ dependence in \eqref{self2}-\eqref{self4} become vanishingly small in the limit of $\tau \rightarrow 0$. This ensures that the system evolves towards a self-similar state as one approaches the time of rupture. This analysis gives physical insights into the dominant forces in operation at the onset of rupture, essentially among the terms with vanishing $\tau$-dependence.

Since the primary aim of this work is to shed light on effects of variable surface viscosity, we restrict analysis in this section only to the case of varying $\beta$. For comparison, we also report the case of zero surface viscosity. Recall that surface viscosity varies with concentration in the linear model by the relation $\eta = 1+\beta (\Gamma-1)$. We consider four test cases for the study as governed by equations \eqref{nlea}-\eqref{nlec}: (i) $Bo=0$ which corresponds to the case of zero surface viscosity as reported by \cite{lister}; (ii) $\beta=1$ where surface viscosity and concentration assume equivalence; (iii) $\beta=0.5$ which corresponds to the case of variable surface viscosity and is a canonical case for $0<\beta<1$; and (iv) $\beta=0$ which corresponds to the case of constant surface viscosity as reported by \cite{matar}.
%%%%%%%%%%%%%%
\begin{figure}
\centering
\includegraphics[trim= 35mm 40mm 26mm 35mm ,clip, width=1.0\textwidth]{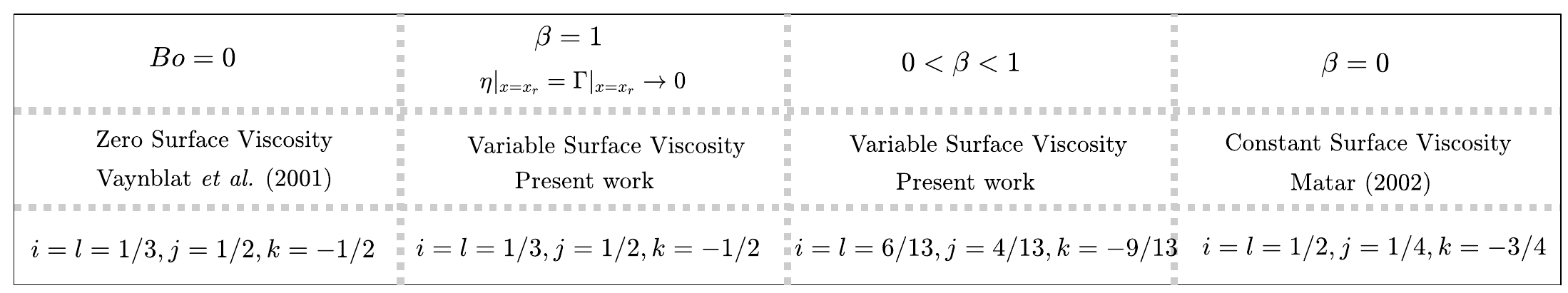}
\caption{Comparison of self-similarity scaling exponents with various surface viscosity models. For reference, surface viscosity in the linear model assumes the form $\eta = 1+\beta(\Gamma-1)$ and $Bo = \tilde{\eta}_0/\mu \tilde{h}_0$ allowing for comparison with existing literature for various values of $Bo$ and $\beta$.}
\label{fig:similarity_comparison}
\end{figure}
%%%%%%%%%%%%%%%%%%%%%%%%
Results for case (ii) are not shown here but will be constrasted with other cases. Further $\beta$ was varied from $0.1$ to $0.9$ in steps of $0.1$ to study the effect of $\beta$ on similarity behaviour. It is found that scaling exponents remained unaffected with $\beta$ and the reason for the same is explained below.

Results for $\beta=0.5$ are shown in figures \ref{selfplot2} and \ref{hvk2}. Following the procedure as explained above, the following set of exponents are found: $i = l = 6/13, j = 4/13, k = -9/13$. These exponents values are different when compared to the constant viscosity case studied by \cite{matar}. In our case, this occurs for $\beta = 0$ and yields the exponents $i=l=1/2,j=1/4,k=-3/4$ which agree with the values reported in \cite{matar}. The exponents with variable surface viscosity are only slightly different from the constant viscosity case, nevertheless they reveal that self-similar behaviour is altered with variable surface viscosity. For case (ii) with $\beta=1$, surface viscosity all along the interface reduces to the expression, $\eta=\Gamma$. Since surfactant concentration at rupture location becomes vanishingly small, surface viscosity too becomes negligible in the vicinity of rupture. The following exponents are found for this case: $i=l=1/3,j=1/2,k=-1/2$ which is identical to the case with zero surface viscosity as obtained in \cite{lister} and corresponds to case (i) in our study with $Bo=0$. It has to be noted that surface viscosity has a finite value away from the rupture point and this leads to deviation of the interface shape from that reported in \cite{lister} in the far-field. A summary of all the cases with relevant scaling exponents in each case is given in figure \ref{fig:similarity_comparison}.

It may be intuitive to conclude that if a concentration-dependent surface viscosity is present, the self-similar scalings will revert to those obtained for $Bo = 0$, since $\Gamma\rightarrow 0$ near the rupture location. Though the same is suggested by \cite{matar}, our analysis shows that this is not entirely true and obvious, for there is a finite surfactant concentration gradient at the rupture location, even though $\Gamma\rightarrow 0$. Our self-similarity analysis for the variable surface viscosity model results in scalings that lie between the regimes $Bo = 0$ (zero surface viscosity) and $\beta = 0$ (constant surface viscosity). The reason underlying this result is further illustrated by fig. \ref{fig:blu} showing the temporal evolution of the surface viscosity at the rupture point for $\beta = 0$ and $\beta=0.1$ (concentration driven surface viscosity). Even though the surface viscosity starts with a higher value than the constant viscosity case, the effect decreases near the rupture point for $\beta>0$. The interface develops towards a more cusp-like profile (closer to profiles for $Bo = 0$) decreasing the surface viscosity even below than that of the constant surface viscosity case. The same may also be understood by comparing fig. \ref{hvk1} that exhibits a flatter profile with fig. \ref{hvk2} that exhibits a profile with sharper cusps. For the extreme case of $\beta =1$, surface viscosity effects are suppressed completely at the rupture point and a cusp-like profile of a clean interface, $\Gamma=0$ (or of a particle-laden interface with no surface viscosity, $Bo=0$) is recovered. This also means that however large the initial surface viscosity we choose, we can get a cusp-like profile if we have $\beta=1$. Self-similarity is not seen in simulations with NPM in the jammed limit due to the nature of the model. 

The scaling exponents discussed above and given in figure \ref{fig:similarity_comparison} also reveal the dominant balance of forces in the vicinity of rupture. This can be determined by substituting the exponents into the similarity equations (\ref{self2}-\ref{self4}) for various cases. With $Bo=0$ (case (i)) and $\beta=1$ (case (ii)), the dominant balance is between inertia, van der Waals and viscous forces and the interface forms a cusp during rupture. In the case of $0\le\beta <1$ (cases (iii) and (iv)), the dominant balance is between inertia, van der Waals and surface viscous forces. Viscous forces become sub-dominant as one approaches the rupture point. In all the above cases, capillary and Marangoni forces play negligible role near rupture consistent with observations in earlier studies. The above dominant balance was further verified by comparing the magnitude of each term in the evolution equation \eqref{nlec}.

It is interesting to note that for cases (iii) and (iv), another set of consistent scaling exponents are also found. For case (iii), these are $i=l=3/4, j=1/2, k=-9/8$ and for case (iv), we get $i=l=1,j=1/2,k=-3/2$. The latter set of exponents were not reported by \cite{matar} perhaps ignored due to the fact that a large value of Peclet number was used in their study. As a result, the algebraic equation arising out of surfactant diffusion term, $1-2j=0$, in equation \eqref{self3} was ignored. Nevertheless, it is found that the above scaling exponents do not affect the dominant balance of forces near rupture. It is likely that these new exponents will change the self-similar profiles reported in figure \ref{fig:self-similarity}, but a detailed investigation on this will be considered in a future study.

In the next section, we investigate nonlinear evolution of the interface at high concentrations using nonlinear surface viscosity model.

%%%%%%%%%%%%%%%%%%%%%%%%%%%%%%%%%%%%%%%%%%%%%%%%%%%%%%%%%%%%%%%%%%%%%%%%%%%%%%%%%%%%
%%%%%%%%%%%%%%%%%%%%%%%%%%%%%%%%%%%%%%%%%%%%%%%%%%%%%%%%%%%%%%%%%%%%%%%%%%%%%%%%%%%%
%%%%%%%%%%%%%%%%%%%%%%%%%%%%%%%%%%%%%%%%%%%%%%%%%%%%%%%%%%%%%%%%%%%%%%%%%%%%%%%%%%%%
%%%%%%%%%%%%%%%%%%%%%%%%%%%%%%%%%%%%%%%%%%%%%%%%%%%%%%%%%%%%%%%%%%%%%%%%%%%%%%%%%%%%
%%%%%%%%%%%%%%%%%%%%%%%%%%%%%%%%%%%%%%%%%%%%%%%%%%%%%%%%%%%%%%%%%%%%%%%%%%%%%%%%%%%%
%%%%%%%%%%%%%%%%%%%%%%%%%%%%%%%%%%%%%%%%%%%%%%%%%%%%%%%%%%%%%%%%%%%%%%%%%%%%%%%%%%%%
%%%%%%%%%%%%%%%%%%%%%%%%%%%%%%%%%%%%%%%%%%%%%%%%%%%%%%%%%%%%%%%%%%%%%%%%%%%%%%%%%%%%
%%%%%%%%%%%%%%%%%%%%%%%%%%%%%%%%%%%%%%%%%%%%%%%%%%%%%%%%%%%%%%%%%%%%%%%%%%%%%%%%%%%%
%%%%%%%%%%%%%%%%%%%%%%%%%%%%%%%%%%%%%%%%%%%%%%%%%%%%%%%%%%%%%%%%%%%%%%%%%%%%%%%%%%%%
%%%%%%%%%%%%%%%%%%%%%%%%%%%%%%%%%%%%%%%%%%%%%%%%%%%%%%%%%%%%%%%%%%%%%%%%%%%%%%%%%%%%
%%%%%%%%%%%%%%%%%%%%%%%%%%%%%%%%%%%%%%%%%%%%%%%%%%%%%%%%%%%%%%%%%%%%%%%%%%%%%%%%%%%%
%%%%%%%%%%%%%%%%%%%%%%%%%%%%%%%%%%%%%%%%%%%%%%%%%%%%%%%%%%%%%%%%%%%%%%%%%%%%%%%%%%%%
%%%%%%%%%%%%%%%%%%%%%%%%%%%%%%%%%%%%%%%%%%%%%%%%%%%%%%%%%%%%%%%%%%%%%%%%%%%%%%%%%%%%
\subsection{`Jammed' limit analysis}\label{sec:NLAjammed}
\begin{figure}
\centering
\includegraphics[trim= 12mm 82mm 20mm 90mm ,clip, width=0.45\textwidth]{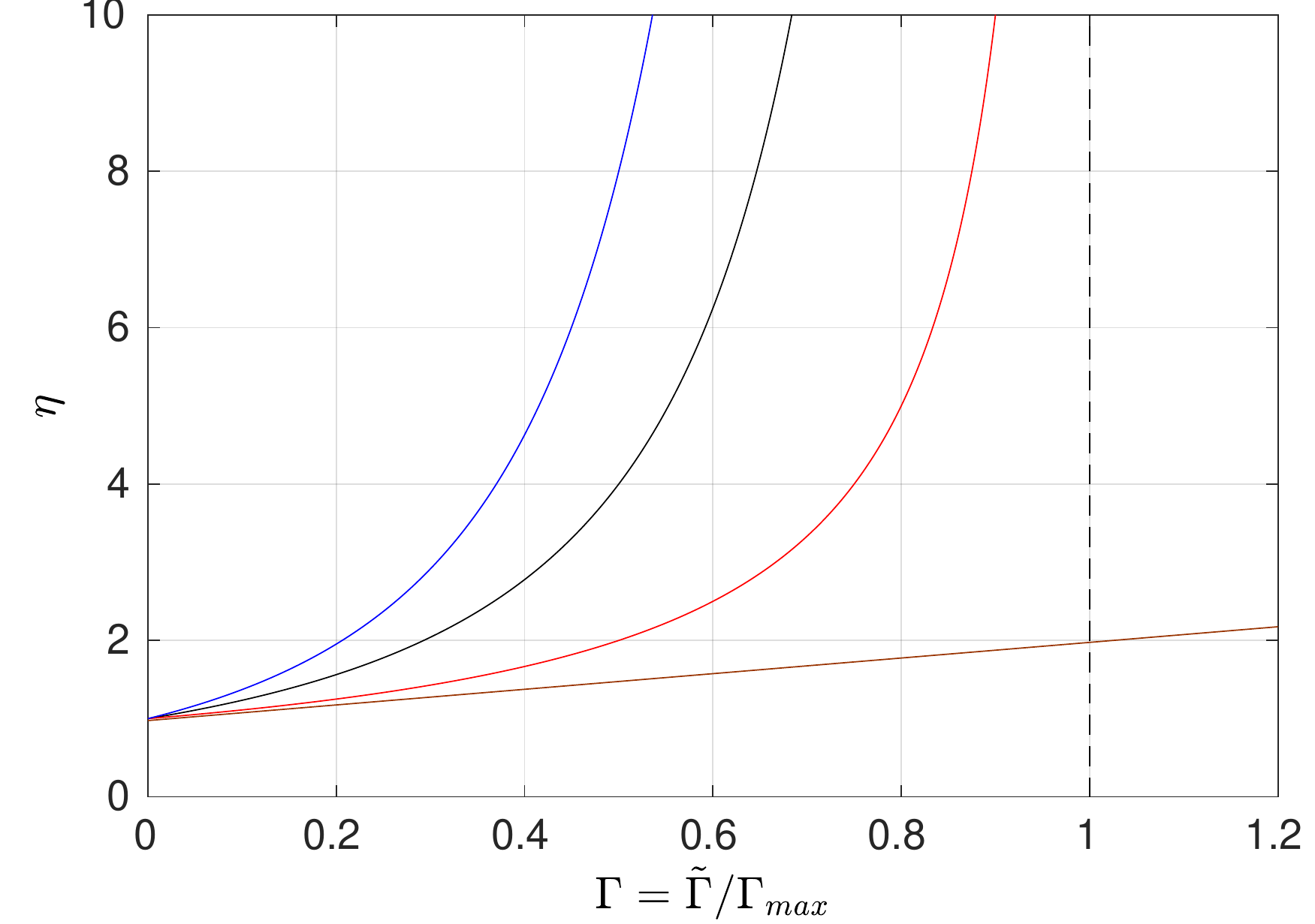}
\caption{Equation (\ref{eta2}) plotted for three different values of exponent, $\alpha=1,2,3$. The higher the value of the exponent, faster the divergence of surface viscosity with concentration. In the dilute limit, the three models approach the linear model, albeit with different slopes.}
\label{fig8:a}
\end{figure}
\begin{figure}
\centering
\subfigure[]{\label{fig9:a}\includegraphics[trim= 12mm 84mm 24mm 84mm ,clip, width=0.45\textwidth]{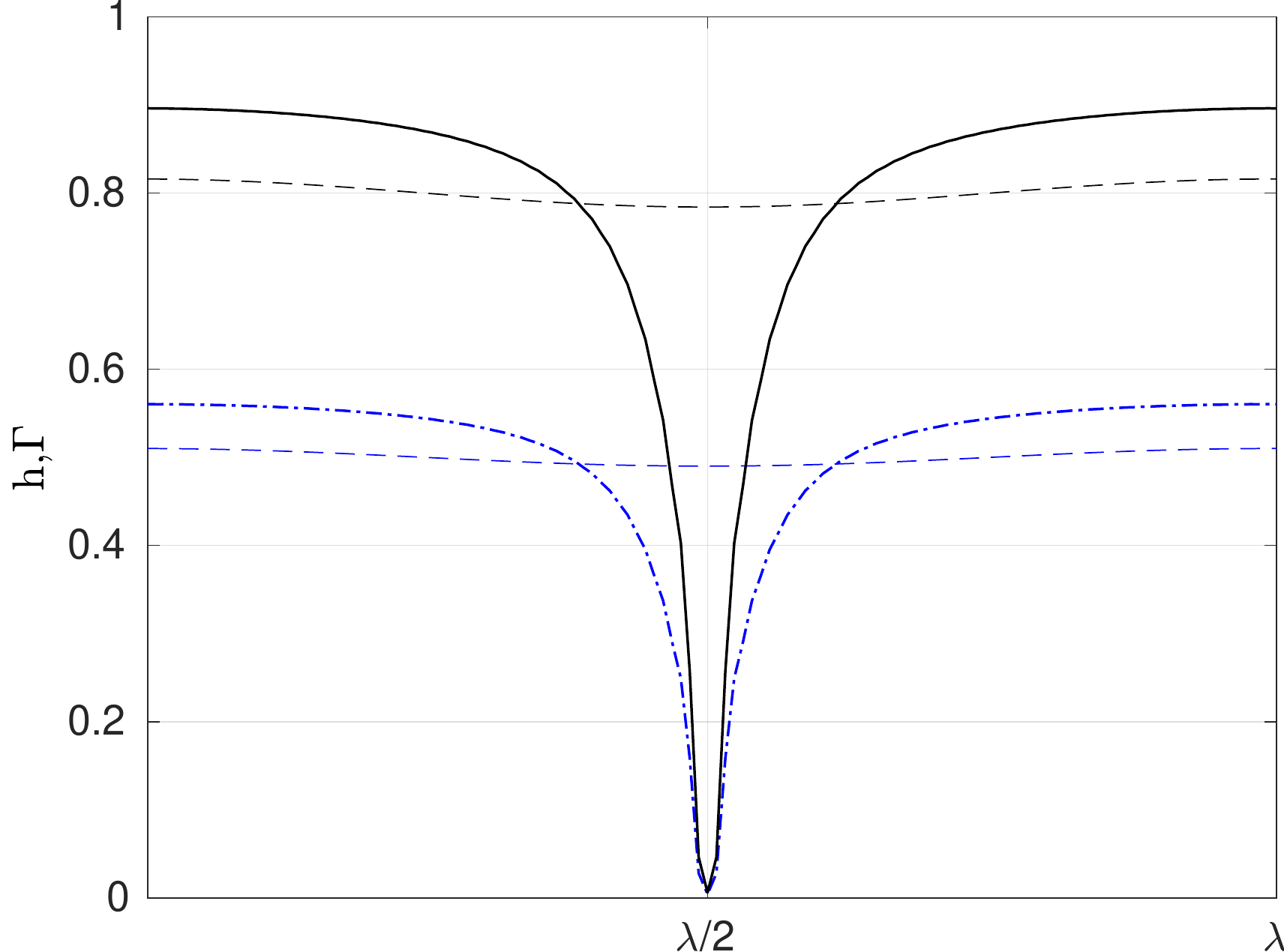}}
\subfigure[]{\label{fig9:b}\includegraphics[trim= 12mm 84mm 24mm 84mm ,clip, width=0.45\textwidth]{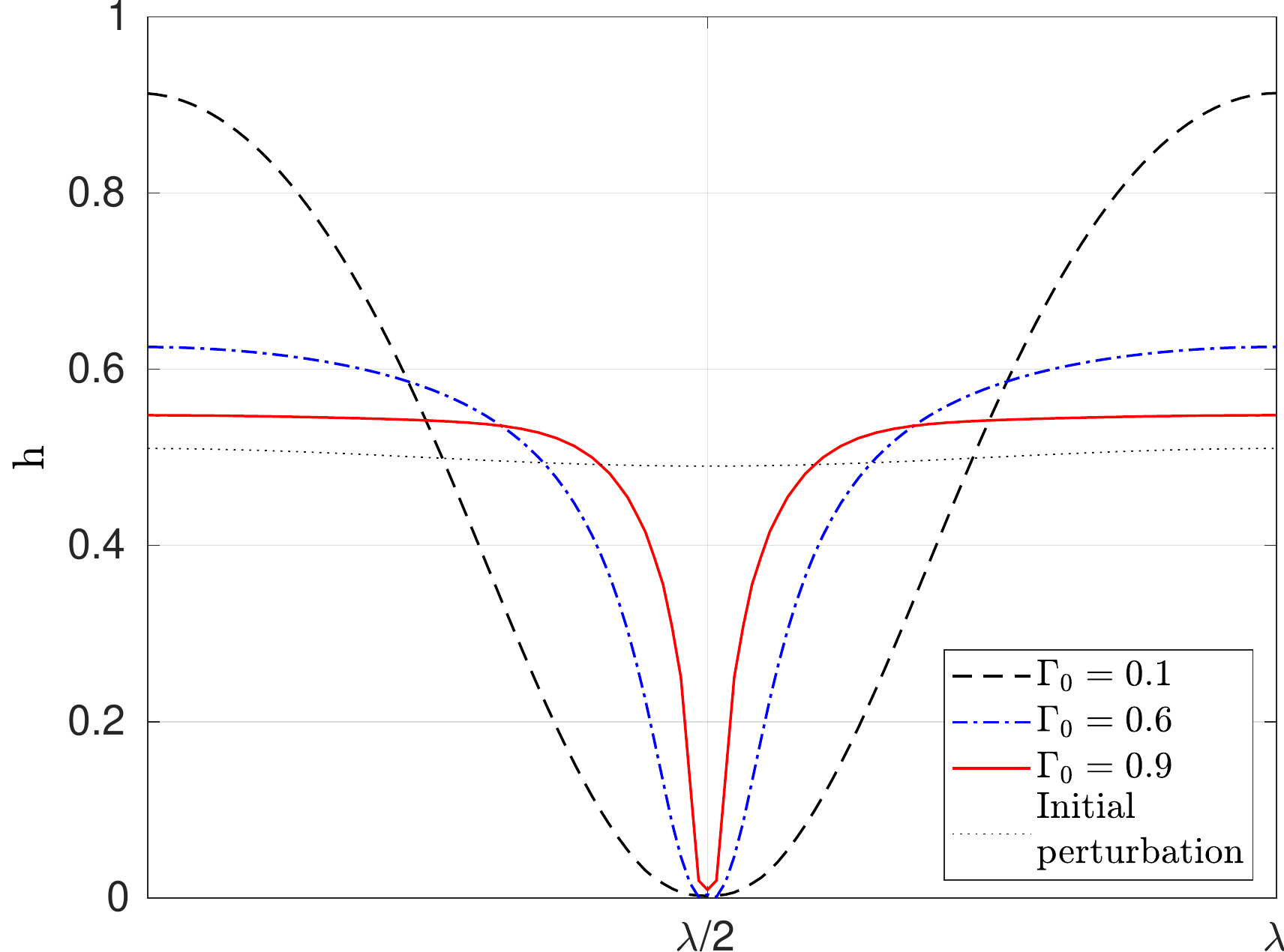}}
\subfigure[]{\label{fig10:a}\includegraphics[trim= 12mm 76mm 24mm 84mm ,clip, width=0.45\textwidth]{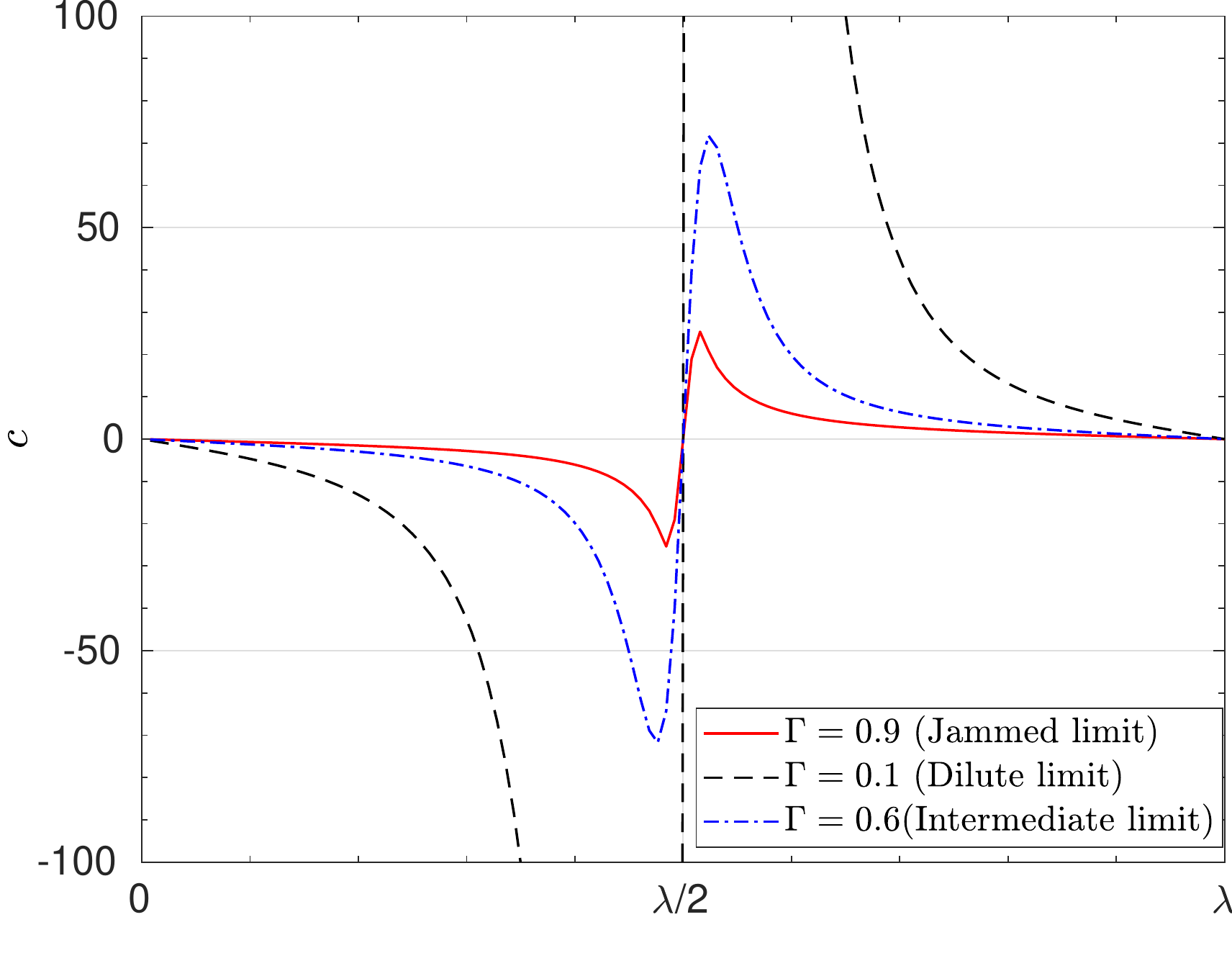}}
\subfigure[]{\label{fig10:b}\includegraphics[trim= 8mm 76mm 24mm 84mm ,clip, width=0.45\textwidth]{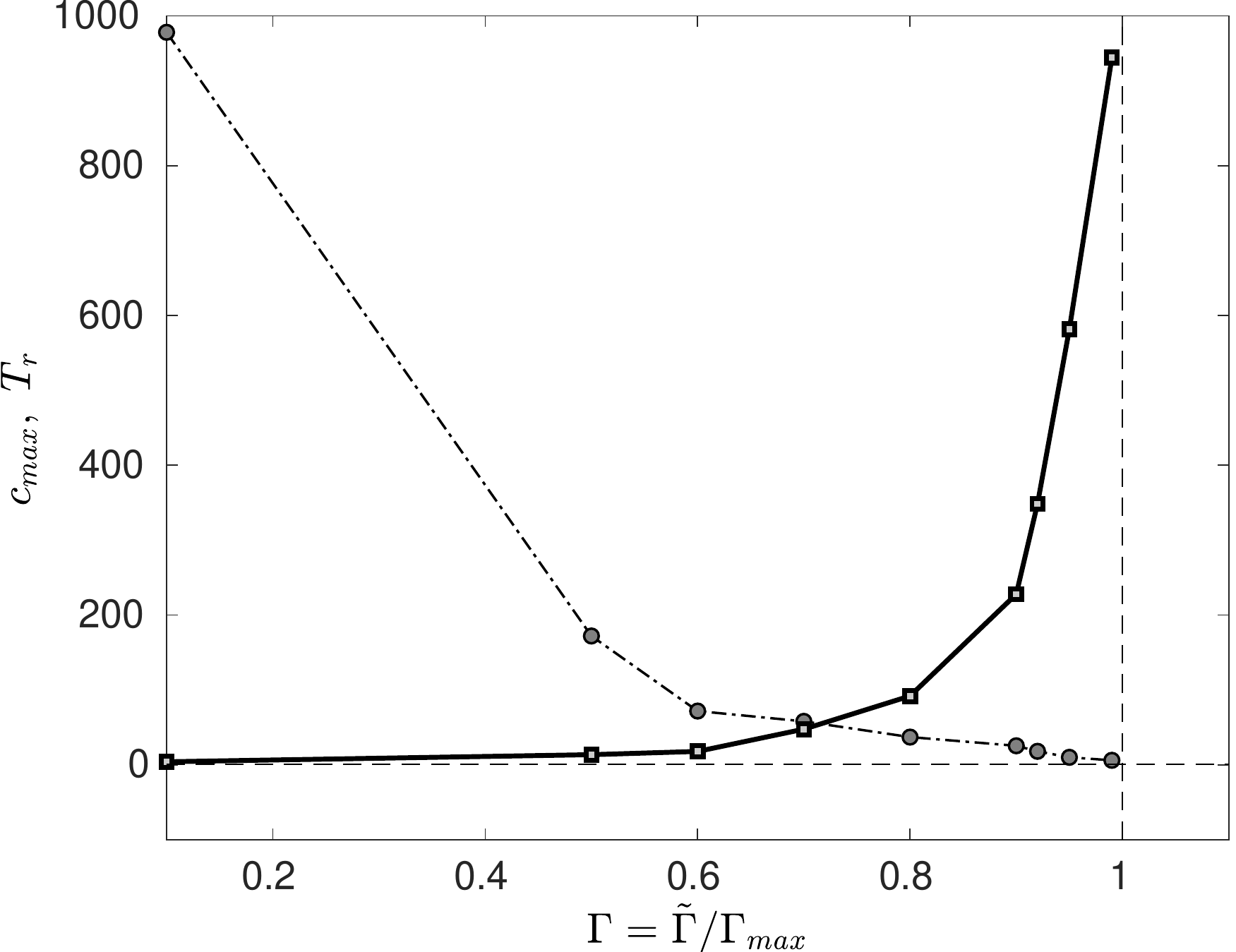}}
\caption{ Evolution of film profiles using NPM obtained by solving equations \eqref{nlea}, \eqref{nleb} and \eqref{nled}. (a) Height profiles (solid) and surfactant profile (dot-dash) at the instance of rupture for $\Gamma_0^{(nl)}=0.8$. The dashed lines represent the initial perturbation. (b) Height profiles for three different values of surfactant concentration ranging from dilute ($\Gamma_0^{(nl)}=0.1$), intermediate ($\Gamma_0^{(nl)}=0.6$) and jammed limit ($\Gamma_0^{(nl)}=0.9$) occurring at three different rupture times, $T_r = 4.051, 17.866, 227.843$ respectively. The interface profile is remarkably different in the three cases, so is the rupture time. (c) Corresponding Tangential velocities at rupture, $c(x,T_{rup})$, for varying $\Gamma_0^{(nl)}$, plotted just before rupture. (d) Maximum tangential velocity $c_{max}$ (dot-dash) and rupture time, $T_r$ (solid) plotted for various values of $\Gamma_0^{(nl)}$. Other parameters are fixed at $\Pen=1,\mathbb{D}=10^{-3},\Rey=10^{-2},\textit{M}=10^{-1},\textit{Bo}=1$.}
\end{figure}
The rigidification of the interface in the limit of jammed state concentration of surfactants is perhaps the most interesting aspect of this work. The notion that films at jammed limit are very stable as suggested by linear stability analysis can be visualised better in nonlinear simulations using the nonlinear phenomenological model (NPM) for surface viscosity given by (\ref{eta2}). Figure \ref{fig8:a} illustrates the variation of $\eta$ for NPM (\ref{eta2}) for various values of $\alpha$. Higher values of $\alpha$ causes $\eta$ to diverge at lower values of $\Gamma$. To compare film profiles for both the models in dilute limit, the values of $\alpha,~\beta,~\tilde{\Gamma}_0$ and $\tilde{\Gamma}_{max}$ are chosen so as to satisfy the relation (\ref{etanonlinear}). Both the profiles appear to be in agreement in this limit, as shown in Appendix \ref{etapp}, figure \ref{fig:appb}. Having validated the code using NPM, we now study the effect of high surfactant concentration for which NPM is ideally suited. The height profile and the surfactant concentration profile for the particular value $\Gamma_0^{(nl)}=0.8$ and $\alpha=2$ of NPM are shown in fig. \ref{fig9:a}. The film stiffens and tends to be immobile all along the interface except in a narrow region at the centre of the domain where height and concentration profiles are at a minimum. Unlike in earlier cases, the film approaches rupture in a very narrow region. This evolution quickly deviates from a linear evolution (not shown) which mainly follows a smooth amplifying sinusoidal structure. A better way to demonstrate the difference in film shape in NPM is to plot the film height profile for different initial surfactant concentrations. Figure \ref{fig9:b} compares film shapes at rupture for three different initial concentrations: the dilute limit at $\Gamma_0^{(nl)}=0.1$, an intermediate limit at $\Gamma_0^{(nl)}=0.6$ and a near-jammed state limit at $\Gamma_0^{(nl)}=0.9$. The profiles in the dilute limit exhibit a large radius of curvature and is similar to what is in earlier sections. In the near-jammed limit, rupture occurs over a very narrow region within the interface. It has to be noted that the interface flattens at the point of rupture but has a very small radius of curvature. The case of $\Gamma_0^{(nl)} = 1$ is a singular limit and cannot be solved numerically. This is consistent with the perturbation solution carried out in section \S\ref{sec:LSAjammed}. Moreover, rupture time diverges with concentration as shown in figure \ref{fig10:b}.

A particularly interesting result is the evolution of tangential velocity, $c(x,t)$, on the interface with varying concentration as shown in figure \ref{fig10:a}. It is clear that the maximum tangential velocities decreases and asymptotes towards zero with increasing concentration, suggesting a transition from a mobile interface for $\Gamma_0^{(nl)}=0$ to an immobile interface at jammed state. Such a surface viscosity-driven transition from partially mobile to immobile interfaces have also been reported by \cite{iva}. For drainage of a thin liquid film between two gas bubbles, interfacial resistance arising from surface viscosity have been reported. Consequently, the kinetics of film break-up is also found to slow down drastically with increase in concentration, and diverges to infinity as $\Gamma_0^{(nl)}\to 1$, suggesting the formation of a solid-like rigid interface which does not deform.

A key result of the paper is summarised in figure \ref{fig10:b} which shows the transition of the interface from mobile to immobile limit associated with a divergence of rupture time as surfactant concentration approaches the jamming limit. This result offers a simply way to model fluid-fluid systems with variable surface viscosity effect to describe a smooth transition from free-slip regime to no-slip region such as in the case of a surfactant-laden sedimenting drop (see \cite{leal}). Recent experiments by \cite{basa2018} on an evaporating droplet containing polystyrene and soft micro-gel particles reveal that micro-gel particles form an immobile interface. As a result, mean-squared displacement of adsorbed polystyrene particles reveal a transition from liquid-like to solid-like behaviour. This effect can potentially be modelled using a nonlinear surface viscosity model used in the current study.

%%%%%%%%%%%%%%%%%%%%%%%%%%%%%%%%%%%%%%%%%%%%%%%%%%%%%%%%%%%%%%%%%%%%%%%%%%%%%%%%%%%%%%%%%%%%%%%%%%%%%%%%%%%%%%%%%%%%%%%%%%%%%%%%%%%%%%%%%%%%%%%%%%%%
%%%%%%%%%%%%%%%%%%%%%%%%%%%%CONCLUSIONS%%%%%%%%%%%%%%%%%%%%%%%%%%%%%%%%%%%%%%%%%%%%%%%%%%%%%%%%%%%%%%%%%%%%%%%%%%%%%%%%%%%%%%%%%%%%%%%%%%%%%%%%%%%%%%%%%%%%%CONCLUSIONS%%%%%%%%%%%%%%%%%%%%%%%%%%%%%%%%%%%%%%%%%%%%%%%%%%%%%%%%%%%%%%%%%%%%%%%%%%%%%%%%%%%%%%%%%%%%%%%%%%%%%%%%%%%%%%%%%%%%%%%%%%%%%%%%%%%%%%%%%%%%%%%%%%%%%%%
\section{Summary and Discussions}\label{sec5}
We have formulated a unified model that includes various interfacial effects due to the presence of surface active agents in a thin free film. Apart from intermolecular forces and Marangoni effect, it is shown that surface viscosity effects play a key role in determining the stability of a thin film. It is well known that Marangoni effects due to surfactant driven surface tension gradients stabilizes a thin film, but this stabilization does not explain the long shelf-life of certain surfactant-stabilized emulsions and Pickering emulsions. Marangoni stabilization only tends to delaying rupture by a small factor. We have shown that concentration-dependent surface viscosity mediated stabilization is a missing element in most earlier studies and offers a theoretical tool to analyse long shelf-life of such emulsions. As a canonical problem, stability of a free film is studied and the theory can be easily extended to bounded films and other interfacial problems. To better illustrate the dependence of surface viscosity on surfactant concentration, two distinct phenomenological models are employed. In the linear model, surface viscosity varies linearly with surfactant concentration and is expected to be suited for dilute limits. The nonlinear model allows us to probe the role of surface viscosity profile in greater detail. In the nonlinear model (NPM), surface viscosity varies nonlinearly with concentration such that it diverges at a critical concentration which is termed as the `jammed limit'. The two models are shown to concur in the dilute limit and this lends validity to the nonlinear model at all concentrations.

In the linear model, surface viscosity stabilizes the film with increasing Boussinesq number but does not affect the cut-off wavenumber. On the other hand, NPM drastically alters the stability characteristics of the thin film. The growth rate is found to scale with $\delta = (1-\Gamma_0^{(nl)})^\alpha$ where $0<\Gamma_0^{(nl)}<1$ is the non-dimensional concentration scaled in terms of the `jammed limit' concentration, $\Gamma_{max}$. The parameter $\delta$ can thus be made arbitrarily small near the jamming limit. This leads to a dramatic stabilization of the thin film and significantly delays rupture in the nonlinear simulations. Using standard perturbation techniques, analysis of the dispersion relation yields a simple criteria for film stability. It is found that when $\Gamma_0^{(nl)} > 3 \mathbb{D}/\textit{M}$, instability of the thin film is completely suppressed. This suggests that surfactant concentration can be arbitrarily increased beyond a critical value ($=3\mathbb{D}/\textit{M}$) to achieve permanent stabilization of the film. But in reality, such concentrations are difficult to achieve for realistic parameter values of $\textit{M}$ and $\mathbb{D}$. A key finding of the present paper is that rupture times can be arbitrarily increased by varying surfactant concentration. This stabilization is achieved via surface viscosity effects which increase nonlinearly with surfactant concentration. Large surface viscosities render the interface immobile which helps counteract the rupture process by slowing down film evolution. 

The findings of this work have relevance to many other experimental and theoretical works. Experiments with rising drop/sedimenting bubbles \citep{leal} reveal that surfactants enhance the drag force on a drop/bubble. This effect has been modelled theoretically using a spherical cap approximation \citep{leal, sadh} wherein surfactants are assumed to be in a jammed state on the leeward side of the drop/bubble and are assumed to be absent on the rest of the surface. A no-slip condition is employed on the spherical cap and a stress-free condition on the rest of the surface. NPM employed here leads to a smooth transition from the no-slip to the stress-free condition on the interface and prevents a jump in the interface conditions as was employed by \cite{sadh}. Our results are also consistent with \cite{iva} who found that interface mobility reduces with increasing surfactant concentration. This has direct significance to the stabilty of Pickering emulsions \citep{wasan, lang}. There is sufficient anecdotal evidence in literature on Pickering emulsions \citep{basav} which show that removing adsorbed particles (by changing the pH of the aqueous phase) reduces emulsion stability dramatically. The treatment of the surface active agents as `surfactants' also allows one to look at proteins in food systems as surface active agents \citep{murr}. A few studies have found that truly stable foams can only be formed when coagulated proteins form a rigid solid network on the interface \citep{wals, mart}. Our work serves as a mathematical deduction of the various phenomena described above by modelling the surface viscosity.

A number of interesting features remain unexplored. Though nonlinear surface viscosity effects sucessfully explain the transion from mobile to immobile interfaces, they do not account for elasticity found in particle-laden interfaces. Liquid marbles \citep{quere} and particle-coated bubbles can sustain non-spherical shapes upon compression \citep{stone}. This could be due to anistropic surface tension or elasticity or both. Studies with particle-covered Landau-Levich dip-coating flows \citep{harish2013b} reveal that elastic effects play an important role in explaining experimental power-law dependence between coating thickness and withdrawal speed. It is therefore imperative that future studies incorporate a combination of variable surface viscosity and elasticity to better predict characteristics of particle/surfactant-covered interfacial systems, especially when jamming is expected occur.
\\
\\

An earlier version of this work was presented by one of the authors (A.C.) at the AIChE Annual Meeting 2018 and A.C. thanks Dr. Ashutosh Ricchariya (LVPEI, Hyderabad) and Department of Biotechnology, Govt. of India for travel support. H.N.D and S.K. thank Pavan Pujar from NIT Surathkal who carried out preliminary analysis of this work. The authors gratefully acknowledge Prof. James J. Feng (UBC, Vancouver) and Dr. K. Badarinath (IIT Hyderabad) for their critical reading of the manuscript and constructive suggestions. A.C was supported by a doctoral fellowship from the Ministry of Human Resource Development, Govt. of India and H.N.D thanks partial support from an Early Career Research Award (ECR/2015/000086).

%%%%%%%%%%%%%%%%%%%%%
\appendix%%%%%%%%%%%%%%%%%%%%%%%%%%%
%%%%%%%%%%%%%%%%%%%%%%
\section{Interfacial boundary conditions and $x$-momentum equation}
\subsection{Normal stress balance}
\label{nsapp}
For a 1D interface, the normal stress balance assumes the form: 
\begin{equation}
-\bar{n}\cdot||\mathsfbi{T}||\cdot\bar{n} = 2\kappa\sigma + 2\kappa\left(k^s + \mu^s\right)\bar{\nabla_s}\cdot\bar{u}
\end{equation}
We resolve the $\nabla_s$ curvilinear gradient operator and the tensor products to get a PDE in cartesian coordinates. This is done analytically and validated by the open-source technical computing software, \textit{Mathematica}. The dimensional equation turns out to be:
 \begin{eqnarray}
-\tilde{P} + \frac{1}{(1+\tilde{h}^{2}_{\tilde{x}})^{5/2}}\Big[-\tilde{\sigma}(1+\tilde{h}^{2}_{\tilde{x}})\tilde{h}_{\tilde{x}\tilde{x}} - \mu\big\{    \mathbb{A}\big\}\big\{  \mathbb{B}                  \big\}\Big] = 0\\
where,~~\mathbb{A}=2\sqrt{1+\tilde{h}^{2}_{\tilde{x}}} + 2\tilde{h}^{2}_{\tilde{x}}\sqrt{1+\tilde{h}^{2}_{\tilde{x}}} + \frac{\tilde{\eta}}{\mu}\tilde{h}_{\tilde{x}\tilde{x}}, \nonumber\\
\mathbb{B}= -\tilde{v}_{\tilde{z}} + \tilde{h}_{\tilde{x}}\nonumber\big\{  \tilde{u}_{\tilde{z}} + \tilde{h}_{\tilde{x}}\left(\tilde{v}_{\tilde{z}} + \tilde{v}_{\tilde{x}}\right)\big\}.
\end{eqnarray}
Here the surface dilatational $(k^s)$ and surface shear $(\mu^s)$ viscosities are combined into a parameter $\eta$, since they occur in additive pairs in the equations. This is only observed for a 1D interface. Inserting relevant scalings (\ref{scale}-\ref{sigma}) as discussed in the main text, we observe the pressure-term appears at leading order followed by the capillary-term at $O(\epsilon)$. Hence, the Disjoining pressure number $(\mathbb{D}^{-1})$ (or $\sigma$) is rescaled to retain capillary effects. We obtain the final simplified non-dimensional PDE, which reads:
\begin{equation}
-P + \Big[-\mathbb{D}^{-1}h_{xx} - \epsilon^2\mathbb{D}^{-1}\textit{M}(\Gamma-1) - \big\{  2\left(1+\epsilon^2 h_{x}^{2} \right)+\epsilon^2 \textit{Bo}\frac\eta h_{xx}   \big\}\big\{  -v_z +h_x u_z +\epsilon^2h_{x}^{2}v_z  +\epsilon^3h_{x}^{2}v_x              \big\}\Big] = 0 
\end{equation}
Considering only the leading order terms, we get (\ref{nsbc}) used in deriving the film evolution equations.
%%%%%%%%%%%%%%%%%%%%
\subsection{Tangential stress balance}
\label{tsapp}
For a 1D interface, the tangential stress balance assumes the form: 
\begin{equation}
-\bar{t}\cdot||\mathsfbi{T}||\cdot\bar{n} = \bar{t}\cdot\bar{\nabla_s}\sigma + \left(k^s + \mu^s\right)\bar{t}\cdot\bar{\nabla_s}\bar{\nabla_s}\cdot\bar{u} + \bar{t}\cdot\bar{\nabla_s}\left(k^s + \mu^s\right)\bar{\nabla_s}\cdot\bar{u}
\end{equation}
The dimensional equation turns out to be:
 \begin{eqnarray}
&\tilde{\sigma}_{\tilde{x}}(1+\tilde{h}^{2}_{\tilde{x}})^{2} + \mu\sqrt{1+\tilde{h}^{2}_{\tilde{x}}}\bigg[-\tilde{u}_{\tilde{z}} + \tilde{h}^4_{\tilde{x}}\tilde{u}_{\tilde{z}} - 4\tilde{h}_{\tilde{x}}\tilde{v}_{\tilde{z}}\left(1+\tilde{h}^{2}_{\tilde{x}}\right) + \tilde{v}_{\tilde{x}}\left(\tilde{h}^4_{\tilde{x}} - 1\right)       \bigg] \nonumber\\
&+ \tilde{\eta}_{\tilde{x}}\left(1+\tilde{h}^{2}_{\tilde{x}}\right)\bigg[  -\tilde{v}_{\tilde{z}} + \tilde{h}_{\tilde{x}}\left( \tilde{u}_{\tilde{z}} + \tilde{h}_{\tilde{x}}\tilde{v}_{\tilde{z}} + \tilde{v}_{\tilde{x}}\right) \bigg] + \tilde{\eta} \bigg[   \tilde{h}^5_{\tilde{x}}\tilde{v}_{\tilde{z}\tilde{z}} + \tilde{h}_{\tilde{x}\tilde{x}}\left(\tilde{u}_{\tilde{z}} + \tilde{v}_{\tilde{x}}\right) \nonumber\\
&+ \tilde{h}^2_{\tilde{x}}\left( \tilde{u}_{\tilde{z}\tilde{z}} - \tilde{h}_{\tilde{x}\tilde{x}}\left(\tilde{u}_{\tilde{z}} + \tilde{v}_{\tilde{x}}\right) + \tilde{v}_{\tilde{x}\tilde{z}}  \right) - \tilde{v}_{\tilde{x}\tilde{z}} + \tilde{h}^4_{\tilde{x}}\left( \tilde{u}_{\tilde{z}\tilde{z}} + 2\tilde{v}_{\tilde{x}\tilde{z}}\right) + \tilde{h}^3_{\tilde{x}}\left( \tilde{u}_{\tilde{x}\tilde{z}} + \tilde{v}_{\tilde{x}\tilde{x}} \right) \nonumber\\
&+  \tilde{h}_{\tilde{x}}\left(4\tilde{h}_{\tilde{x}\tilde{x}}\tilde{v}_{\tilde{z}} - \tilde{v}_{\tilde{z}\tilde{z}}\right) + \tilde{u}_{\tilde{x}\tilde{z}} + \tilde{v}_{\tilde{x}\tilde{x}} \bigg] = 0
\end{eqnarray}
Inserting relevant scalings and simplifying, we obtain the final non-dimensional PDE:
\begin{eqnarray}\label{fots}
&\mathbb{D}^{-1}\sigma_x - u_z +(\epsilon h_x)^4 u_z -4\epsilon^2 h_x v_z\left( 1+ (\epsilon h_x)^2\right) - \epsilon^2 v_x + \epsilon^6 h^4_x v_x \nonumber\\
&+ \epsilon^2 \mathbf{\textit{Bo}}\eta_x\left(  -v_z + h_x u_z +\epsilon^2 h_x v_z + \epsilon^2 v_x  \right) + \epsilon^2\mathbf{\textit{Bo}}\eta\Big( \epsilon^4 h^5_x + h_{xx} u_z + \epsilon^2 h_{xx} v_x - v_{xz} \nonumber \\
&-\epsilon^2 h_{xx} h^2_x u_z -\epsilon^4 h_{xx} h^2_x v_x + \epsilon^2 h^2_x v_{xz} \left(1+\epsilon^2  h^2_x \right) + h^2_x u_{zz}\left(1+\epsilon^2  h^2_x \right) + h_x u_{xz}\left(1+\epsilon^2  h^2_x \right) \nonumber\\
&+ \epsilon^2 h_x v_{xx}\left(1+\epsilon^2  h^2_x \right) + 4\epsilon^2 h_x h_{xx} v_z - h_x v_{zz}  \Big) = 0.
\end{eqnarray}
The above equation reduces to (\ref{tsbc}) when the leading order terms are considered. The first order corrections of this equation used to derive (\ref{nlec}) or (\ref{nled}) can also be realized when $O(\epsilon^2)$ terms are considered. As can be observed clearly, the surface viscosity enters the equations in the first order corrections of tangential stress boundary condition, and has highly nonlinear coupling with the height and various velocity gradient terms.
%%%%%%%%%%%%%%%%%%%
\subsection{$x$-momentum equation}
\label{xmapp}
The non-dimensional $x$-momentum equation valid to $O(\epsilon^2)$ take the form,
\begin{eqnarray}
u_{zz} = \epsilon^2\mathbf{\Rey}\left(u_{t} + uu_x + vu_z\right) + \epsilon^2\left(P+\phi\right)_x - \epsilon^2u_{xx}
\end{eqnarray}
Expanding all terms in powers of $\epsilon^2$, the leading order equation becomes
\begin{equation}\label{fox}
 u^{(0)}_{zz} = 0,
\end{equation}
and the first order equation at $O(\epsilon^2)$ assumes the form
\begin{equation}\label{fox1}
u^{(1)}_{zz} = \epsilon^2\mathbf{\Rey}\left(u^{(0)}_{t} + u^{(0)}u^{(0)}_x + v^{(0)}u^{(0)}_z\right) + \epsilon^2\left(P^{(0)}+\phi^{(0)}\right)_x - \epsilon^2u^{(0)}_{xx},
\end{equation}
where superscripts $0$ and $1$ denote leading and first order correction terms respectively in the asymptotic expansion (\ref{nsx}). Integrating (\ref{fox1}) and using symmetry boundary condition at $z=0$ (see eqn. \ref{sym}), we obtain an expression for $u^{(1)}_{z}$ which when compared to the first order correction of (\ref{fots}), gives the third nonlinear evolution equation (\ref{nlec}) (or (\ref{nled})).
%%%%%%%%%%%%%%%%%%%%%%%%%%%%%%%%%%%%%%%%%%%%%%%%%%%%%%%
%%%%%%%%%%%%%%%%%%%%%%%%%%%%%%%%%%%%%%%%%%%%%%%%%%%%%%%
\section{Notes on the validity of scalings for $\tilde{\eta}_s$}
\label{etapp}
The linear model adds small corrections to the surface viscosity to exhibit weak relationship with surfactant concentration. We may consider it to be apt when the surfactants are sparsely distributed on the interface (dilute limit):
\begin{eqnarray}
\label{etal}
\tilde{\eta}^{(l)}=\left.\tilde{\eta}_{0}\right|_{\tilde{\Gamma}=\tilde{\Gamma}_{0}} +\left.\frac{\partial\tilde{\eta}}{\partial\tilde{\Gamma}}\right|_{\tilde{\Gamma}=\tilde{\Gamma}_{0}}\left(\tilde{\Gamma}-\tilde{\Gamma}_{0}\right)\nonumber\\
\Rightarrow\tilde{\eta}^{(l)}=\tilde{\eta}_{0}\Bigg[1+\beta\left(\frac{\tilde{\Gamma}-\tilde{\Gamma}_{0}}{\tilde{\Gamma}_{0}}\right)\Bigg],
\end{eqnarray}
where, $\tilde{\eta}_{0}$ is the surface viscosity at a surfactant concentration within dilute limit $\tilde{\Gamma}=\tilde{\Gamma}_{0}$. The definition of $\beta$ as given in Table \ref{table:dimnumbers} can also be realised in the above equation. The surface viscosity for NPM is defined as,
\begin{eqnarray}\label{etanl}
\tilde{\eta}^{(nl)}=\left.\tilde{\eta}_{ref}\right|_{\tilde{\Gamma}=0} \left(1-\frac{\tilde{\Gamma}}{\tilde{\Gamma}_{max}}\right)^{-\alpha}.
\end{eqnarray}
which diverges at $\tilde{\Gamma}\to\tilde{\Gamma}_{max}$ and reduces to a linear model when expanded in the limit $\tilde{\Gamma}\ll\tilde{\Gamma}_{max}$:
\begin{eqnarray}\label{etanl2}
\tilde{\eta}^{(nl)}=\tilde{\eta}_{ref}\left(1+\alpha\frac{\tilde{\Gamma}}{\tilde{\Gamma}_{max}}\right).
\end{eqnarray}
Here, $\tilde{\eta}_{ref}$ is the surface viscosity for a clean interface ($\Gamma\rightarrow 0$). From (\ref{etal}), we also get the surface viscosity for a clean interface as,
\begin{eqnarray}\label{linpm}
\left.\tilde{\eta}^{(l)}\right|_{\tilde{\Gamma}=0} = \tilde{\eta}_{0}\left(1-\beta\right) = \tilde{\eta}_{ref}
\end{eqnarray}
\begin{figure}
\centering
\includegraphics[trim= 12mm 78mm 24mm 84mm ,clip, width=0.55\textwidth]{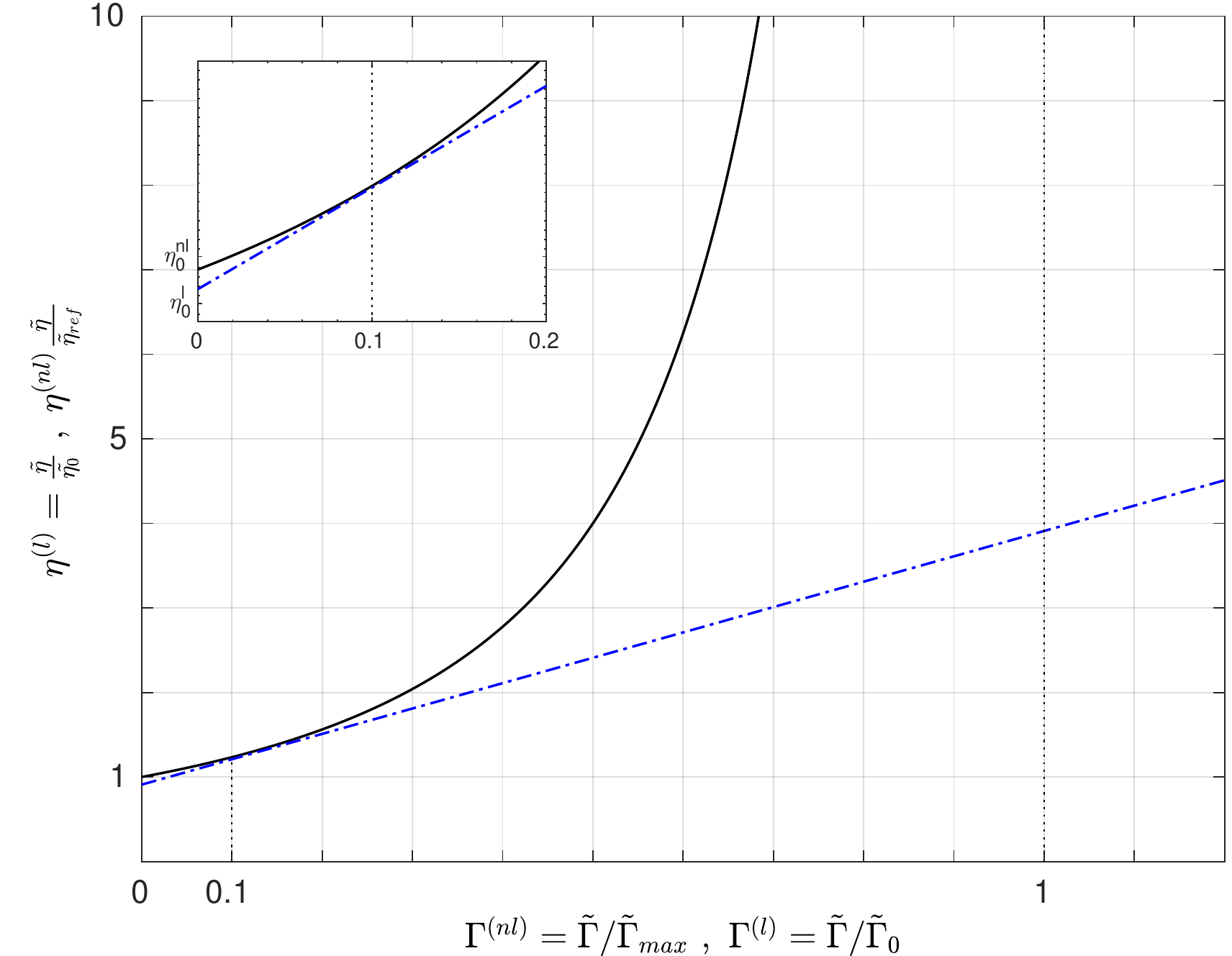}
\caption{Graphical illustration of linear and nonlinear surface viscosity models given in equations \eqref{eta1} and \eqref{eta2}. The dilute limit of nonlinear model (NPM) approaches the linear model as shown in the figure. This comparison serves to define a suitable `dilute' limit of NPM in order to compare growth rate and film evolution curves for the two models. For the specific case of $\beta=10^{-3}$ and $\alpha=2$, the two models have the same `effective' non-dimensional surface viscosity for $\Gamma_0^{(nl)}=0.1$. [Inset] Zoomed in version for $\Gamma^{(nl)}=0-0.2$ (dilute limit).}
\label{fig:app}
\end{figure}
Comparing (\ref{etal}) and (\ref{etanl2}), we get an equation to relate the model parameters at a dilute concentration. This is visualized in fig. \ref{fig:app}. Another necessary condition is satisfied by equating the slopes (or derivatives of (\ref{etal}) and (\ref{etanl2})):
\begin{eqnarray}\label{etanapp}
\tilde{\eta}_{ref}\frac{\alpha}{\tilde{\Gamma}_{max}} = \tilde{\eta}_{0}\frac{\beta}{\tilde{\Gamma}_{0}}
\end{eqnarray}
Using (\ref{linpm}) in the above relation, we obtain the equivalence condition (\ref{etanonlinear}). Note the advantage of defining the linear model for an arbitrary $\tilde{\Gamma}_{0}$ and not for $\tilde{\Gamma}_{0}=0$ (clean interface). This definition renders the linear model valid about any value of reference surfactant concentration, $\tilde{\Gamma}_0$ when $\beta$ at that concentration is known. Further, to relate the linear model with NPM, (\ref{etanapp}) will change accordingly, as now the NPM cannot be expanded in the limit of $\tilde{\Gamma}\ll\tilde{\Gamma}_{max}$.

\begin{figure}
\centering
\includegraphics[trim= 12mm 84mm 24mm 82mm ,clip, width=0.55\textwidth]{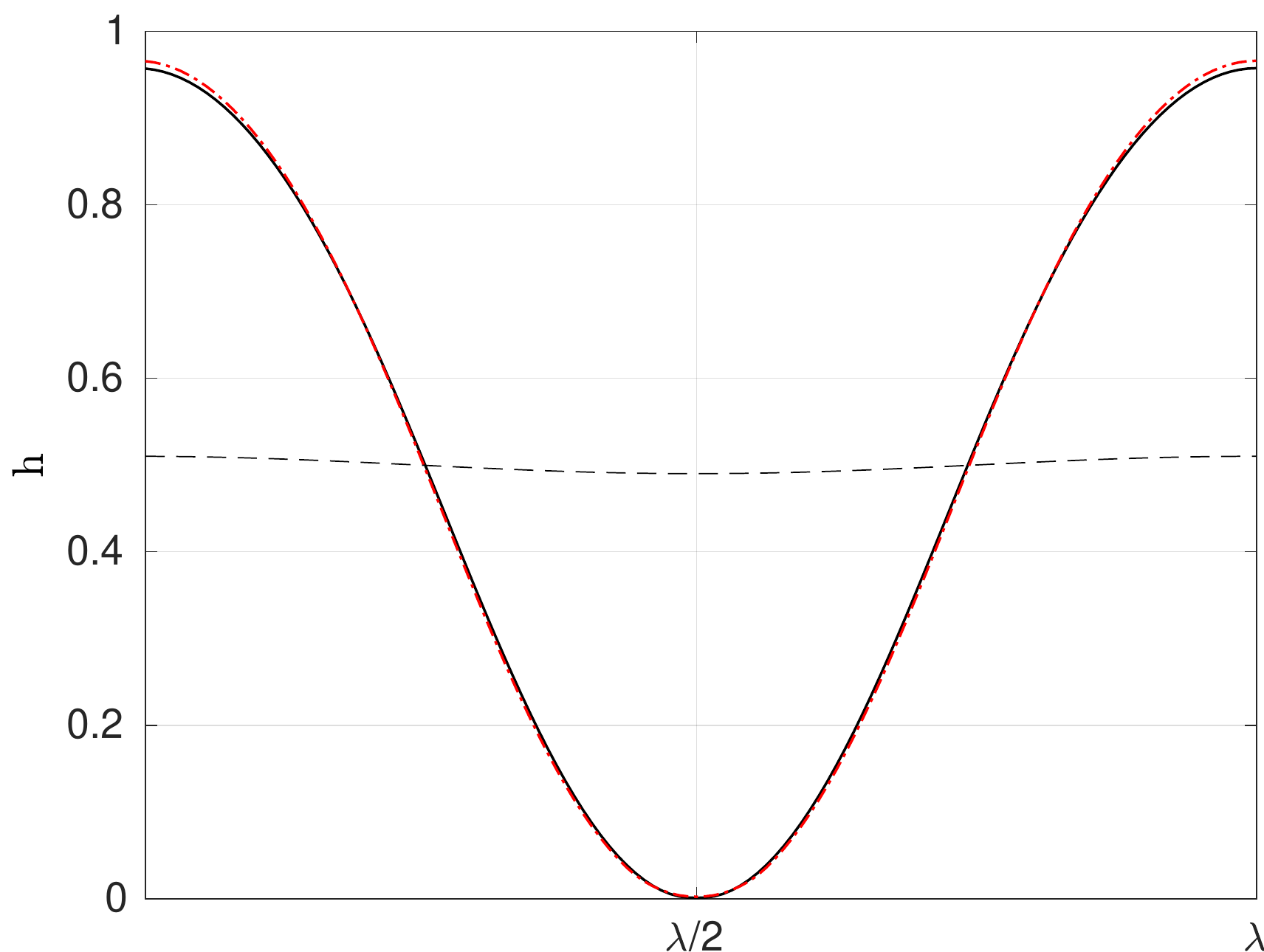}
\caption{Comparison of height profile at rupture between the linear model and NPM in the dilute limit. For the linear model, $\Gamma_0^{(l)}=1.8,\beta=0.1,T_r=3.588$. For NPM, $\Gamma_0^{(nl)}=0.1,\alpha=2,T_r=3.613$. Other parameters are fixed $\Pen=1,\mathbb{D}=10^{-3},\Rey=10^{-2},\textit{M}=10^{-1},\textit{Bo}=1$.}
\label{fig:appb}
\end{figure}

Having estabilished the criteria for comparing the linear and nonlinear models in the dilute limit, we show that this comparison is robust and yields identical film profiles even in the nonlinear regime as shown in figure \ref{fig:appb} which shows the comparison of height profiles from the two models at the instance of rupture.

%%%%%%%%%%%%%%%%%%%%%%%%%%%%%%%%%%%%%%%%%

\bibliography{references}
\bibliographystyle{jfm}

\end{document}